\documentclass[prd,nofootinbib,preprint,superscriptaddress,twocolumn,10pt]{revtex4}
\pdfoutput=1
\usepackage{aas_macros}
\usepackage{newtxtext,newtxmath}
\usepackage{hyperref}
\usepackage{booktabs}
\usepackage{mathtools}
\usepackage{cleveref}
\usepackage{makecell}
\usepackage{subfigure}
\usepackage{bm}
\usepackage{ulem}
\usepackage{orcidlink}
\newcommand{\be}{\begin{equation}}
\newcommand{\ee}{\end{equation}}
\newcommand{\ba}{\begin{eqnarray}}
\newcommand{\ea}{\end{eqnarray}}
\newcommand{\bi}{\begin{itemize}}
\newcommand{\ei}{\end{itemize}}

\usepackage{color}

\newcommand{\reffig}{Fig.\ref}

\usepackage{xcolor}
\DeclareRobustCommand{\VAN}[3]{#2}
\let\VANthebibliography\thebibliography
\def\thebibliography{\DeclareRobustCommand{\VAN}[3]{##3}\VANthebibliography}
\usepackage{graphicx}
\usepackage{grffile}

\begin{document}
\title{Weak Lensing Reconstruction by Counting Galaxies: Improvement with DES Y3 Galaxies}

\author{Jian Qin\orcidlink{0000-0003-0406-539X}}
\email{qinjian@sjtu.edu.cn}
\affiliation{Department of Astronomy, School of Physics and Astronomy, Shanghai Jiao Tong University, Shanghai, 200240,China}
\affiliation{Key Laboratory for Particle Astrophysics and Cosmology
(MOE)/Shanghai Key Laboratory for Particle Physics and Cosmology,China}

\author{Pengjie Zhang\orcidlink{0000-0003-2632-9915}}
\email{zhangpj@sjtu.edu.cn}
\affiliation{Department of Astronomy, School of Physics and Astronomy, Shanghai Jiao Tong University, Shanghai, 200240,China}
\affiliation{Tsung-Dao Lee Institute, Shanghai Jiao Tong University, Shanghai, 200240, China}
\affiliation{Key Laboratory for Particle Astrophysics and Cosmology
(MOE)/Shanghai Key Laboratory for Particle Physics and Cosmology,China}

\author{Yu Yu\orcidlink{0000-0002-9359-7170}}
\email{yuyu22@sjtu.edu.cn}
\affiliation{Department of Astronomy, School of Physics and Astronomy, Shanghai Jiao Tong University, Shanghai, 200240,China}
\affiliation{Key Laboratory for Particle Astrophysics and Cosmology
(MOE)/Shanghai Key Laboratory for Particle Physics and Cosmology,China}

\author{Haojie Xu}
\affiliation{Shanghai Astronomical Observatory, Chinese Academy of Sciences, Shanghai 200030, China}
\affiliation{Department of Astronomy, School of Physics and Astronomy, Shanghai Jiao Tong University, Shanghai, 200240,China}
\affiliation{Key Laboratory for Particle Astrophysics and Cosmology
(MOE)/Shanghai Key Laboratory for Particle Physics and Cosmology,China}

\author{Ji Yao}
\affiliation{Shanghai Astronomical Observatory, Chinese Academy of Sciences, Shanghai 200030, China}
\affiliation{Department of Astronomy, School of Physics and Astronomy, Shanghai Jiao Tong University, Shanghai, 200240,China}
\affiliation{Key Laboratory for Particle Astrophysics and Cosmology
(MOE)/Shanghai Key Laboratory for Particle Physics and Cosmology,China}

\author{Yuan Shi}
\affiliation{Department of Astronomy, School of Physics and Astronomy, Shanghai Jiao Tong University, Shanghai, 200240,China}
\affiliation{Key Laboratory for Particle Astrophysics and Cosmology
(MOE)/Shanghai Key Laboratory for Particle Physics and Cosmology,China}

\author{Huanyuan Shan}
\affiliation{Shanghai Astronomical Observatory, Chinese Academy of Sciences, Shanghai 200030, China}

\begin{abstract}

In \citep{Qin+}, we attempted to reconstruct the weak lensing convergence map $\hat{\kappa}$ from cosmic magnification by linearly weighting the DECaLS galaxy overdensities in different magnitude bins of $grz$ photometry bands.
The $\hat{\kappa}$ map is correlated with cosmic shear at 20-$\sigma$ significance.
However, the low galaxy number density in the DECaLS survey prohibits the measurement of $\hat{\kappa}$ auto-correlation.
In this paper, we apply the reconstruction method to the Dark Energy Survey Year 3 (DES Y3) galaxies from the DES Data Release 2 (DR2). 
With greater survey depth and higher galaxy number density, convergence-shear cross-correlation signals are detected with $S/N\approx 9,16,20$ at $0.4<z_\kappa<0.6,0.6<z_\kappa<0.8$ and $0.8<z_\kappa<1.0$ respectively.
More remarkably, the $\hat{\kappa}-\hat{\kappa}$ correlations of the $0.4<z_\kappa<0.6$ and $0.6<z_\kappa<0.8$ bins show reasonably good agreement with predictions based on theoretical interpretation of $\hat{\kappa}-\gamma$ measurement.
This result takes a step further towards the cosmological application of our lensing reconstruction method.

\end{abstract}
\maketitle

\section{Introduction}

Weak gravitational lensing, which probes directly the matter distribution of the Universe, provides powerful insight into dark energy, dark matter and gravity at cosmological scales \citep{2001PhR...340..291B, 2015RPPh...78h6901K,BS2001, HJ2008, van2010a, FuFan2014, Kil2015}. 
One effect of weak lensing is to distort the shapes of galaxy images, commonly referred as cosmic shear \cite{2006ewg3.rept.....P,2006astro.ph..9591A}.
Numerous ongoing telescopes have the main goal of measuring cosmic shear, such as the Dark Energy Survey \citep[DES,][]{DES2016}, the Kilo-Degree Survey \citep[KiDS,][]{KiDS2013}, 
the Hyper Suprime-Cam Subaru Strategic Program survey \citep[HSC-SSP,][]{HSC2018}, 
Vera C. Rubin Observatory's Legacy Survey of Space and Time \citep[LSST,][]{2009arXiv0912.0201L}, Euclid \citep{2011arXiv1110.3193L}, 
and the China Space Station Telescope (CSST) \citep{2019ApJ...883..203G,2024MNRAS.527.5206Y}.
With these powerful surveys, cosmic shear is making significant contributions to the precision cosmology \citep[e.g.,][]{Hamana2020,2021A&A...645A.104A,2021A&A...645A.105G, 2022A&A...665A..56L,PhysRevD.105.023514,PhysRevD.105.023515, 2023arXiv230400702L}.

Another, however less explored, weak lensing effect is cosmic magnification. 
It affects the large-scale structure by altering the observed flux of sources and changing the solid angle of sky patches \citep{BN1992, BN1995}.
This effect can be observed through changes in number count, magnitude, and size \citep{2010MNRAS.405.1025M,2011MNRAS.411.2113J,Schmidt_Leauthaud_Massey_Rhodes_George_Koekemoer_Finoguenov_Tanaka_2012,2014ApJ...780L..16H,Duncan_Heymans_Heavens_Joachimi_2016,2018MNRAS.476.1071G}.
Similar to how the cosmic shear signal can be contaminated by galaxy intrinsic alignment, the cosmic magnification signal is overwhelmed by galaxy intrinsic clustering.
In observations, cosmic magnification is typically detected through cross-correlations of two samples within the same sky area but at significantly different redshifts. 
Lower redshift lenses include luminous red galaxies (LRGs) and clusters \citep[e.g.,][]{pub.1059915677,2019MNRAS.484.1598B,2016MNRAS.457.3050C,2020MNRAS.495..428C}. 
Higher redshift sources include quasars \citep{2005ApJ...633..589S,2012ApJ...749...56B}, Lyman break galaxies \citep{2012MNRAS.426.2489M,2017A&A...608A.141T}, and submillimetre galaxies \citep{2021A&A...656A..99B,Crespo_GonzalezNuevo_Bonavera_Cueli_Casas_Goitia_2022}.
Recently, the cross-correlation between cosmic magnification and cosmic shear has been detected in HSC \citep{Liu_Liu_Gao_Wei_Li_Fu_Futamase_Fan_2021} and DES $\times$ DESI \citep{2023MNRAS.524.6071Y}.

These measurements of cosmic magnification are indirect, relying on cross-correlation. 
However, it is feasible to directly extract the magnification signal from multiple galaxy overdensity maps of different brightness  \citep{2005PhRvL..95x1302Z,2011MNRAS.415.3485Y,ABS,YangXJ15,YangXJ17,Zhang18,2021RAA....21..247H,2024MNRAS.527.7547M}. 
The key aspect here to consider is the flux dependence characteristic of magnification bias. 
The main contamination to address is galaxy intrinsic clustering. 
Although galaxy bias is complex \citep[e.g.,][]{Bonoli09,Hamaus10,Baldauf10}, the primary component to eliminate is the deterministic bias. \cite{2021RAA....21..247H} introduced a modified internal linear combination (ILC) method that can remove the average galaxy bias in a model-independent way.

 In our recent work \citep{Qin+}, we extend the methodology of \citep{2021RAA....21..247H} to utilize multiple photometry bands and apply it to the DECaLS galaxies from the DESI imaging surveys DR9. 
 It marks the first instance of a reconstructed lensing map from magnification covering a quarter of the sky. Galaxy intrinsic clustering is suppressed by a factor of $\mathcal{O}(10)$ and a convergence-shear cross-correlation signal is detected with $S/N \simeq 20$.
However, the DECaLS survey's low galaxy number density prohibits the measurement of the $\hat{\kappa}$ auto-correlation.

In this paper, we apply the same reconstruction method to the DES Y3 galaxies from the DES Data Release 2 (DR2\cite{2018ApJS..239...18A}).
DES has greater survey depth and higher galaxy number.  
This allows us to investigate the auto-correlation of the reconstructed convergence map, which is not feasible in our DECaLS work \citep{Qin+}.

The paper is organized as follows. 
In Sec. \ref{sec:method2}, we present the reconstruction method and the modeling of the correlation functions. 
In Sec. \ref{sec:data3}, we describe the data and how we process the data, which include the galaxy samples for lensing reconstruction, the imaging systematics mitigation, the galaxy samples with shear measurement and the correlation measurements.  
Sec. \ref{sec:result4} contains details of analyses for the correlation functions including the fitting to the model, the internal tests of the analysis, and a comparison between the convergence-shear and convergence-convergence correlation analyses.
Summary and discussions are given in Sec. \ref{sec:summary5}.

\section{Method}\label{sec:method2}
The main steps of the lensing convergence reconstruction and the cross-correlation analysis are introduced in our DECaLS paper \citep{Qin+}. 
In this section, we review the key steps of the method.
Additionally, we introduce the convergence-convergence correlation analysis, 
which further validates the reconstruction.

\subsection{Lensing convergence map reconstruction }
Our goal is to perform lensing reconstruction for DES Y3 galaxies using the techniques described by \cite{2005PhRvL..95x1302Z,2011MNRAS.415.3485Y,2021RAA....21..247H}. 
We categorize galaxies into $N_F$ flux bins for each band of $g,r,z$.
In weak lensing limit, the galaxy number overdensity for each flux bin can be expressed as \citep[e.g.][]{scranton2005detection,2011MNRAS.415.3485Y}:
\begin{equation} \label{delta}
\delta^L_{i}=
b_i\delta_m+g_i\kappa+\delta_i^S\ .
\end{equation}
In this equation, $\delta_m$ represents the underlying matter overdensity, while $b_i$ is the deterministic bias for the $i$-th flux bin. 
The term $\delta_i^S$ accounts for galaxy stochasticity. 
$\kappa$ is the lensing convergence, with $g$ being the magnification coefficient. 
This coefficient is influenced by galaxy selection criteria and observational conditions \citep{Wietersheim-Kramsta_Joachimi_van, JElvinPoole2022DarkES}. 
For flux-limited samples, $g$ is determined by the logarithmic slope of the galaxy luminosity function \citep[e.g.][]{10.1093/mnras/stad1594,2024MNRAS.527.1760W}:
\begin{equation}
    g=2(\alpha - 1)\ ,\ \ \ \alpha=-\frac{d\ln{n(F)}}{d\ln{F}}-1\ .
    \label{eq:prefactorg}
\end{equation}
For sources at redshift $z_s$, $\kappa$ provides a direct measure of the underlying matter overdensity $\delta_m$ as follows \citep[e.g.][]{Liu_Liu_Gao_Wei_Li_Fu_Futamase_Fan_2021,2023MNRAS.524.6071Y}:
\begin{equation}
\kappa\left(\boldsymbol{\theta}, z_{\mathrm{s}}\right)=\frac{3 H_{0}^{2} \Omega_{\mathrm{m}}}{2 c^{2}} \int_{0}^{\chi_{\mathrm{s}}} \frac{D\left(\chi_{\mathrm{s}}-\chi\right) D(\chi)}{D\left(\chi_{\mathrm{s}}\right)} \delta_{\mathrm{m}}(z, \boldsymbol{\theta})(1+z) d \chi\ .
\end{equation}
In this equation, $\chi$ and $\chi_{s}$ denote the radial comoving distances to the lens at redshift $z$ and to the source at redshift $z_{\mathrm{s}}$, respectively.
$D(\chi)$ is the comoving angular diameter distance, which equals $\chi$ in a flat universe.

A linear estimator for the convergence $\kappa$ can be expressed as \citep{2021RAA....21..247H}:
\be\label{eq:linear combination}
\hat{\kappa}=\sum_{i} w_{i}\delta_{i}^{\rm L}\ , 
\ee 
The weights $w_i$ are determined by satisfying three conditions:
\begin{equation}
\label{eqn:wq}
    {\rm Vanishing\ multiplicative\ error:} \sum_i w_i g_i=1\ ,
\end{equation}
\begin{equation}
\label{eqn:we}
{\rm Eliminating\ intrinsic\ clustering:} \sum_i w_i =0\ ,
\end{equation}
\begin{equation}
\label{eqn:ws}
    {\rm Minimizing\ shot\ noise:}\  N_{\rm shot}=\sum_{i,j}
w_{i}w_{j}\frac{\bar{n}_{ij}}{\bar{n}_{i}\bar{n}_{j}}\ .
\end{equation}
Here, $\bar{n}_{i}$ is the average galaxy surface number density for the $i$-th flux bin, and $\bar{n}_{ij}$ is that between the $i$-th and $j$-th flux bins.
The first condition (Eq. \ref{eqn:wq}) ensures that the estimator $\hat{\kappa}$ is free from multiplicative bias, given correct magnification coefficient $g_i$. The second condition (Eq. \ref{eqn:we}) is to eliminate the average intrinsic galaxy clustering. The third condition minimizes the shot noise. These conditions determine the weights as follows:
\begin{equation}
\label{eqn:W}
     {\bf W}=\frac{C {\bf N}^{-1} {\bf g}-B{\bf N}^{-1} {\bf I}}{AC-B^2}\ .
\end{equation}
Here, $A\equiv {\bf g}^T{\bf N}^{-1}{\bf g}$, $B\equiv {\bf g}^T{\bf N}^{-1}{\bf I}$, and $C\equiv {\bf I}^T{\bf N}^{-1}{\bf I}$. 
The shot noise matrix is $N_{ij}\equiv \bar{n}_{ij}/\bar{n}_i\bar{n}_j$, and ${\bf I}\equiv [1,1,\cdots,1]^T$.

\subsection{Validation through the  convergence-shear cross-correlation}
The reconstructed map is then:
\be\label{eq:k+em}
\hat{\kappa}=A\kappa+\epsilon\delta_{\rm m}+\kappa^{\rm S}\ .
\ee
Two key issues need to be addressed. The first is the residual galaxy clustering in the reconstructed $\hat{\kappa}$. The deterministic bias $b_i$, after applying weights, becomes:
\begin{equation}
    \epsilon=\sum_i w_i b_i\ .
\end{equation}
$\epsilon\neq 0$ in general.
The second issue is a potential multiplicative error in the overall amplitude of $\hat{\kappa}$, which can result from measurement error \footnote{We aim to create flux-limited samples from the DES Y3 galaxies to ensure that $g$, derived from the logarithmic slope of the luminosity function (Eq. \ref{eq:prefactorg}), remains unbiased.
However, this is not always achievable, as observational conditions may introduce additional galaxy selection effects, such as photo-$z$ related selection.
A comprehensive account of these observational conditions would allow for unbiased estimation of $g$, which is beyond the scope of this work.
Therefore, in this study, we use the $g$ estimated by Eq. \ref{eq:prefactorg} as an approximation.} in $g$. 
This is quantified by a dimensionless parameter $A=\sum w_i \hat g_i$.
If errors in $g$ exist, $A\neq 1$.
The last term $\kappa^{\rm S}$ in Eq. \ref{eq:k+em}, which accounts for factors such as stochastic galaxy bias and shot noise, will be referred to as \textit{the stochastic term} throughout this paper for clarity. 
The first two terms in Eq. \ref{eq:k+em} will then be referred to as \textit{the deterministic term}.
An ideal reconstruction would achieve $A=1$ and $\epsilon=0$, but our current estimator only satisfies $\sum_i w_i=0$.\footnote{This issue can be addressed using principal component analysis of the galaxy cross-correlation matrix in the hyperspace of galaxy properties \citep{Zhou_Zhang_Chen_2023,2023arXiv230615177M}. However, this method requires robust clustering measurements, which are not applicable to the DES Y3 galaxies we use.} Additionally, due to uncertainties in estimating $g$ \citep{Wietersheim-Kramsta_Joachimi_van,JElvinPoole2022DarkES}, we need to evaluate $A$ using the data and verify whether $A=1$.

Motivated by  \cite{Liu_Liu_Gao_Wei_Li_Fu_Futamase_Fan_2021}, we perform cross-correlations between our $\hat{\kappa}$ map and cosmic shear catalogs in the same patch of sky, but of multiple redshift bins. In this cross-correlation, the stochastic terms do not contribute\footnote{{The stochastic galaxy bias defined here refers to the component of galaxy clustering that is uncorrelated with the cosmic density field at the two-point level. Therefore, this stochastic term does not correlate with cosmic shear.}}. 
Then, the convergence-shear cross-correlation signal is
\be\label{eq:xikg_decom}
{\xi}^{\kappa\gamma}_{j,\rm th}(\theta)=
A\xi^{\kappa\gamma}_j(\theta)+\epsilon\xi^{m\gamma}_j(\theta)\ .
\ee
Here, $\theta$ is the angular separation, and $j$ denotes the $j$-th source redshift bin of the cosmic shear catalog. 
$\xi^{m\gamma}_j$ is the matter-shear cross-correlation.
Since $A$ and $\epsilon$ are independent of $j$, we can simultaneously constrain cosmological parameters along with $A$ and $\epsilon$ by measuring $\langle\hat{\kappa}\gamma\rangle$ across various source redshift bins. For this study, which primarily aims to test the feasibility of our reconstruction method, we fix the cosmology to the best-fit Planck 2018 flat $\Lambda$CDM model \citep{Planck2018parameters}, with key cosmological parameters $\Omega_m = 0.315$, $\Omega_\Lambda = 1 - \Omega_m$, $n_s = 0.965$, $h = 0.674$, and $\sigma_8 = 0.811$.

The theoretical computation of $\xi^{\kappa \gamma}$ and $\xi^{m\gamma}$ is performed using the Limber approximation \citep{Limber1953}.
The correlation functions are linked to their respective power spectra by:
\be\label{eq:kg:hankel}
\xi^{\kappa(m)\gamma}(\theta) = \int_0^\infty \frac{d\ell \ell}{2\pi} C_\ell^{\kappa(m)\gamma} J_2(\ell \theta) \ .
\ee
Here, $J_2$ is the 2nd order Bessel function. $C_\ell^{\kappa\gamma}$ and $C_\ell^{m\gamma}$ represent the convergence-shear and matter-shear cross power spectra, respectively. In a flat Universe, these are given by:
\be
    C_\ell^{\kappa(m)\gamma} = \int d\chi W^{\kappa(m)}(\chi) W^\gamma(\chi) \frac{1}{\chi^2} P_{m}\left(k = \frac{\ell}{\chi}; z\right) \ .
\label{eqn:ckg}
\ee
Here, $\chi$ is the comoving radial distance and $P_m$ is the 3-dimensional matter power spectrum. The projection kernels $W^{\kappa}$, $W^{\gamma}$, and $W^m$ are defined as:

\begin{align}
    & W^{\kappa(\gamma)}(\chi)  = \frac{3H_0^2\Omega_{m}}{2a(\chi)c^2}
    \int_\chi^{\infty} d\chi^\prime n_{\kappa(\gamma)}(\chi^\prime)\frac{\chi^\prime-\chi}{\chi^\prime}\ , \nonumber \\
    & W^m(\chi)  = n_m(\chi)\ ,\\
    & n_{\kappa,\gamma,m}(\chi)  = n_{\kappa,\gamma,m}(z)\frac{H(z)}{c} \ , \nonumber
\end{align}
where $n_\kappa=n_m$ is the normalized redshift distribution of galaxies for $\kappa$ reconstruction,
and $n_{\gamma}$ is that for cosmic shear. 

By fitting against ${\xi}_{j,\rm th}^{\kappa \gamma}$ at multiple shear redshifts, we can constrain $A$ and $\epsilon$, which probe the convergence map's detection significance ($A/\sigma_A$) and systematic errors ($\epsilon$).

\subsection{Convergence-convergence correlation}
A further validation of the $\hat{\kappa}$ maps is to check the correlation $\langle \hat{\kappa}_i \hat{\kappa}_j \rangle$.
The theoretical expectation is
\begin{equation}\label{eq:xik1k2_decom}
\begin{aligned}
{\xi}^{\kappa\kappa}_{ij,\rm th} &= \xi_{ij}^{\rm D} + \xi_{ij}^{\rm S} \ , \\ 
\xi_{ij}^{\rm D} &= A_iA_j\xi^{\kappa_i\kappa_j} + \epsilon_i\epsilon_j\xi^{m_im_j} + A_i\epsilon_j\xi^{\kappa_i m_j} + A_j\epsilon_i\xi^{\kappa_j m_i} \ .
\end{aligned}
\end{equation}
where $i$ and $j$ represent different redshift bins of the reconstructed convergence map. 
These correlations are related to the power spectra \citep{Limber1953} by
\be\label{eq:kk:hankel}
\xi^{\kappa_i(m_i)\kappa_j(m_j)}(\theta)=\int_0^\infty\frac{d\ell\ell}{2\pi}C_\ell^{\kappa_i(m_i)\kappa_j(m_j)}J_0(\ell\theta)\ .
\ee
Here $J_0$ is the 0th order Bessel function.
The angular power spectra are related to the 3-dimensional power spectrum by
\be
    C_\ell^{\kappa_i(m_i)\kappa_j(m_j)} =\int d\chi W^{\kappa_i(m_i)}(\chi)W^{\kappa_j(m_j)}(\chi)\frac{1}{\chi^2}P_{m}\left(k = \frac{\ell}{\chi};z\right) \ .
\ee

In this case, the stochastic term, denoted by $\xi_{ij}^{\rm S}$ in Eq. \ref{eq:xik1k2_decom}, may not be ignored.
Accurately modeling these terms is challenging, and we do not attempt it in this work.
Along with the fitting results of $A$ and $\epsilon$ from the convergence-shear cross-correlation analyses,
we can then predict $\xi_{ij}^{\rm D}$.
Comparing with the $\langle \hat{\kappa}_i \hat{\kappa}_j \rangle$ measurements,
we can evaluate the impact of $\xi_{ij}^{\rm S}$.
This provides an estimation of the stochastic term, inaccessible to the $\hat{\kappa}-\gamma$ analysis.

\begin{table*}
    \centering
    \begin{tabular}{ccccccccc}
    \hline
    \hline
{Flux Bin} &  Magnitude Range &     Galaxy Number      &  $\bar{n}({\rm arcmin}^{-2})$ &  $\alpha$ &     $g$ &     $w$ & band & photo-$z$ bin \\
\midrule
1  &  22.3-22.5 &  2566484 &         0.15 &   1.33 &  0.66 &  0.03 &    g &  0.4-0.6 \\
2  &  22.1-22.3 &  2566484 &         0.15 &   1.46 &  0.92 &  0.05 &    g &  0.4-0.6 \\
3  &  21.8-22.1 &  2566484 &         0.15 &   1.87 &  1.73 &  0.08 &    g &  0.4-0.6 \\
4  &  16.1-21.8 &  2566484 &         0.15 &   2.53 &  3.06 &  0.13 &    g &  0.4-0.6 \\
\hline

5  &  21.7-22.0 &  5066477 &         0.30 &   0.76 & -0.48 &  0.09 &    r &  0.4-0.6 \\
6  &  21.4-21.7 &  5066477 &         0.30 &   0.82 & -0.36 &  0.16 &    r &  0.4-0.6 \\
7  &  21.0-21.4 &  5066477 &         0.30 &   1.22 &  0.44 &  0.24 &    r &  0.4-0.6 \\
8  &  15.7-21.0 &  5066477 &         0.30 &   1.84 &  1.67 &  0.26 &    r &  0.4-0.6 \\
\hline

9  &  21.2-21.5 &  5331458 &         0.31 &   0.69 & -0.62 & -0.14 &    z &  0.4-0.6 \\
10 &  20.8-21.2 &  5331458 &         0.31 &   0.76 & -0.49 & -0.22 &    z &  0.4-0.6 \\
11 &  20.2-20.8 &  5331458 &         0.31 &   0.85 & -0.31 & -0.32 &    z &  0.4-0.6 \\
12 &  13.6-20.2 &  5331458 &         0.31 &   1.48 &  0.95 & -0.35 &    z &  0.4-0.6 \\

\hline
\hline

1  &  23.3-23.5 &  2995791 &         0.18 &   0.76 & -0.48 & -0.03 &    g &  0.6-0.8 \\
2  &  23.1-23.3 &  2995791 &         0.18 &   1.23 &  0.45 & -0.02 &    g &  0.6-0.8 \\
3  &  22.7-23.1 &  2995791 &         0.18 &   1.52 &  1.03 & -0.00 &    g &  0.6-0.8 \\
4  &  15.9-22.7 &  2995791 &         0.18 &   2.35 &  2.70 &  0.04 &    g &  0.6-0.8 \\
\hline

5  &  22.8-23.0 &  5137054 &         0.30 &   0.93 & -0.13 &  0.14 &    r &  0.6-0.8 \\
6  &  22.4-22.8 &  5137054 &         0.30 &   1.13 &  0.26 &  0.21 &    r &  0.6-0.8 \\
7  &  22.0-22.4 &  5137054 &         0.30 &   1.13 &  0.26 &  0.25 &    r &  0.6-0.8 \\
8  &  17.3-22.0 &  5137054 &         0.30 &   1.79 &  1.58 &  0.28 &    r &  0.6-0.8 \\
\hline

9  &  22.1-22.5 &  6068182 &         0.36 &   0.48 & -1.04 & -0.12 &    z &  0.6-0.8 \\
10 &  21.6-22.1 &  6068182 &         0.36 &   0.39 & -1.22 & -0.20 &    z &  0.6-0.8 \\
11 &  21.0-21.6 &  6068182 &         0.36 &   0.55 & -0.90 & -0.28 &    z &  0.6-0.8 \\
12 &  14.8-21.0 &  6068182 &         0.36 &   1.47 &  0.95 & -0.28 &    z &  0.6-0.8 \\

\hline
\hline
1  &  24.3-24.5 &  11254810 &         0.66 &   1.10 &  0.20 & -0.01 &    g &  0.8-1.0 \\
2  &  24.0-24.3 &  11254810 &         0.66 &   1.12 &  0.24 & -0.03 &    g &  0.8-1.0 \\
3  &  23.5-24.0 &  11254810 &         0.66 &   1.03 &  0.06 & -0.16 &    g &  0.8-1.0 \\
4  &  15.8-23.5 &  11254810 &         0.66 &   1.93 &  1.87 & -0.17 &    g &  0.8-1.0 \\
\hline

5  &  23.3-23.5 &   7805040 &         0.46 &   0.91 & -0.17 &  0.09 &    r &  0.8-1.0 \\
6  &  23.1-23.3 &   7805040 &         0.46 &   1.15 &  0.31 &  0.19 &    r &  0.8-1.0 \\
7  &  22.7-23.1 &   7805040 &         0.46 &   1.52 &  1.03 &  0.30 &    r &  0.8-1.0 \\
8  &  19.5-22.7 &   7805040 &         0.46 &   2.37 &  2.74 &  0.46 &    r &  0.8-1.0 \\
\hline

9  &  22.2-22.5 &   7640938 &         0.45 &   0.62 & -0.77 & -0.12 &    z &  0.8-1.0 \\
10 &  21.9-22.2 &   7640938 &         0.45 &   0.77 & -0.47 & -0.15 &    z &  0.8-1.0 \\
11 &  21.5-21.9 &   7640938 &         0.45 &   1.16 &  0.31 & -0.18 &    z &  0.8-1.0 \\
12 &  13.8-21.5 &   7640938 &         0.45 &   1.89 &  1.78 & -0.21 &    z &  0.8-1.0 \\
\hline

    \end{tabular}
    \caption{ Summary of the galaxy sub-samples used for lensing reconstruction.
    Throughout the paper we use $z_\kappa$ to denote the photometric redshift of galaxies 
    used for lensing reconstruction, 
    and use $z_\gamma$ for photometric redshift of galaxies for cosmic shear measurements. 
    In addition to the information presented in the table, it should be noted that there are overlaps between the sub-samples in terms of the galaxies they contain.
    For example, sample 1 and sample 5 for $0.8<z_{\kappa}<1.0$ share a common set of $1.7\times10^6$ galaxies, which we denote as $n_{15}=3.4\times10^6$.
    Similarly $n_{16}=2.7\times10^6$ and $n_{17}=1.4\times10^6$.
    However $n_{ij}=0$ if sample i and j are from the same photometry band (e.g. $n_{12}=n_{13}=0$).
    }
    \label{tab:sub-sample}
\end{table*}

\begin{figure}
\includegraphics[width=0.5\textwidth]{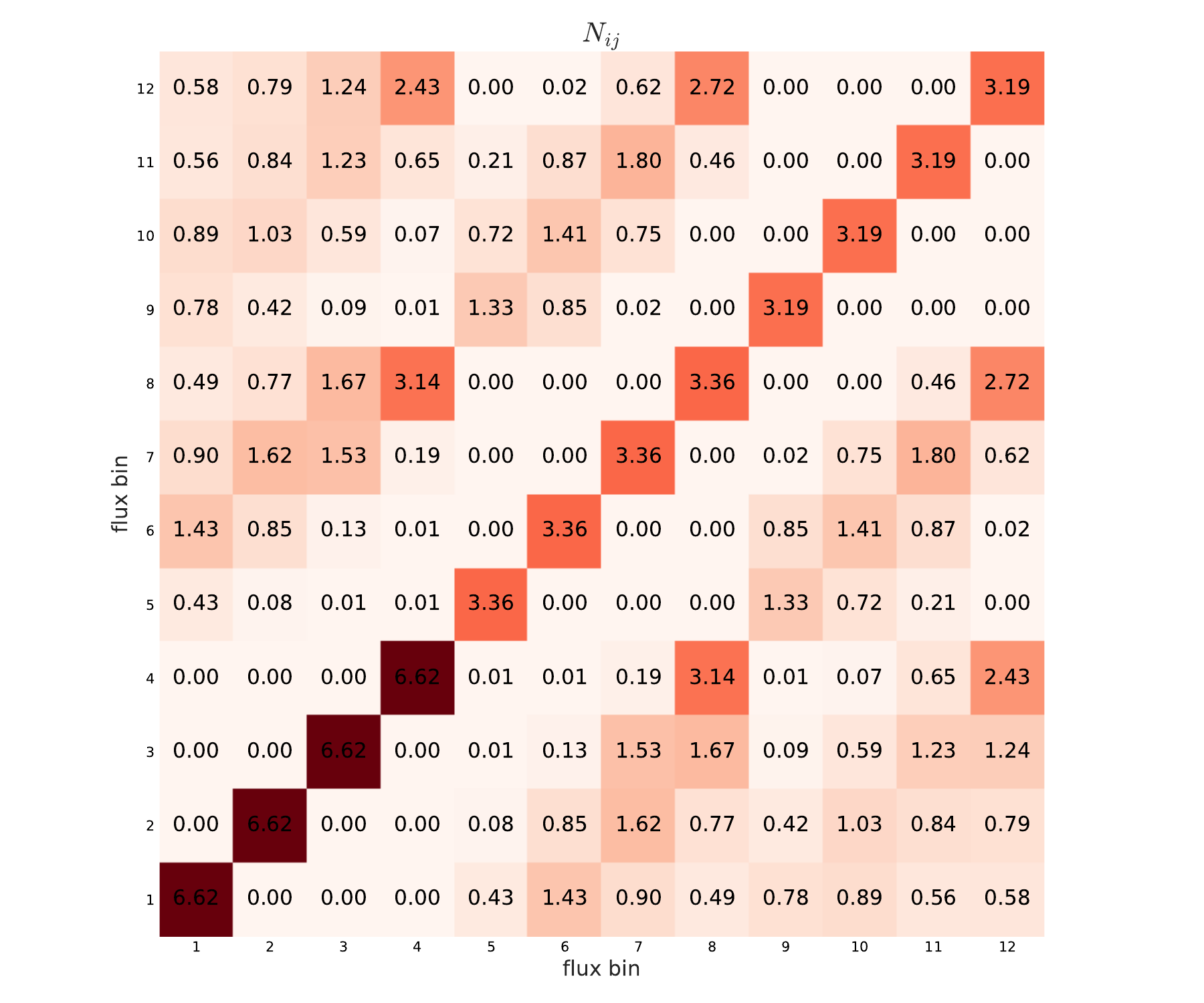}
\caption{The shot noise matrix $N_{ij}\equiv\frac{\bar{n}_{ij}}{\bar{n}_i\bar{n}_j}$ derived from the galaxy number density of the 12 flux bins/sub-samples, shown for the photo-$z$ bin $0.4<z_{\kappa}<0.6$ as an example.
The diagonal elements $N_{ii}\equiv\frac{1}{\bar{n}_i}$ are measured from the average number density of each flux bin (the fourth column in Table \ref{tab:sub-sample}).
The off-diagonal elements $N_{ij}$ are measured from the overlap between the sub-samples in terms of the number of galaxies they contain (see the discussion in the caption of Table \ref{tab:sub-sample}).
This overlap arises form the fact that the relative ranking of galaxy brightness varies in different photometry bands.
 }
\label{fig:nij}
\end{figure}


\section{Data analysis}\label{sec:data3}

We apply our method of lensing reconstruction to the DES Y3 galaxies from the DES Data Release 2 (DR2\cite{2018ApJS..239...18A}).

\subsection{Data}
The DES galaxy samples used for lensing reconstruction are created in accordance with the selection criteria outlined in section 2.1 of \cite{yang2021extended}.
Extended imaging objects are required to have been observed at least once in each optical band to ensure reliable photo-$z$ estimation.
After removing objects with Galactic latitude $|\mathrm{b}|<25.0^{\circ}$ to avoid high stellar density regions and objects whose fluxes are affected by bright stars, large galaxies, or globular clusters, 
we obtain the DES Y3 galaxy sample with a sky coverage of $\sim$4800 deg$^2$.

\begin{figure} 
\centering
\includegraphics[width=0.9\columnwidth]{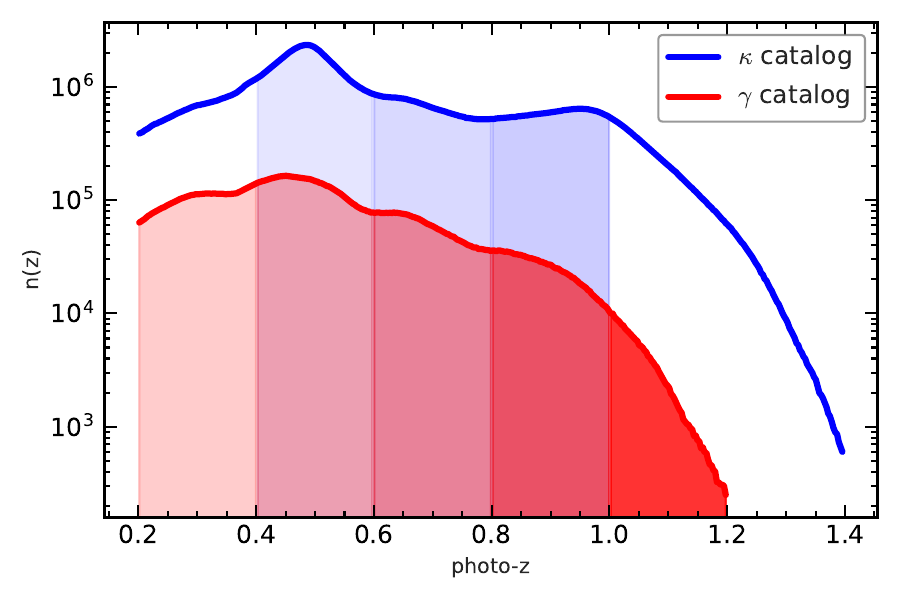}
\caption{The photometric redshift distributions of the DES Y3 galaxies used for lensing reconstruction (blue line) and the DES Y3 galaxies with shear measurement (red line). 
The blue shaded regions denote photo-$z$ ranges $0.4<z_{\kappa}<0.6$, $0.6<z_{\kappa}<0.8$, and $0.8<z_{\kappa}<1.0$ employed for lensing reconstruction in our study. 
The red shaded regions represent the photo-$z$ bins of the shear catalog, [0.2, 0.4, 0.6, 0.8, 1.0, 1.2], defined for cross-correlation analysis.
From this figure, we can see that there are many more galaxies available for lensing magnification reconstruction than for cosmic shear measurements.
\label{fig:nz}}
\end{figure}

For the convergence-shear cross-correlation analysis, 
we utilize the DES Y3 shear measurements covering the same footprint as the reconstructed convergence map.
The detailed shear measurements procedure, along with the estimated $m$ and $c$, are described in \cite{2023MNRAS.524.6071Y}.
For both the shear and convergence reconstruction galaxies, we employ the photometric redshift based on \cite{zhou2021clustering}.
Note that shear measurements require high imaging quality to ensure reliable shape measurements.
Consequently, the number of shear galaxies is significantly smaller than that of the convergence reconstruction galaxies (see Fig. \ref{fig:nz}).

\subsection{Reconstruction}

To reconstruct the convergence map, we consider three photometric redshift bins: $0.4 < z_\kappa < 0.6$, $0.6 < z_\kappa < 0.8$, and $0.8 < z_\kappa < 1.0$, selected from the DES Y3 galaxies. 
 As an approximation for the flux-limit selection, for each band, we select galaxies with a magnitude cut approximately 0.5 lower than the peak of the galaxy number counts as a function of magnitude. 
We then divide the galaxies in each band uniformly into $N_F = 4$ flux bins, resulting in a total of 12 flux bins at each redshift. 
For each flux bin, we convert the 3D galaxy number density distribution into a 2D sky map at a resolution of $N_{\rm side} = 4096$.
We then downgrade the resolution to $N_{\rm side} = 1024$ and define the coverage fraction parameter $f_{\beta}$ for each pixel. Pixels with $f_{\beta}$ exceeding the threshold $f = 0.9$ are selected.
The $f_{\beta}$ definition, steps of measuring galaxy number overdensity, and calculation of the $\alpha$, $g$, $w$ factors for each flux bin are described in \cite{Qin+}.
A summary of the flux bins and the results of the $\alpha$, $g$, and $w$ is presented in Table \ref{tab:sub-sample}. 
The measured shot noise matrix $N_{ij} \equiv \frac{\bar{n}_{ij}}{\bar{n}_i \bar{n}_j}$ for $0.4 < z_p < 0.6$ is shown as an example in \reffig{fig:nij}.
The reconstructed convergence maps are then obtained by performing a weighted sum over the galaxy overdensity maps (Eq.\eqref{eq:linear combination}).
For consistency tests described in Sec.\ref{sec:robustest}
, we also utilize additional sets of magnitude cuts, $N_F$, and $f^{\rm threshold}$.


Following \cite{Qin+,HaojieXu2022UsingAT}, we mitigate imaging systematics, such as stellar contamination and Galactic extinction, in the reconstructed convergence $\hat\kappa$ maps  using the Random Forest (RF) mitigation method developed by \cite{10.1093/mnras/stab3252}. The effects of this mitigation on the measured lensing correlation functions are detailed in Appendix \ref{sec:imgweight}.

\subsection{Correlation functions}

To validate the reconstruction, we analyze the $\hat{\kappa}-\gamma$ and $\hat{\kappa}-\hat{\kappa}$ correlation functions.
We select shear galaxies within the photometric redshift range $0.2 < z_\gamma < 1.2$, resulting in about 18 million galaxy shear samples.
They are then divided into five photo-$z$ bins: [0.2, 0.4, 0.6, 0.8, 1.0, 1.2].
For each photo-$z$ bin of $\hat{\kappa}$ and $\gamma$, the correlation functions is estimated as follows:
\be
\hat{\xi}_{ij}^{\kappa\gamma}(\theta) = \langle \hat\kappa_i\gamma_{t,j}\rangle_\theta / (1 + \bar{m}) \ .
\ee
\be
\hat{\xi}^{\kappa\kappa}_{ij}(\theta) = \langle \hat\kappa_i \hat\kappa_j \rangle_\theta \ .
\ee
Here, the average $\langle \cdots \rangle_\theta$ is done over all pixel-galaxy pairs within the angular separation $\theta$, $\bar{m}$ represents the average shear multiplicative bias, and $\gamma_t$ is the tangential component of the shear.

For both the convergence-shear and convergence-convergence correlation functions, the calculations are performed using the publicly available code TreeCorr\footnote{https://github.com/rmjarvis/TreeCorr}.
We use $N = 100$ Jackknife patches to estimate the covariance matrix and rescale the covariance matrix following \cite{percival2014clustering,Wang_Zhao_Zhao_Philcox_Alam_Tamone_deMattia_Ross_Raichoor_Burtin2020} to obtain an unbiased estimation.

\begin{figure} 
\centering
\includegraphics[width=\columnwidth]{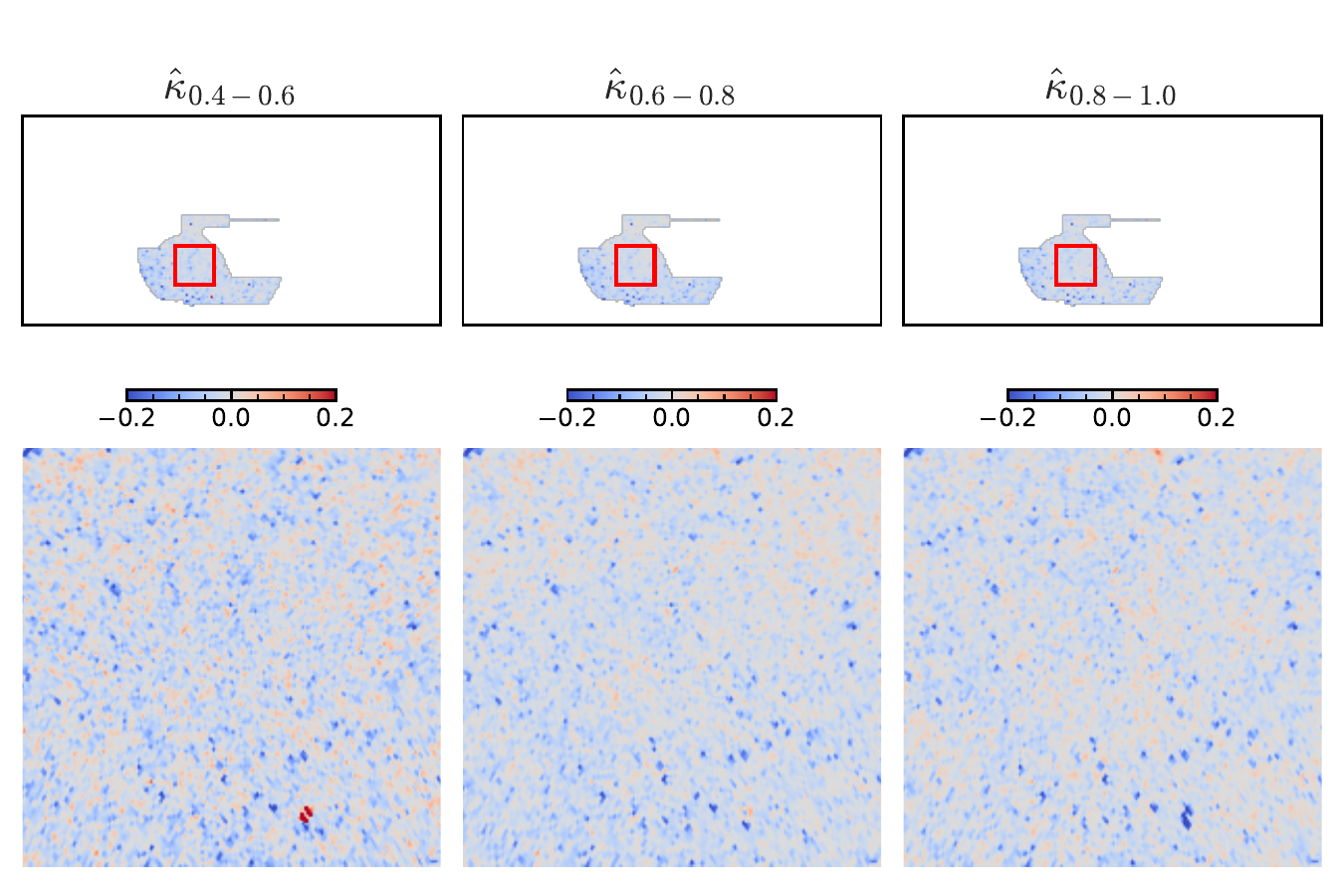}
\caption{The reconstructed lensing convergence maps.
The columns from left to right represent the three photo-$z$ bins of the convergence.
The lower panels correspond to zoomed-in views of the areas indicated by the red box in the upper panels.
The resolution is downgraded from $N_{\rm side}=1024$ to $N_{\rm side}=256$ for clearer demonstration of the lensing features.
\label{fig:kapmap}}
\end{figure}

\begin{figure*} 
\centering
\includegraphics[width=1.9\columnwidth]{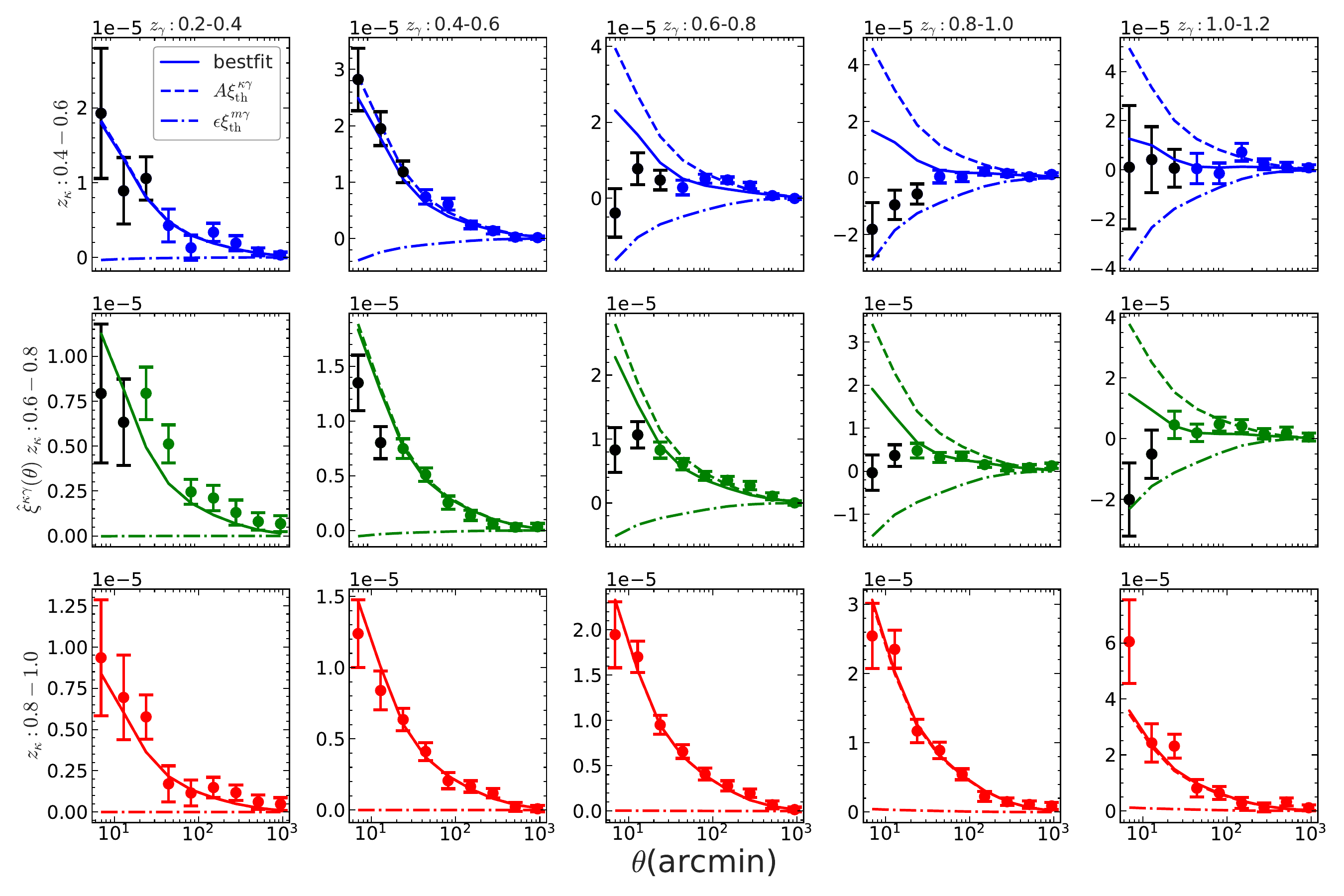}
\caption{Measured cross correlation function between the reconstructed lensing convergence and the shear.
The panels display the results for the three photo-$z$ bins of the convergence from top to bottom, 
and for the five photo-$z$ bins of the shear from left to right. 
The blue solid lines represent the best-fit to the model (Eq.\eqref{eq:xikg_decom}).
The dashed lines indicate the two components of the best-fit, the signal term $A\xi^{\kappa\gamma}$ and the intrinsic clustering term $\epsilon\xi^{m\gamma}$.
Parameters $A$ and $\epsilon$ are defined to quantify the quality of the lensing reconstruction (Eq.\eqref{eq:k+em}).
During the fitting, they are kept same across the shear redshift bins.
For clarity, only results after the imaging systematics mitigation are shown here. 
Comparisons of measurements before and after mitigation are shown in \reffig{fig:xi_kg_imgW}, 
while fitting results are summarized in Table \ref{tab:robust}.
\label{fig:xi}}
\end{figure*}

For theoretical calculation of $\xi^{\kappa\gamma}$, $\xi^{m\gamma}$, $\xi^{\kappa_i\kappa_j}$, $\xi^{m_im_j}$, $\xi^{\kappa_i m_j}$ and $\xi^{\kappa_j m_i}$, 
we apply the Core Cosmology Library (CCL, \citealt{pyccl}) and use  HaloFit \citep{Smith2003, Takahashi2012, Mead2015} to calculate the nonlinear matter power spectrum.
The galaxy redshift distribution $n(z)$ are calculated for each tomographic bin, combining the photo-$z$ distribution and a Gaussian photo-$z$ error PDF with $\sigma_z=0.05$\footnote{We conducted tests using different assumed values of $\sigma_z$ and found that the impact is negligible.}.
\reffig{fig:nz} displays the photo-$z$ distribution of the DES Y3 galaxies used for lensing construction and the DES shear catalog.
The differences in the theoretical templates' shape and redshift dependence allow us to separate the measured correlations into the signal term ($\xi^{\kappa\gamma}$ or $\xi^{\kappa\kappa}$) and the intrinsic clustering term ($\xi^{m\gamma}$, $\xi^{mm}$, or $\xi^{\kappa m}$).


\begin{table*}
\begin{center}
\begin{tabular}{llllllll}
\hline\hline
{}&{} &                           $A$ &                      $\epsilon$ & $\chi^2_{\rm min}$ & $\sqrt{\chi^2_{\rm null}-\chi^2_{\rm min}}$ & $A/\sigma_A$ & $\rho=\frac{\sigma_{A\epsilon}}{\sqrt{\sigma_A\sigma_\epsilon}}$ \\
\midrule

            \Xhline{3\arrayrulewidth}
            $0.4<z_\kappa<0.6$ & baseline & \textbf{3.24$\pm$0.36}(3.96$\pm$0.44) & \textbf{-0.09$\pm$0.02}(-0.13$\pm$0.02) & \textbf{35.2}(41.6) & \textbf{10.5}(10.3) & \textbf{9.0}(9.0) & \textbf{0.8}(0.9) \\
            \Xhline{3\arrayrulewidth}
             & f=0.5 & \textbf{3.49$\pm$0.34}(4.07$\pm$0.44) & \textbf{-0.10$\pm$0.02}(-0.13$\pm$0.02) & \textbf{33.0}(33.3) & \textbf{11.6}(10.4) & \textbf{10.3}(9.2) & \textbf{0.8}(0.9) \\
             & f=0.7 & \textbf{3.58$\pm$0.36}(4.09$\pm$0.44) & \textbf{-0.11$\pm$0.02}(-0.14$\pm$0.03) & \textbf{36.5}(30.2) & \textbf{11.7}(10.3) & \textbf{9.9}(9.3) & \textbf{0.8}(0.8) \\
             & $N_F=3$ & \textbf{3.88$\pm$0.39}(4.31$\pm$0.48) & \textbf{-0.12$\pm$0.02}(-0.15$\pm$0.03) & \textbf{44.1}(35.4) & \textbf{11.5}(10.0) & \textbf{9.9}(9.0) & \textbf{0.8}(0.9) \\
             & $N_F=5$ & \textbf{3.25$\pm$0.32}(3.94$\pm$0.41) & \textbf{-0.10$\pm$0.02}(-0.12$\pm$0.02) & \textbf{39.3}(36.6) & \textbf{11.8}(11.2) & \textbf{10.2}(9.6) & \textbf{0.8}(0.9) \\
             & $m_{g,r,z}-0.5$ & \textbf{2.95$\pm$0.33}(3.13$\pm$0.42) & \textbf{-0.07$\pm$0.02}(-0.08$\pm$0.02) & \textbf{35.7}(33.5) & \textbf{11.0}(9.3) & \textbf{8.9}(7.5) & \textbf{0.8}(0.8) \\
             & $m_g-0.5$ & \textbf{3.45$\pm$0.35}(3.98$\pm$0.40) & \textbf{-0.10$\pm$0.02}(-0.13$\pm$0.02) & \textbf{33.2}(32.5) & \textbf{11.3}(11.4) & \textbf{9.9}(9.9) & \textbf{0.8}(0.8) \\
             & $m_r-0.5$ & \textbf{3.02$\pm$0.33}(3.27$\pm$0.40) & \textbf{-0.07$\pm$0.02}(-0.08$\pm$0.02) & \textbf{44.9}(38.3) & \textbf{11.8}(10.0) & \textbf{9.2}(8.2) & \textbf{0.8}(0.8) \\
             & $m_z-0.5$ & \textbf{3.59$\pm$0.37}(3.96$\pm$0.44) & \textbf{-0.11$\pm$0.02}(-0.13$\pm$0.02) & \textbf{34.2}(29.7) & \textbf{11.3}(10.2) & \textbf{9.7}(9.0) & \textbf{0.8}(0.9) \\
            \Xhline{3\arrayrulewidth}
            $0.6<z_\kappa<0.8$ & baseline & \textbf{1.76$\pm$0.11}(1.93$\pm$0.13) & \textbf{-0.10$\pm$0.02}(-0.12$\pm$0.02) & \textbf{36.5}(33.5) & \textbf{18.4}(17.4) & \textbf{16.0}(14.8) & \textbf{0.7}(0.8) \\
            \Xhline{3\arrayrulewidth}
             & f=0.5 & \textbf{1.75$\pm$0.12}(1.87$\pm$0.13) & \textbf{-0.10$\pm$0.02}(-0.11$\pm$0.02) & \textbf{38.9}(33.8) & \textbf{17.6}(16.9) & \textbf{14.6}(14.4) & \textbf{0.7}(0.8) \\
             & f=0.7 & \textbf{1.76$\pm$0.11}(1.86$\pm$0.13) & \textbf{-0.10$\pm$0.02}(-0.11$\pm$0.02) & \textbf{40.2}(34.0) & \textbf{18.3}(16.9) & \textbf{16.0}(14.3) & \textbf{0.7}(0.8) \\
             & $N_F=3$ & \textbf{1.75$\pm$0.11}(1.87$\pm$0.13) & \textbf{-0.11$\pm$0.02}(-0.12$\pm$0.02) & \textbf{38.5}(36.1) & \textbf{17.6}(16.3) & \textbf{15.9}(14.4) & \textbf{0.7}(0.8) \\
             & $N_F=5$ & \textbf{1.74$\pm$0.11}(1.85$\pm$0.13) & \textbf{-0.09$\pm$0.02}(-0.10$\pm$0.02) & \textbf{41.3}(35.9) & \textbf{19.1}(17.7) & \textbf{15.8}(14.2) & \textbf{0.7}(0.8) \\
             & $m_{g,r,z}-0.5$ & \textbf{1.95$\pm$0.13}(2.06$\pm$0.15) & \textbf{-0.09$\pm$0.02}(-0.09$\pm$0.02) & \textbf{36.2}(30.7) & \textbf{18.4}(16.8) & \textbf{15.0}(13.7) & \textbf{0.7}(0.8) \\
             & $m_g-0.5$ & \textbf{1.90$\pm$0.12}(2.08$\pm$0.14) & \textbf{-0.11$\pm$0.02}(-0.14$\pm$0.02) & \textbf{35.7}(34.9) & \textbf{18.7}(17.2) & \textbf{15.8}(14.9) & \textbf{0.7}(0.8) \\
             & $m_r-0.5$ & \textbf{1.60$\pm$0.12}(1.72$\pm$0.14) & \textbf{-0.08$\pm$0.02}(-0.09$\pm$0.02) & \textbf{32.9}(31.8) & \textbf{15.7}(14.8) & \textbf{13.3}(12.3) & \textbf{0.8}(0.8) \\
             & $m_z-0.5$ & \textbf{2.17$\pm$0.13}(2.37$\pm$0.15) & \textbf{-0.10$\pm$0.02}(-0.11$\pm$0.02) & \textbf{41.2}(31.4) & \textbf{20.4}(19.8) & \textbf{16.7}(15.8) & \textbf{0.8}(0.8) \\
            \Xhline{3\arrayrulewidth}
            $0.8<z_\kappa<1.0$ & baseline & \textbf{1.19$\pm$0.06}(1.17$\pm$0.07) & \textbf{0.01$\pm$0.04}(0.06$\pm$0.04) & \textbf{26.5}(41.8) & \textbf{28.4}(23.4) & \textbf{19.8}(16.8) & \textbf{0.7}(0.7) \\
            \Xhline{3\arrayrulewidth}
             & f=0.5 & \textbf{1.20$\pm$0.06}(1.18$\pm$0.06) & \textbf{0.02$\pm$0.05}(0.02$\pm$0.05) & \textbf{25.1}(40.3) & \textbf{27.2}(28.0) & \textbf{20.0}(19.7) & \textbf{0.7}(0.7) \\
             & f=0.7 & \textbf{1.18$\pm$0.05}(1.18$\pm$0.05) & \textbf{0.02$\pm$0.04}(0.05$\pm$0.04) & \textbf{24.0}(39.2) & \textbf{28.8}(29.4) & \textbf{23.6}(23.6) & \textbf{0.6}(0.6) \\
             & $N_F=3$ & \textbf{1.28$\pm$0.05}(1.26$\pm$0.05) & \textbf{-0.01$\pm$0.05}(0.03$\pm$0.04) & \textbf{22.5}(37.7) & \textbf{29.1}(31.7) & \textbf{25.6}(25.2) & \textbf{0.6}(0.6) \\
             & $N_F=5$ & \textbf{1.17$\pm$0.05}(1.16$\pm$0.06) & \textbf{0.00$\pm$0.04}(0.04$\pm$0.04) & \textbf{24.7}(42.8) & \textbf{28.6}(29.6) & \textbf{23.4}(19.3) & \textbf{0.6}(0.7) \\
             & $m_{g,r,z}-0.5$ & \textbf{0.97$\pm$0.06}(0.96$\pm$0.06) & \textbf{-0.02$\pm$0.05}(0.02$\pm$0.04) & \textbf{26.8}(42.1) & \textbf{21.5}(22.1) & \textbf{16.2}(16.0) & \textbf{0.6}(0.6) \\
             & $m_g-0.5$ & \textbf{1.21$\pm$0.06}(1.19$\pm$0.05) & \textbf{0.01$\pm$0.04}(0.05$\pm$0.04) & \textbf{26.3}(41.8) & \textbf{28.0}(30.1) & \textbf{20.2}(23.8) & \textbf{0.6}(0.6) \\
             & $m_r-0.5$ & \textbf{1.15$\pm$0.06}(1.13$\pm$0.05) & \textbf{-0.03$\pm$0.05}(0.00$\pm$0.04) & \textbf{25.1}(46.4) & \textbf{25.6}(27.3) & \textbf{19.2}(22.6) & \textbf{0.6}(0.7) \\
             & $m_z-0.5$ & \textbf{0.96$\pm$0.06}(0.96$\pm$0.06) & \textbf{0.02$\pm$0.05}(0.05$\pm$0.05) & \textbf{29.2}(44.4) & \textbf{21.0}(21.8) & \textbf{16.0}(16.0) & \textbf{0.7}(0.7) \\

\end{tabular}
\end{center}
    \caption{Results of the convergence-shear cross-correlation analysis for the three photo-$z$ bins of convergence.
    Besides the baseline, the impacts of $f^{\rm threshold}, N_{F}$, the three bands magnitude cuts on the analysis are tested.
    The baseline sets are $f^{\rm threshold}=0.9, N_{F}=4$ and magnitude cuts shown in Table \ref{tab:sub-sample}.
    The results are presented from top to bottom for the photo-$z$ bins $0.4 < z_\kappa < 0.6$, $0.6 < z_\kappa < 0.8$, and $0.8 < z_\kappa < 1.0$.    And the d.o.f. = 28, 33, and 43 for the three $z_\kappa$ bins, respectively.
    The fitting results remain consistent across different $f^{\rm threshold}$, $N_{F}$, and magnitude cuts.
    The main table shows the results after the mitigation, while in parentheses are the results before the mitigation.
    The $S/N$ of the detection of the weak lensing signal is enhanced after the mitigation.
   The last column shows the correlation coefficient between $A$ and $\epsilon$, where $\sigma^2_{A\epsilon}$, $\sigma^2_A$, and $\sigma^2_\epsilon$ are the (co)variances.
      }    \label{tab:robust}
\end{table*}

\begin{figure}
\centering
\includegraphics[width=\columnwidth]{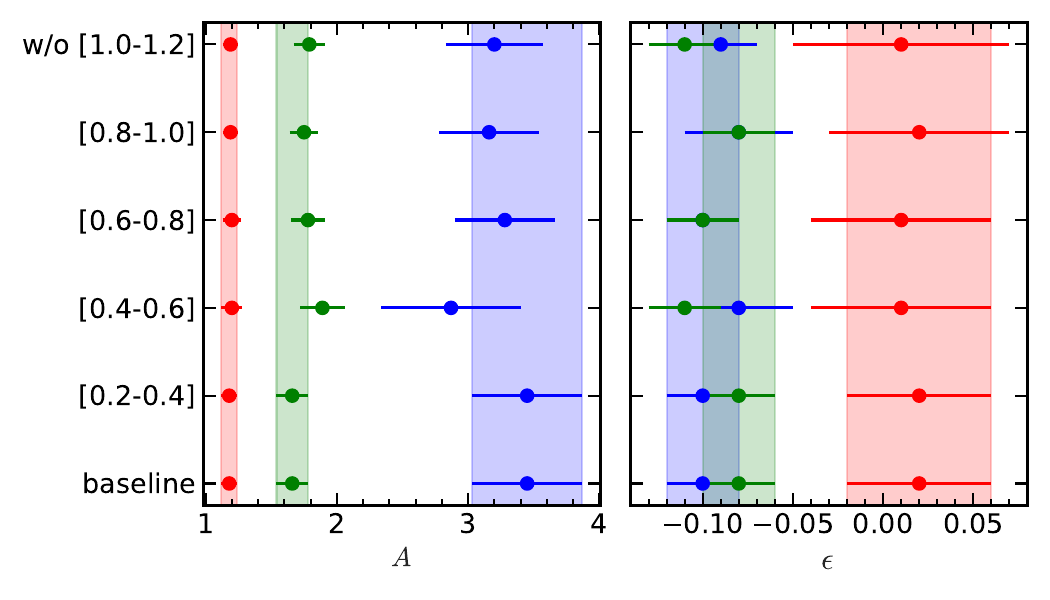}
\caption{Impact of the shear redshifts on the constraints of $A$ (the left panel) and $\epsilon$ (the right panel).
We drop one shear redshift bin at a time, and perform the cross correlation analysis and constrain the parameters with the remaining bins. 
The y-axis denote the shear redshift bin that is dropped.
The shadow regions shows the $1 \sigma$ error of the constraints for the baseline set (i.e., all the shear bins are included).
The colors red, green, and blue indicate the convergence map reconstructed in the photo-$z$ bins $0.4 < z_\kappa < 0.6$, $0.6 < z_\kappa < 0.8$, and $0.8 < z_\kappa < 1.0$, respectively.
The constraining results are consistent with the baseline set for all the cases, 
which indicates the robustness of the cross-correction analysis against different shear samples.
}
\label{fig:Ae_dropZg}
\end{figure}

\begin{table}
\centering
\begin{tabular}{llllll}
\hline
\hline
    {} & baseline & A=0 & A=1 & $\epsilon$=0 \\
$\chi^2_{\rm min}$  &    35.2 &  117.5 &  81.1 &          61.2 \\
AIC                 &    39.3 &  119.5 &  83.2 &          63.3 \\
\hline
$\chi^2_{\rm min}$  &    36.5 &  286.9 &  83.2 &          67.5 \\
AIC                 &    40.6 &  288.9 &  85.3 &          69.6 \\
\hline
$\chi^2_{\rm min}$  &    26.5 &  483.1 &  38.1 &          26.6 \\
AIC                 &    30.6 &  485.2 &  40.1 &          28.6 \\
\hline
\end{tabular}
\caption{Akaike Information Criterion (AIC) Results for Convergence-Shear Cross-Correlation Analysis. 
The baseline model follows Eq. \eqref{eq:xikg_decom}. 
Other models set $A=0$, $A=1$, or $\epsilon=0$. 
Results, from top to bottom, correspond to lensing convergence in photo-$z$ bins: $0.4 < z_\kappa < 0.6$, $0.6 < z_\kappa < 0.8$, and $0.8 < z_\kappa < 1.0$. 
In all cases, except where $\epsilon=0$ for $0.8 < z_\kappa < 1.0$, alternative models are ruled out, with AIC scores exceeding the baseline by more than 10. 
This suggests the robustness of the weak lensing signal detection and the minimal residual galaxy clustering contamination for $0.8 < z_\kappa < 1.0$.}
\label{tab:AIC}
\end{table}

\section{Results}\label{sec:result4}

We present visualizations of the reconstructed convergence maps of the three $z_\kappa$ bins in \reffig{fig:kapmap}.
The maps reveal lensing features of the large-scale structure. 
We analyze the derived correlation functions from these maps to validate the lensing reconstruction.

\subsection{Convergence-shear cross correlation}
\label{sec:result-kap-shear}
\subsubsection{Baseline results}
Figure \ref{fig:xi} displays the measured convergence-shear cross-correlation $\hat{\xi}^{\kappa\gamma}$. Significant cross-correlation signals are detected for the convergence maps reconstructed in the three photo-$z$ bins, spanning the five photo-$z$ bins of the shear. 
For each photo-$z$ bin of the convergence, we fit the cross-correlation measurements to the theoretical model (Eq. \eqref{eq:xikg_decom}), keeping $A$ and $\epsilon$ fixed across the shear redshift bins.
For map $\hat{\kappa}_i$, the $\chi^2$ for the fitting is approximated by
\begin{equation}
\label{eq:chi_1}
\begin{aligned}
    \chi^2 &\approx \sum\limits_{j\alpha\beta}\big[\hat{\xi}^{\kappa\gamma}_j(\theta_\alpha)-{\xi}^{\kappa\gamma}_{j,\rm th}(\theta_\alpha)\big] \times \textbf{Cov}_{j}^{-1} \\
    &\times \big[\hat{\xi}^{\kappa\gamma}_j(\theta_\beta)-{\xi}^{\kappa\gamma}_{j,\rm th}(\theta_\beta)\big] \ .
\end{aligned}
\end{equation}
Here, \textbf{Cov} is the data covariance matrix\footnote{Here we have neglected correlations between  $\xi^{\kappa\gamma}$ of different shear redshift bins ($i\ne j$) arising from the four-point correlation ($\langle \gamma_{i}\kappa\gamma_{j}\kappa\rangle-\langle \gamma_{i}\kappa\rangle \langle \gamma_{j}\kappa\rangle)\sim \langle \gamma_{i}\kappa\rangle \langle \gamma_{j}\kappa\rangle+\langle \gamma_{i}\gamma_{j}\rangle\langle \kappa\kappa\rangle $. For the current data, it is negligible comparing to the shape measurement error in $\gamma$ and shot noise in $\kappa$. }.

The best fit is
\ba
\big(A^{\text{bestfit}}\ \epsilon^{\text{bestfit}}\big)^{\mathbf{T}}=\bm{F}^{-1}\bigg[\sum_j\bm{\xi}_j^{\mathbf{T}}\textbf{Cov}_j^{-1}\bm{\hat{\xi}}_j\bigg]\ ,
\ea
where
\begin{equation}
    \bm{\xi}_j\equiv\begin{pmatrix}
    \xi_{j,{\rm th}}^{\kappa\gamma}(\theta_1) & \xi_{j,{\rm th}}^{\kappa m}(\theta_1) \\ 
    \xi_{j,{\rm th}}^{\kappa\gamma}(\theta_2) & \xi_{j,{\rm th}}^{\kappa m}(\theta_2)\\
    \vdots & \vdots\\
    \xi_{j,{\rm th}}^{\kappa\gamma}(\theta_n) & \xi_{j,{\rm th}}^{\kappa m}(\theta_n)
    \end{pmatrix}; \ \
\hat{\bm{\xi}}_j\equiv
\begin{pmatrix}
\hat{\xi}_j^{\kappa\gamma}(\theta_1)  \\ 
\hat{\xi}_j^{\kappa\gamma}(\theta_2) \\
\vdots\\
\hat{\xi}_j^{\kappa\gamma}(\theta_n) 
\end{pmatrix}\ .
\end{equation}
$\bm{F}$ is the fisher matrix of $A$ and $\epsilon$, 
\ba
\bm{F} =& 
\begin{pmatrix} 
\sum\limits_{j\alpha\beta}\xi^{\kappa\gamma}_{j,\alpha}\textbf{Cov}_{j}^{-1}\xi^{\kappa\gamma}_{j,\beta} & \sum\limits_{j\alpha\beta}\xi^{\kappa\gamma}_{j,\alpha}\textbf{Cov}_{j}^{-1}\xi^{\kappa m}_{j,\beta}\nonumber \\ 
\sum\limits_{j\alpha\beta}\xi^{\kappa\gamma}_{j,\alpha}\textbf{Cov}_{j}^{-1}\xi^{\kappa m}_{j,\beta} & \sum\limits_{j\alpha\beta}\xi^{\kappa m}_{j,\alpha}\textbf{Cov}_{j}^{-1}\xi^{\kappa m}_{j,\beta} \nonumber
\end{pmatrix}\
\ea
The associated errors and covariance matrix of the constraints are given by $\bm{F}^{-1}$. 

We use $\chi^2_{\rm min}$ and ${\chi^2_{\rm null}}$to demonstrate the goodness of fit to the model, where the $\chi^2_{\rm null}$ and $\chi^2_{\rm min}$ is  defined by
\ba
\label{eq:chi_1}
    \chi^2_{\rm null} = \sum\limits_{j\alpha\beta}
    \big[\hat{\xi}^{\kappa\gamma}_j(\theta_\alpha)\big]
    \textbf{Cov}_{j}^{-1} \big[\hat{\xi}^{\kappa\gamma}_j(\theta_\beta)\big] \ ,
\ea
\begin{equation}
\label{eq:chi_1}
\begin{aligned}
    \chi^2_{\rm min} &= \sum\limits_{j\alpha\beta}
    \big[\hat{\xi}^{\kappa\gamma}_j(\theta_\alpha)
    -{\xi}_{j,\rm th}^{\rm bestfit}(\theta_\alpha)\big]
    \times \textbf{Cov}_{j}^{-1} \\
    &\times \big[\hat{\xi}^{\kappa\gamma}_j(\theta_\beta)
    -{\xi}_{j,\rm th}^{\rm bestfit}(\theta_\beta)\big] \ ,
\end{aligned}
\end{equation}
where
\be
\xi_{\rm th}^{\rm bestfit}=
A^{\rm bestfit}\xi^{\kappa\gamma}+\epsilon^{\rm bestfit}\xi^{m\gamma}\ .
\ee

To achieve reasonable fitting results / $\chi^2_{\rm min}$, we cut the first 3 angular bins of the $\langle\hat\kappa\gamma\rangle$  measurements for $0.4<z_{\kappa}<0.6$ and the first 2 for $0.6<z_{\kappa}<0.8$. These small scales are expected to be affected by factors such as the baryonic feedback, which are not adequately captured by our model. We leave the investigation of these scales to future studies. Note that, for the joint analysis of convergence-convergence and convergence-shear presented in Appendix \ref{sec:approx_model}, we use the same scale cuts.

Table \ref{tab:robust} summarizes the fitting results.
The degrees of freedom ($\mathrm{d.o.f.}$) for the fitting are 28, 33, and 43 for the convergence maps at $0.4 < z_{\kappa} < 0.6$, $0.6 < z_{\kappa} < 0.8$, and $0.8 < z_{\kappa} < 1.0$, respectively.
The two-parameter fitting returns a reasonable $\chi^2_{\rm min}/\mathrm{d.o.f.} \sim 1$, supporting our theoretical modeling.

According to $A$, we detect a convergence-shear cross-correlation signal with $S/N \equiv A/\sigma_A \gtrsim 10$ for the three convergence maps.
The $S/N$ reaches 20 for the convergence map at $0.8 < z_{\kappa} < 1.0$.
Deviation from one is detected in $A$ and increases as the redshift decreases, starting from $A \sim 1.2$ for $0.8 < z_{\kappa} < 1.0$ and reaching $\sim 3$ for $0.4 < z_{\kappa} < 0.6$.
This tendency is consistent with the results of our reconstruction for the DECaLS galaxies \citep{Qin+}, where $A$ is found to be $\sim 0.9$ for $0.9 < z_{\kappa} < 1.2$ and $\sim 2.5$ for $0.5 < z_{\kappa} < 0.8$.
Why $A\neq 1$ and why $A$ has such a redshift dependence are certainly key issues for future investigation.

According to $\epsilon$, galaxy intrinsic clustering is suppressed by a factor of approximately 10 to 100, reaching levels of $\sim-0.10 \pm 0.02$ for $0.4 < z_{\kappa} < 0.6$ and $0.6 < z_{\kappa} < 0.8$, and $-0.01 \pm 0.04$ for $0.8 < z_{\kappa} < 1.0$.
The best-fit to the cross-correlation measurements is shown in \reffig{fig:xi}, with the two components $A\xi^{\kappa\gamma}$ and $\epsilon\xi^{m\gamma}$ also highlighted.
The intrinsic clustering term $\epsilon\xi^{m\gamma}$ is negligible for $z_{\kappa} > z_{\gamma}$ (the lower left panels of \reffig{fig:xi}) and becomes significant for $z_{\kappa} < z_{\gamma}$ (the upper right panels of \reffig{fig:xi}).
This pattern matches the behavior of the theoretical templates, where matter only correlates with shear when it is at a lower redshift than the shear galaxies.
However, our reconstruction suppresses the average galaxy intrinsic clustering, resulting in a small $\epsilon$, thereby suppressing $\epsilon\xi^{m\gamma}$.
This makes the signal term $A\xi^{\kappa\gamma}$ the dominant term in the measured cross-correlations.


With non-independent data points and reasonable covariance estimation, $\chi^2_{\rm null} = 145.2, 375.0, 833.1$ and $\chi^2_{\min} = 35.2, 36.5, 26.5$ with respect to the null and to the model, respectively.
According to the data-driven signal-to-noise ratio $\sqrt{\chi^2_{\rm null}}$, 
we achieve $12.0\sigma, 19.4\sigma, 28.9\sigma$ detection of non-zero cross-correlation signals for the three photo-$z$ bins of $\hat{\kappa}$.
Based on the fitting-driven signal-to-noise ratio $\sqrt{\chi^2_{\rm null} - \chi^2_{\rm min}}$, the significance is approximately $10.5\sigma, 18.4\sigma, 28.4\sigma$.
The lensing signal is detected at lower significance ($A/\sigma_A=9, 16, 20$).
Table \ref{tab:robust} also compares results before and after imaging systematics mitigation, showing consistency in the fitting results.
After mitigation, both $\sqrt{\chi^2_{\rm null} - \chi^2_{\rm min}}$ and $A/\sigma_A$ exhibit increases.

\subsubsection{Internal test of the convergence-shear cross-correlation analysis}\label{sec:robustest}
We evaluate the influence of various factors in our analysis, as presented in Table \ref{tab:robust}- \ref{tab:AIC} and \reffig{fig:Ae_dropZg}
\begin{itemize}
    \item 
    Impact of $f^{\rm threshold}$.
    A higher $f^{\rm threshold}$ implies a stricter selection of pixels in the galaxy number overdensity map.
    Table \ref{tab:robust} indicates that the effect of $f^{\rm threshold}$ is minimal, and the fitting results remain consistent across $f^{\rm threshold} = 0.5, 0.7, 0.9$.
    \item 
    Impact of $N_F$.  The constraints on $A$ and $\epsilon$ are stable across $N_F = 3, 4, 5$. The parameter $\epsilon = \Sigma_i w_i b_i$ depends on the galaxy biases in the chosen flux bins and, consequently, on the number of flux bins $N_F$. The stability of $\epsilon$ across different $N_F$ values suggests that our method for eliminating galaxy intrinsic clustering is robust.
    \item 
    Impact of the magnitude cut.
    We select galaxies that are 0.5 magnitudes brighter than the baseline set.
    The first scenario involves selecting galaxies that are 0.5 magnitudes brighter across all three bands, denoted as $m_{g,r,z}-0.5$.
    The other three scenarios involve selecting galaxies that are 0.5 magnitudes brighter in each band individually, denoted as $m_{g}-0.5$, $m_{r}-0.5$, and $m_{z}-0.5$.
    A brighter magnitude cut of 0.5 results in a reduction of approximately 30$-$40\% in the number of galaxies for a particular photometry band.
    Consequently, the $S/N$ of the cross-correlation signal is reduced for a brighter magnitude cut.
    Nevertheless, the constraints on $A$ and $\epsilon$ remain stable across different magnitude cuts,
    as shown in Table \ref{tab:robust}.
        \item 
    Impact of imaging systematics.
    Since they are unlikely to cause density fluctuations correlated with cosmic shear, 
    their main impact is to introduce spurious fluctuations from contamination such as stellar density, Galactic extinction, sky brightness, seeing, and airmass.
    Table \ref{tab:robust} shows that the fitting results are consistent before and after the mitigation.
     However errorbars in $A$ and $\epsilon$ shrink after the mitigation.
    \item 
    Impact of the shear catalog. 
    Since parameters $A$ and $\epsilon$ characterize the reconstructed lensing convergence, they should be independent of the shear catalog used. 
    We investigate the impact of selecting different shear catalogs by systematically excluding one shear photo-$z$ bin at a time. 
    The results are presented in Fig. \ref{fig:Ae_dropZg}. 
    For all the investigated cases, the constraints on $A$ and $\epsilon$ align with the baseline set within the $1\sigma$ error range, indicating the robustness of the analysis against different shear samples.
     \item
    Akaike information criterion (AIC) analysis. 
    We adjust the theoretical template to compare different models.
    Besides the baseline model described by Eq. \eqref{eq:xikg_decom}, we also consider alternative models where either $A$ or $\epsilon$ is fixed.
    We compare four scenarios: the baseline model, fixing $A=0$, fixing $A=1$, and fixing $\epsilon=0$.
    For each model, we repeat the fitting process and then calculate the AIC, which can be expressed, to second-order, as:
    $$
    \mathrm{AIC}_{\mathrm{c}}=-2 \ln p\left(\lambda_{\text {bestfit }} \mid D\right)+2 N_\lambda+\frac{2 N_\lambda+1}{N_D-N_\lambda-1} .
    $$
    Here, $N_\lambda$ is the number of parameters, $N_D$ is the number of data points, $p$ is the likelihood which is Gaussian in our case, and $p\left(\lambda_{\text {bestfit }} \mid D\right)$ is the value for the best-fit parameters.
    The \textit{best} model has the smallest AIC. A model with an AIC larger by 10 or more should be ruled out, while a difference of less than 2 suggests the models are indistinguishable.
    Table \ref{tab:AIC} shows the results of $\chi^2_{\rm min}$ (the minimum $\chi^2$ value obtained during fitting) and AIC.
    In all cases, except where $\epsilon=0$ for $0.8 < z_\kappa < 1.0$, alternative models are ruled out by AIC.
    This suggests the robustness of the weak lensing signal detection (ruling out $A=0$),
    the significance of the multiplicative bias in the amplitude of $\hat{\kappa}$ (ruling out $A=1$),
    and the necessity of the intrinsic clustering term in the model (ruling out $\epsilon=0$).

\end{itemize}

\begin{figure}
\centering
\includegraphics[width=1.05\columnwidth]{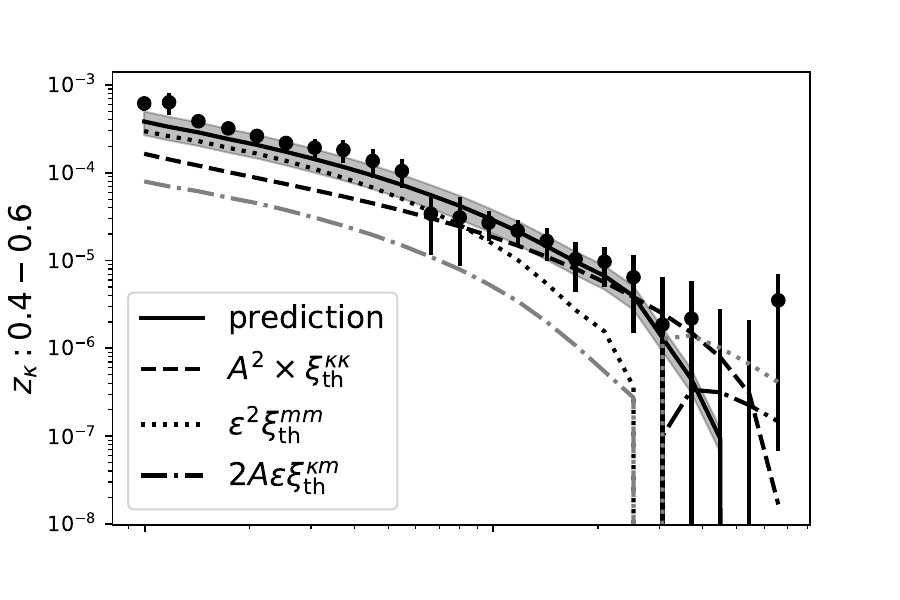}
\vspace{-0.85cm}
\centering
\vspace{-0.85cm}
\includegraphics[width=1.05\columnwidth]{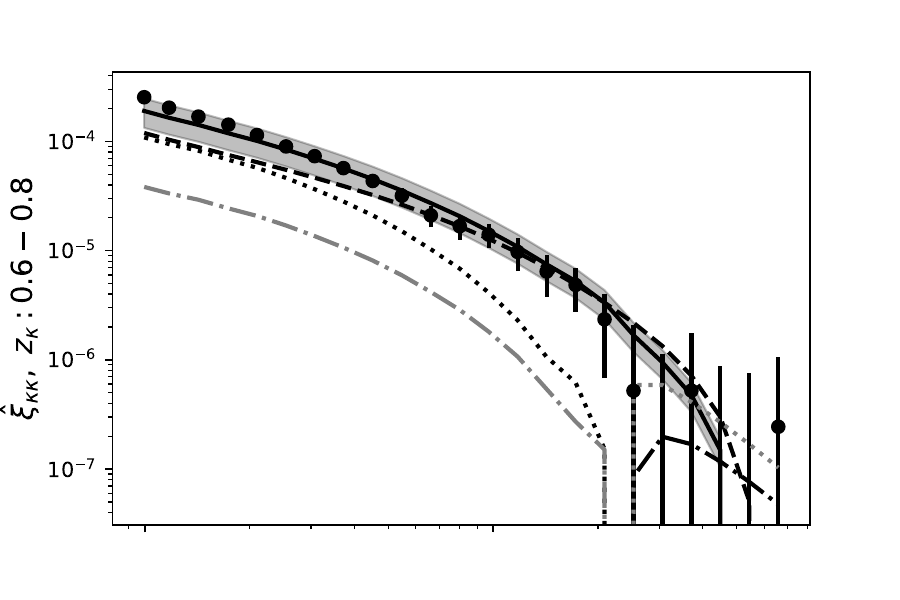}
\vspace{-0.85cm}
\centering
\vspace{-0.85cm}
\includegraphics[width=1.05\columnwidth]{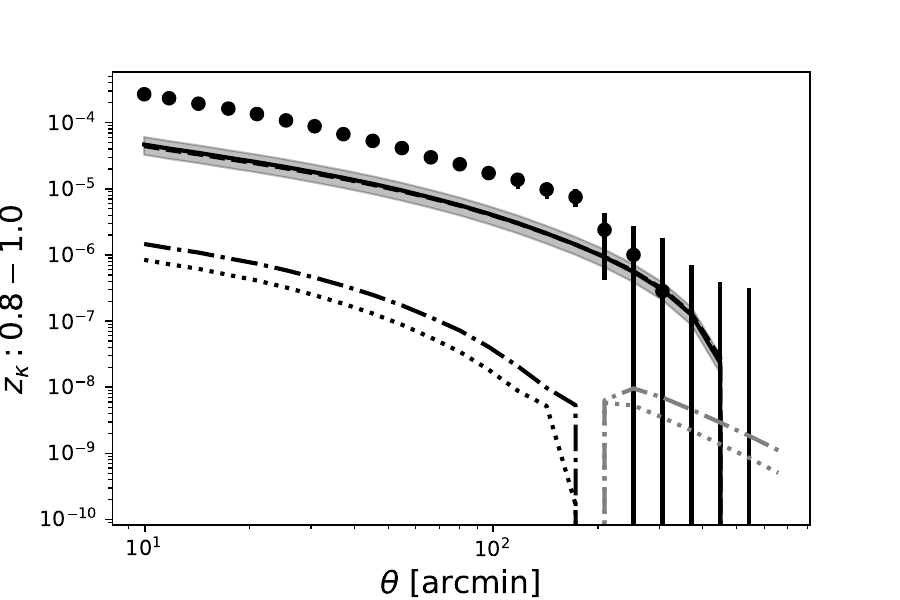}
\caption{The auto-correlation measurements of the reconstructed lensing convergence and their comparison to the prediction.
Three photo-$z$ bins are shown in the three panels.
The errors of the measurements are estimated by the jackknife method.
The black solid line represents the prediction from the deterministic terms, with parameters $A$ and $\epsilon$ derived from the convergence-shear cross-correlation analysis.
The shaded region indicates the 1$\sigma$ error of the prediction, calculated through error propagation of the fitting results for $A$ and $\epsilon$.
The three components of the prediction, i.e., the $A^2\xi^{\kappa\kappa}$, $\epsilon^2\xi^{mm}$, and $2A\epsilon\xi^{m\kappa}$,
are shown with dashed, dotted, and dash-dotted lines, respectively.
The light-grey lines indicate where the values are negative.
For the photo-$z$ bin $0.4 < z_\kappa < 0.6$ and $0.6 < z_\kappa < 0.8$, the predictions match the measurements well.
The $A^2\xi^{\kappa\kappa}$ term dominates the prediction at large scales, while the $\epsilon^2\xi^{mm}$ term dominates at small scales.
For the photo-$z$ bin $0.8 < z_\kappa < 1.0$, the measurements are significantly higher than the prediction at all scales.
the stochastic term, which are not included in the prediction, are expected to contribute to this discrepancy.
For clarity, only the results after the imaging systematics mitigation are shown here.
The measurements before/after the imaging systematics mitigation are compared in \reffig{fig:clkk_imgW}.
}
\label{fig:clkk}
\end{figure}

\begin{figure}
\centering
\includegraphics[width=1.05\columnwidth]{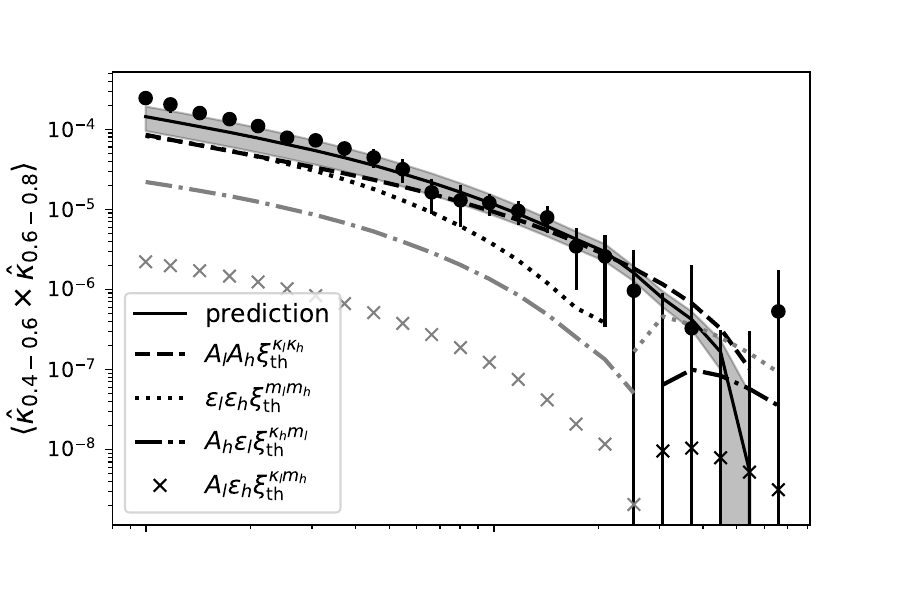}
\vspace{-0.85cm}
\centering
\vspace{-0.85cm}
\includegraphics[width=1.05\columnwidth]{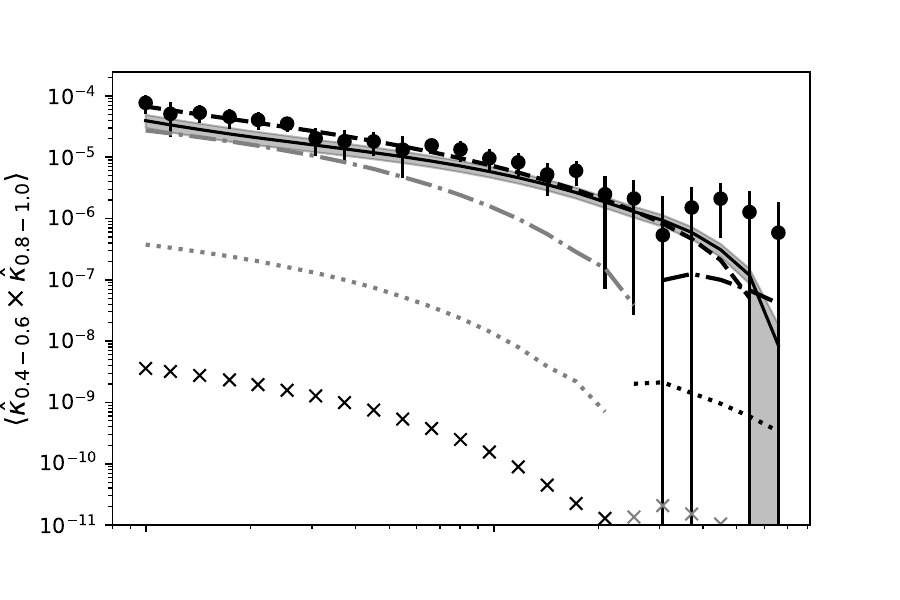}
\vspace{-0.85cm}
\centering
\vspace{-0.85cm}
\includegraphics[width=1.05\columnwidth]{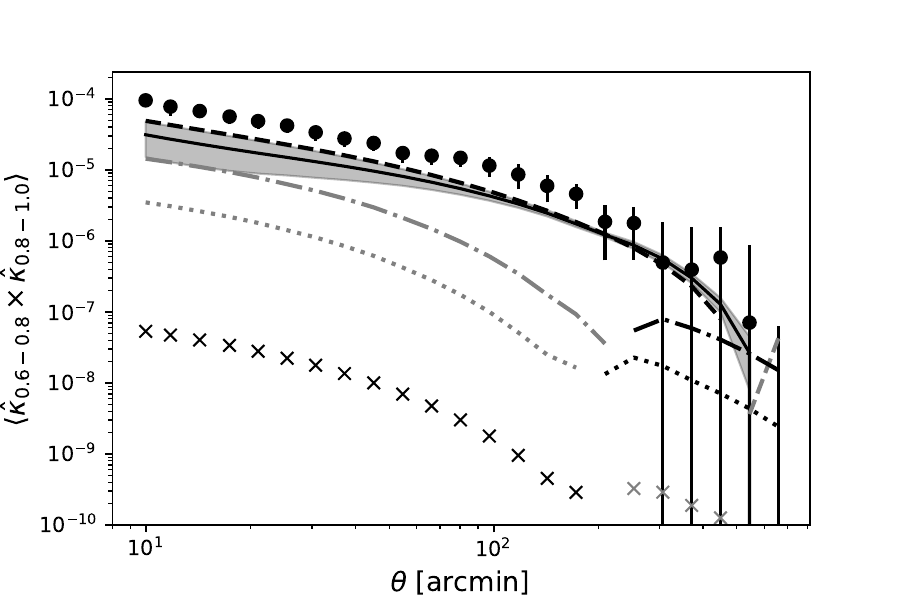}

\caption{Same as \reffig{fig:clkk}, but for the cross-correlation between different photo-$z$ bins of the reconstructed lensing convergence.
The three panels show different combinations of the photo-$z$ bins.
The prediction consists of four components: $A_lA_h\xi^{\kappa_l\kappa_h}$, $\epsilon_l\epsilon_h\xi^{m_lm_h}$, $A_l\epsilon_h\xi^{\kappa_l m_h}$, and $A_h\epsilon_l\xi^{\kappa_h m_l}$, 
where the indices $l$/$h$ denote the lower/higher redshift bins. 
The light-grey lines indicate where the values are negative.
For the $(0.4 < z_\kappa < 0.6) \times (0.6 < z_\kappa < 0.8)$ case, the predictions match the measurements well,
consistent with the auto-correlation results for the two bins. 
The $A_lA_h\xi^{\kappa_l\kappa_h}$ term dominates the prediction at large scales, while the $\epsilon_l\epsilon_h\xi^{m_lm_h}$ term becomes important at small scales.
For the other two cases, the predictions are systematically lower,
which are expected to result from the influence of stochastic terms for $0.8 < z_\kappa < 1.0$. 
Notably, the discrepancy is smaller than in the auto-correlation for $0.8 < z_\kappa < 1.0$, 
which aligns with expectations as the deterministic terms perform well for the lower two redshift bins.
}
\label{fig:clk1k2}
\end{figure}

\subsection{Convergence-convergence correlations}

\reffig{fig:clkk} shows the measured $\hat\xi^{\kappa\kappa}_{ij}$ for $i=j$, representing the auto-correlation of the reconstructed lensing convergence.
We also compare the measurements with the deterministic terms $\xi^{\rm D}$ (see Eq.\eqref{eq:xik1k2_decom}) combined with the fitting results of $A$ and $\epsilon$ from the convergence-shear cross-correlation analysis.

For the photo-$z$ bins $0.4 < z_{\kappa} < 0.6$ and $0.6 < z_\kappa < 0.8$, the deterministic terms align well with the measurements.
The lensing term $A^2\xi^{\kappa\kappa}$ dominates the prediction at large scales, while the intrinsic clustering term $\epsilon^2\xi^{mm}$ becomes more significant at small scales.
This implies that the $\hat\kappa$ map has sub-dominant contamination from $\kappa^{\rm S}$.

However, for $0.8 < z_{\kappa} < 1.0$, the measurements exceed the deterministic terms across all scales, indicating the significance of stochastic terms for this bin.
Another potential factor for this discrepancy is residual imaging systematics.
The mitigation has a more significant impact on the auto-correlation than on the convergence-shear cross-correlation, as detailed in Appendix \ref{sec:imgweight}.

The measurements of $\hat\xi^{\kappa\kappa}_{ij}$ and the comparison with $\xi^{\rm D}$ for $i \neq j$ are shown in Figure \ref{fig:clk1k2}.
For the $(0.4 < z_{\kappa} < 0.6) \times (0.6 < z_{\kappa} < 0.8)$ case, predictions align well with measurements.
The lensing term $A_lA_h\xi^{\kappa_l\kappa_h}$ dominates at large scales, while the intrinsic clustering term $\epsilon_l\epsilon_h\xi^{m_lm_h}$ becomes important at small scales.
This is consistent with the auto-correlation behavior for the $0.4 < z_{\kappa} < 0.6$ and $0.6 < z_{\kappa} < 0.8$ bins, where deterministic terms match measurements.
For the $(0.6 < z_{\kappa} < 0.8) \times (0.8 < z_{\kappa} < 1.0)$ and $(0.4 < z_{\kappa} < 0.6) \times (0.8 < z_{\kappa} < 1.0)$ cases, the measurements are systematically higher than $\xi^{\rm D}$.
However, the discrepancy is smaller than in the auto-correlation case for $0.8 < z_{\kappa} < 1.0$, aligning with expectations, as the bias mainly originates from the ${\kappa}$ map at $0.8 < z_{\kappa} < 1.0$.
Despite the issues in $0.8 < z_{\kappa} < 1.0$, the $\hat\kappa$ map for $0.4 < z_{\kappa} < 0.6$ and $0.6 < z_{\kappa} < 0.8$ is shown to have negligible / sub-dominant $\kappa^{\rm S}$.

\section{SUMMARY and discussion}\label{sec:summary5}

Using cosmic magnification, we reconstruct the weak lensing convergence maps from the DES Y3 galaxies.
It is achieved by weighting the overdensity maps of galaxies across different magnitude bins and photometric bands.
To validate the reconstruction, we measure and compare the convergence-shear and convergence-convergence correlations with the theoretical model.
The $\hat{\kappa}-\gamma$ analysis indicates that our method effectively eliminates galaxy intrinsic clustering, achieving $\sim10\sigma$ to $\sim20\sigma$ detection of the $\hat{\kappa}-\gamma$ cross-correlation signal.
The $\hat{\kappa}-\hat{\kappa}$ correlations of the $0.4<z_\kappa<0.6$ and $0.6<z_\kappa<0.8$ bins show reasonably good agreement with predictions based on theoretical interpretation of $\hat{\kappa}-\gamma$ measurement.

Several issues remain prominent for future improvement.
The first is the bias in the overall amplitude of the reconstructed lensing convergence maps.
The bias increases as the redshift decreases, with values ranging from $\sim1.2$ to $\sim3.2$.
The agreement between the convergence-shear and convergence-convergence analyses for the lower two convergence redshift bins indicates that the $\langle\hat{\kappa}\hat{\kappa}\rangle$ measurements also support the existence of the bias (see Appendix \ref{sec:approx_model}).
The amplitude and redshift dependence of the bias are similar to those observed in the reconstructed convergence for the DECaLS galaxies \citep{Qin+}, where tests suggest that the biases are potentially caused by approximations in the galaxy selection function.
This approximation, which treats galaxy selection as flux-limited, is commonly used in cosmic magnification signal detections \citep[e.g.,][]{2017A&A...608A.141T,2021A&A...656A..99B,Crespo_GonzalezNuevo_Bonavera_Cueli_Casas_Goitia_2022}.
Further investigation into the forward modeling \citep{Wietersheim-Kramsta_Joachimi_van,JElvinPoole2022DarkES} of the galaxy selection function is crucial for the effective application of cosmic magnification in weak lensing studies.
Another issue is the discrepancy observed in the $\hat{\kappa}-\hat{\kappa}$ measurements of the $0.8<z_\kappa<1.0$ bin.
A better understanding of galaxy stochasticity \citep[e.g.,][]{2024arXiv240603018Z, 1999ApJ...518L..69T, 2009MNRAS.396.1610B} is needed to address this discrepancy.

Another existing dataset suitable for our reconstruction method is the HSC \citep[][]{HSC2018} survey, which has a depth $\sim 2$ magnitude deeper than DES.
It is expected that the magnification signal, especially the $\hat{\kappa}-\hat{\kappa}$ correlations, from HSC will be further improved.
This is the next step in our research.
By leveraging the DECaLS, DES, and HSC data, along with insights from simulations \citep{2024MNRAS.527.7547M,2024PhRvD.110f3551Z}, we aim to validate and refine our method, ultimately paving the way for its cosmological application to future, even deeper surveys like LSST \citep[LSST,][]{2009arXiv0912.0201L}, Euclid \citep{2011arXiv1110.3193L} and CSST \citep{2019ApJ...883..203G}.




%

\section*{Acknowledgements}
This work is supported the National Key R\&D Program of China (2023YFA1607800, 2023YFA1607801, 2023YFA1607802, 2020YFC2201602), the China Manned Space Project (\#CMS-CSST-2021-A02), and the Fundamental Research Funds for the Central Universities.

\bibliography{mybib}

\begin{thebibliography}{80}
\expandafter\ifx\csname natexlab\endcsname\relax\def\natexlab#1{#1}\fi
\expandafter\ifx\csname bibnamefont\endcsname\relax
  \def\bibnamefont#1{#1}\fi
\expandafter\ifx\csname bibfnamefont\endcsname\relax
  \def\bibfnamefont#1{#1}\fi
\expandafter\ifx\csname citenamefont\endcsname\relax
  \def\citenamefont#1{#1}\fi
\expandafter\ifx\csname url\endcsname\relax
  \def\url#1{\texttt{#1}}\fi
\expandafter\ifx\csname urlprefix\endcsname\relax\def\urlprefix{URL }\fi
\providecommand{\bibinfo}[2]{#2}
\providecommand{\eprint}[2][]{\url{#2}}

\bibitem[{\citenamefont{{Qin} et~al.}(2023)\citenamefont{{Qin}, {Zhang}, {Xu},
  {Yu}, {Yao}, {Ma}, and {Shan}}}]{Qin+}
\bibinfo{author}{\bibfnamefont{J.}~\bibnamefont{{Qin}}},
  \bibinfo{author}{\bibfnamefont{P.}~\bibnamefont{{Zhang}}},
  \bibinfo{author}{\bibfnamefont{H.}~\bibnamefont{{Xu}}},
  \bibinfo{author}{\bibfnamefont{Y.}~\bibnamefont{{Yu}}},
  \bibinfo{author}{\bibfnamefont{J.}~\bibnamefont{{Yao}}},
  \bibinfo{author}{\bibfnamefont{R.}~\bibnamefont{{Ma}}}, \bibnamefont{and}
  \bibinfo{author}{\bibfnamefont{H.}~\bibnamefont{{Shan}}},
  \bibinfo{journal}{arXiv e-prints} \bibinfo{eid}{arXiv:2310.15053}
  (\bibinfo{year}{2023}), \eprint{2310.15053}.

\bibitem[{\citenamefont{{Bartelmann} and
  {Schneider}}(2001{\natexlab{a}})}]{2001PhR...340..291B}
\bibinfo{author}{\bibfnamefont{M.}~\bibnamefont{{Bartelmann}}}
  \bibnamefont{and}
  \bibinfo{author}{\bibfnamefont{P.}~\bibnamefont{{Schneider}}},
  \bibinfo{journal}{\physrep} \textbf{\bibinfo{volume}{340}},
  \bibinfo{pages}{291} (\bibinfo{year}{2001}{\natexlab{a}}),
  \eprint{astro-ph/9912508}.

\bibitem[{\citenamefont{{Kilbinger}}(2015{\natexlab{a}})}]{2015RPPh...78h6901K}
\bibinfo{author}{\bibfnamefont{M.}~\bibnamefont{{Kilbinger}}},
  \bibinfo{journal}{Reports on Progress in Physics}
  \textbf{\bibinfo{volume}{78}}, \bibinfo{eid}{086901}
  (\bibinfo{year}{2015}{\natexlab{a}}), \eprint{1411.0115}.

\bibitem[{\citenamefont{{Bartelmann} and
  {Schneider}}(2001{\natexlab{b}})}]{BS2001}
\bibinfo{author}{\bibfnamefont{M.}~\bibnamefont{{Bartelmann}}}
  \bibnamefont{and}
  \bibinfo{author}{\bibfnamefont{P.}~\bibnamefont{{Schneider}}},
  \bibinfo{journal}{\physrep} \textbf{\bibinfo{volume}{340}},
  \bibinfo{pages}{291} (\bibinfo{year}{2001}{\natexlab{b}}),
  \eprint{astro-ph/9912508}.

\bibitem[{\citenamefont{{Hoekstra} and {Jain}}(2008)}]{HJ2008}
\bibinfo{author}{\bibfnamefont{H.}~\bibnamefont{{Hoekstra}}} \bibnamefont{and}
  \bibinfo{author}{\bibfnamefont{B.}~\bibnamefont{{Jain}}},
  \bibinfo{journal}{Annual Review of Nuclear and Particle Science}
  \textbf{\bibinfo{volume}{58}}, \bibinfo{pages}{99} (\bibinfo{year}{2008}),
  \eprint{0805.0139}.

\bibitem[{\citenamefont{{Van Waerbeke} et~al.}(2010)\citenamefont{{Van
  Waerbeke}, {Hildebrandt}, {Ford}, and {Milkeraitis}}}]{van2010a}
\bibinfo{author}{\bibfnamefont{L.}~\bibnamefont{{Van Waerbeke}}},
  \bibinfo{author}{\bibfnamefont{H.}~\bibnamefont{{Hildebrandt}}},
  \bibinfo{author}{\bibfnamefont{J.}~\bibnamefont{{Ford}}}, \bibnamefont{and}
  \bibinfo{author}{\bibfnamefont{M.}~\bibnamefont{{Milkeraitis}}},
  \bibinfo{journal}{\apjl} \textbf{\bibinfo{volume}{723}}, \bibinfo{pages}{L13}
  (\bibinfo{year}{2010}), \eprint{1004.3793}.

\bibitem[{\citenamefont{{Fu} and {Fan}}(2014)}]{FuFan2014}
\bibinfo{author}{\bibfnamefont{L.-P.} \bibnamefont{{Fu}}} \bibnamefont{and}
  \bibinfo{author}{\bibfnamefont{Z.-H.} \bibnamefont{{Fan}}},
  \bibinfo{journal}{Research in Astronomy and Astrophysics}
  \textbf{\bibinfo{volume}{14}}, \bibinfo{eid}{1061-1120}
  (\bibinfo{year}{2014}).

\bibitem[{\citenamefont{{Kilbinger}}(2015{\natexlab{b}})}]{Kil2015}
\bibinfo{author}{\bibfnamefont{M.}~\bibnamefont{{Kilbinger}}},
  \bibinfo{journal}{Reports on Progress in Physics}
  \textbf{\bibinfo{volume}{78}}, \bibinfo{eid}{086901}
  (\bibinfo{year}{2015}{\natexlab{b}}), \eprint{1411.0115}.

\bibitem[{\citenamefont{{Peacock} et~al.}(2006)\citenamefont{{Peacock},
  {Schneider}, {Efstathiou}, {Ellis}, {Leibundgut}, {Lilly}, and
  {Mellier}}}]{2006ewg3.rept.....P}
\bibinfo{author}{\bibfnamefont{J.~A.} \bibnamefont{{Peacock}}},
  \bibinfo{author}{\bibfnamefont{P.}~\bibnamefont{{Schneider}}},
  \bibinfo{author}{\bibfnamefont{G.}~\bibnamefont{{Efstathiou}}},
  \bibinfo{author}{\bibfnamefont{J.~R.} \bibnamefont{{Ellis}}},
  \bibinfo{author}{\bibfnamefont{B.}~\bibnamefont{{Leibundgut}}},
  \bibinfo{author}{\bibfnamefont{S.~J.} \bibnamefont{{Lilly}}},
  \bibnamefont{and}
  \bibinfo{author}{\bibfnamefont{Y.}~\bibnamefont{{Mellier}}},
  \emph{\bibinfo{title}{{ESA-ESO Working Group on ``Fundamental Cosmology''}}},
  \bibinfo{howpublished}{``ESA-ESO Working Group on ''Fundamental Cosmology``,
  Edited by J.A. Peacock et al. ESA, 2006.''} (\bibinfo{year}{2006}),
  \eprint{astro-ph/0610906}.

\bibitem[{\citenamefont{{Albrecht} et~al.}(2006)\citenamefont{{Albrecht},
  {Bernstein}, {Cahn}, {Freedman}, {Hewitt}, {Hu}, {Huth}, {Kamionkowski},
  {Kolb}, {Knox} et~al.}}]{2006astro.ph..9591A}
\bibinfo{author}{\bibfnamefont{A.}~\bibnamefont{{Albrecht}}},
  \bibinfo{author}{\bibfnamefont{G.}~\bibnamefont{{Bernstein}}},
  \bibinfo{author}{\bibfnamefont{R.}~\bibnamefont{{Cahn}}},
  \bibinfo{author}{\bibfnamefont{W.~L.} \bibnamefont{{Freedman}}},
  \bibinfo{author}{\bibfnamefont{J.}~\bibnamefont{{Hewitt}}},
  \bibinfo{author}{\bibfnamefont{W.}~\bibnamefont{{Hu}}},
  \bibinfo{author}{\bibfnamefont{J.}~\bibnamefont{{Huth}}},
  \bibinfo{author}{\bibfnamefont{M.}~\bibnamefont{{Kamionkowski}}},
  \bibinfo{author}{\bibfnamefont{E.~W.} \bibnamefont{{Kolb}}},
  \bibinfo{author}{\bibfnamefont{L.}~\bibnamefont{{Knox}}},
  \bibnamefont{et~al.}, \bibinfo{journal}{arXiv e-prints}
  \bibinfo{eid}{astro-ph/0609591} (\bibinfo{year}{2006}),
  \eprint{astro-ph/0609591}.

\bibitem[{\citenamefont{{Dark Energy Survey Collaboration}
  et~al.}(2016)\citenamefont{{Dark Energy Survey Collaboration}, {Abbott},
  {Abdalla}, {Aleksi{\'c}}, {Allam}, {Amara}, {Bacon}, {Balbinot}, {Banerji},
  {Bechtol} et~al.}}]{DES2016}
\bibinfo{author}{\bibnamefont{{Dark Energy Survey Collaboration}}},
  \bibinfo{author}{\bibfnamefont{T.}~\bibnamefont{{Abbott}}},
  \bibinfo{author}{\bibfnamefont{F.~B.} \bibnamefont{{Abdalla}}},
  \bibinfo{author}{\bibfnamefont{J.}~\bibnamefont{{Aleksi{\'c}}}},
  \bibinfo{author}{\bibfnamefont{S.}~\bibnamefont{{Allam}}},
  \bibinfo{author}{\bibfnamefont{A.}~\bibnamefont{{Amara}}},
  \bibinfo{author}{\bibfnamefont{D.}~\bibnamefont{{Bacon}}},
  \bibinfo{author}{\bibfnamefont{E.}~\bibnamefont{{Balbinot}}},
  \bibinfo{author}{\bibfnamefont{M.}~\bibnamefont{{Banerji}}},
  \bibinfo{author}{\bibfnamefont{K.}~\bibnamefont{{Bechtol}}},
  \bibnamefont{et~al.}, \bibinfo{journal}{\mnras}
  \textbf{\bibinfo{volume}{460}}, \bibinfo{pages}{1270} (\bibinfo{year}{2016}),
  \eprint{1601.00329}.

\bibitem[{\citenamefont{{de Jong} et~al.}(2013)\citenamefont{{de Jong},
  {Verdoes Kleijn}, {Kuijken}, and {Valentijn}}}]{KiDS2013}
\bibinfo{author}{\bibfnamefont{J.~T.~A.} \bibnamefont{{de Jong}}},
  \bibinfo{author}{\bibfnamefont{G.~A.} \bibnamefont{{Verdoes Kleijn}}},
  \bibinfo{author}{\bibfnamefont{K.~H.} \bibnamefont{{Kuijken}}},
  \bibnamefont{and} \bibinfo{author}{\bibfnamefont{E.~A.}
  \bibnamefont{{Valentijn}}}, \bibinfo{journal}{Experimental Astronomy}
  \textbf{\bibinfo{volume}{35}}, \bibinfo{pages}{25} (\bibinfo{year}{2013}),
  \eprint{1206.1254}.

\bibitem[{\citenamefont{{Aihara} et~al.}(2018)\citenamefont{{Aihara},
  {Arimoto}, {Armstrong}, {Arnouts}, {Bahcall}, {Bickerton}, {Bosch}, {Bundy},
  {Capak}, {Chan} et~al.}}]{HSC2018}
\bibinfo{author}{\bibfnamefont{H.}~\bibnamefont{{Aihara}}},
  \bibinfo{author}{\bibfnamefont{N.}~\bibnamefont{{Arimoto}}},
  \bibinfo{author}{\bibfnamefont{R.}~\bibnamefont{{Armstrong}}},
  \bibinfo{author}{\bibfnamefont{S.}~\bibnamefont{{Arnouts}}},
  \bibinfo{author}{\bibfnamefont{N.~A.} \bibnamefont{{Bahcall}}},
  \bibinfo{author}{\bibfnamefont{S.}~\bibnamefont{{Bickerton}}},
  \bibinfo{author}{\bibfnamefont{J.}~\bibnamefont{{Bosch}}},
  \bibinfo{author}{\bibfnamefont{K.}~\bibnamefont{{Bundy}}},
  \bibinfo{author}{\bibfnamefont{P.~L.} \bibnamefont{{Capak}}},
  \bibinfo{author}{\bibfnamefont{J.~H.~H.} \bibnamefont{{Chan}}},
  \bibnamefont{et~al.}, \bibinfo{journal}{\pasj} \textbf{\bibinfo{volume}{70}},
  \bibinfo{eid}{S4} (\bibinfo{year}{2018}), \eprint{1704.05858}.

\bibitem[{\citenamefont{{LSST Science Collaboration}
  et~al.}(2009)\citenamefont{{LSST Science Collaboration}, {Abell}, {Allison},
  {Anderson}, {Andrew}, {Angel}, {Armus}, {Arnett}, {Asztalos}, {Axelrod}
  et~al.}}]{2009arXiv0912.0201L}
\bibinfo{author}{\bibnamefont{{LSST Science Collaboration}}},
  \bibinfo{author}{\bibfnamefont{P.~A.} \bibnamefont{{Abell}}},
  \bibinfo{author}{\bibfnamefont{J.}~\bibnamefont{{Allison}}},
  \bibinfo{author}{\bibfnamefont{S.~F.} \bibnamefont{{Anderson}}},
  \bibinfo{author}{\bibfnamefont{J.~R.} \bibnamefont{{Andrew}}},
  \bibinfo{author}{\bibfnamefont{J.~R.~P.} \bibnamefont{{Angel}}},
  \bibinfo{author}{\bibfnamefont{L.}~\bibnamefont{{Armus}}},
  \bibinfo{author}{\bibfnamefont{D.}~\bibnamefont{{Arnett}}},
  \bibinfo{author}{\bibfnamefont{S.~J.} \bibnamefont{{Asztalos}}},
  \bibinfo{author}{\bibfnamefont{T.~S.} \bibnamefont{{Axelrod}}},
  \bibnamefont{et~al.}, \bibinfo{journal}{arXiv e-prints}
  \bibinfo{eid}{arXiv:0912.0201} (\bibinfo{year}{2009}), \eprint{0912.0201}.

\bibitem[{\citenamefont{{Laureijs} et~al.}(2011)\citenamefont{{Laureijs},
  {Amiaux}, {Arduini}, {Augu{\`e}res}, {Brinchmann}, {Cole}, {Cropper},
  {Dabin}, {Duvet}, {Ealet} et~al.}}]{2011arXiv1110.3193L}
\bibinfo{author}{\bibfnamefont{R.}~\bibnamefont{{Laureijs}}},
  \bibinfo{author}{\bibfnamefont{J.}~\bibnamefont{{Amiaux}}},
  \bibinfo{author}{\bibfnamefont{S.}~\bibnamefont{{Arduini}}},
  \bibinfo{author}{\bibfnamefont{J.~L.} \bibnamefont{{Augu{\`e}res}}},
  \bibinfo{author}{\bibfnamefont{J.}~\bibnamefont{{Brinchmann}}},
  \bibinfo{author}{\bibfnamefont{R.}~\bibnamefont{{Cole}}},
  \bibinfo{author}{\bibfnamefont{M.}~\bibnamefont{{Cropper}}},
  \bibinfo{author}{\bibfnamefont{C.}~\bibnamefont{{Dabin}}},
  \bibinfo{author}{\bibfnamefont{L.}~\bibnamefont{{Duvet}}},
  \bibinfo{author}{\bibfnamefont{A.}~\bibnamefont{{Ealet}}},
  \bibnamefont{et~al.}, \bibinfo{journal}{arXiv e-prints}
  \bibinfo{eid}{arXiv:1110.3193} (\bibinfo{year}{2011}), \eprint{1110.3193}.

\bibitem[{\citenamefont{{Gong} et~al.}(2019)\citenamefont{{Gong}, {Liu}, {Cao},
  {Chen}, {Fan}, {Li}, {Li}, {Li}, {Zhang}, and {Zhan}}}]{2019ApJ...883..203G}
\bibinfo{author}{\bibfnamefont{Y.}~\bibnamefont{{Gong}}},
  \bibinfo{author}{\bibfnamefont{X.}~\bibnamefont{{Liu}}},
  \bibinfo{author}{\bibfnamefont{Y.}~\bibnamefont{{Cao}}},
  \bibinfo{author}{\bibfnamefont{X.}~\bibnamefont{{Chen}}},
  \bibinfo{author}{\bibfnamefont{Z.}~\bibnamefont{{Fan}}},
  \bibinfo{author}{\bibfnamefont{R.}~\bibnamefont{{Li}}},
  \bibinfo{author}{\bibfnamefont{X.-D.} \bibnamefont{{Li}}},
  \bibinfo{author}{\bibfnamefont{Z.}~\bibnamefont{{Li}}},
  \bibinfo{author}{\bibfnamefont{X.}~\bibnamefont{{Zhang}}}, \bibnamefont{and}
  \bibinfo{author}{\bibfnamefont{H.}~\bibnamefont{{Zhan}}},
  \bibinfo{journal}{\apj} \textbf{\bibinfo{volume}{883}}, \bibinfo{eid}{203}
  (\bibinfo{year}{2019}), \eprint{1901.04634}.

\bibitem[{\citenamefont{{Yao} et~al.}(2024)\citenamefont{{Yao}, {Shan}, {Li},
  {Xu}, {Fan}, {Liu}, {Zhang}, {Yu}, {Wei}, {Hu} et~al.}}]{2024MNRAS.527.5206Y}
\bibinfo{author}{\bibfnamefont{J.}~\bibnamefont{{Yao}}},
  \bibinfo{author}{\bibfnamefont{H.}~\bibnamefont{{Shan}}},
  \bibinfo{author}{\bibfnamefont{R.}~\bibnamefont{{Li}}},
  \bibinfo{author}{\bibfnamefont{Y.}~\bibnamefont{{Xu}}},
  \bibinfo{author}{\bibfnamefont{D.}~\bibnamefont{{Fan}}},
  \bibinfo{author}{\bibfnamefont{D.}~\bibnamefont{{Liu}}},
  \bibinfo{author}{\bibfnamefont{P.}~\bibnamefont{{Zhang}}},
  \bibinfo{author}{\bibfnamefont{Y.}~\bibnamefont{{Yu}}},
  \bibinfo{author}{\bibfnamefont{C.}~\bibnamefont{{Wei}}},
  \bibinfo{author}{\bibfnamefont{B.}~\bibnamefont{{Hu}}}, \bibnamefont{et~al.},
  \bibinfo{journal}{\mnras} \textbf{\bibinfo{volume}{527}},
  \bibinfo{pages}{5206} (\bibinfo{year}{2024}), \eprint{2304.04489}.

\bibitem[{\citenamefont{{Hamana} et~al.}(2020)\citenamefont{{Hamana},
  {Shirasaki}, {Miyazaki}, {Hikage}, {Oguri}, {More}, {Armstrong}, {Leauthaud},
  {Mandelbaum}, {Miyatake} et~al.}}]{Hamana2020}
\bibinfo{author}{\bibfnamefont{T.}~\bibnamefont{{Hamana}}},
  \bibinfo{author}{\bibfnamefont{M.}~\bibnamefont{{Shirasaki}}},
  \bibinfo{author}{\bibfnamefont{S.}~\bibnamefont{{Miyazaki}}},
  \bibinfo{author}{\bibfnamefont{C.}~\bibnamefont{{Hikage}}},
  \bibinfo{author}{\bibfnamefont{M.}~\bibnamefont{{Oguri}}},
  \bibinfo{author}{\bibfnamefont{S.}~\bibnamefont{{More}}},
  \bibinfo{author}{\bibfnamefont{R.}~\bibnamefont{{Armstrong}}},
  \bibinfo{author}{\bibfnamefont{A.}~\bibnamefont{{Leauthaud}}},
  \bibinfo{author}{\bibfnamefont{R.}~\bibnamefont{{Mandelbaum}}},
  \bibinfo{author}{\bibfnamefont{H.}~\bibnamefont{{Miyatake}}},
  \bibnamefont{et~al.}, \bibinfo{journal}{\pasj} \textbf{\bibinfo{volume}{72}},
  \bibinfo{eid}{16} (\bibinfo{year}{2020}), \eprint{1906.06041}.

\bibitem[{\citenamefont{{Asgari} et~al.}(2021)\citenamefont{{Asgari}, {Lin},
  {Joachimi}, {Giblin}, {Heymans}, {Hildebrandt}, {Kannawadi}, {St{\"o}lzner},
  {Tr{\"o}ster}, {van den Busch} et~al.}}]{2021A&A...645A.104A}
\bibinfo{author}{\bibfnamefont{M.}~\bibnamefont{{Asgari}}},
  \bibinfo{author}{\bibfnamefont{C.-A.} \bibnamefont{{Lin}}},
  \bibinfo{author}{\bibfnamefont{B.}~\bibnamefont{{Joachimi}}},
  \bibinfo{author}{\bibfnamefont{B.}~\bibnamefont{{Giblin}}},
  \bibinfo{author}{\bibfnamefont{C.}~\bibnamefont{{Heymans}}},
  \bibinfo{author}{\bibfnamefont{H.}~\bibnamefont{{Hildebrandt}}},
  \bibinfo{author}{\bibfnamefont{A.}~\bibnamefont{{Kannawadi}}},
  \bibinfo{author}{\bibfnamefont{B.}~\bibnamefont{{St{\"o}lzner}}},
  \bibinfo{author}{\bibfnamefont{T.}~\bibnamefont{{Tr{\"o}ster}}},
  \bibinfo{author}{\bibfnamefont{J.~L.} \bibnamefont{{van den Busch}}},
  \bibnamefont{et~al.}, \bibinfo{journal}{\aap} \textbf{\bibinfo{volume}{645}},
  \bibinfo{eid}{A104} (\bibinfo{year}{2021}), \eprint{2007.15633}.

\bibitem[{\citenamefont{{Giblin} et~al.}(2021)\citenamefont{{Giblin},
  {Heymans}, {Asgari}, {Hildebrandt}, {Hoekstra}, {Joachimi}, {Kannawadi},
  {Kuijken}, {Lin}, {Miller} et~al.}}]{2021A&A...645A.105G}
\bibinfo{author}{\bibfnamefont{B.}~\bibnamefont{{Giblin}}},
  \bibinfo{author}{\bibfnamefont{C.}~\bibnamefont{{Heymans}}},
  \bibinfo{author}{\bibfnamefont{M.}~\bibnamefont{{Asgari}}},
  \bibinfo{author}{\bibfnamefont{H.}~\bibnamefont{{Hildebrandt}}},
  \bibinfo{author}{\bibfnamefont{H.}~\bibnamefont{{Hoekstra}}},
  \bibinfo{author}{\bibfnamefont{B.}~\bibnamefont{{Joachimi}}},
  \bibinfo{author}{\bibfnamefont{A.}~\bibnamefont{{Kannawadi}}},
  \bibinfo{author}{\bibfnamefont{K.}~\bibnamefont{{Kuijken}}},
  \bibinfo{author}{\bibfnamefont{C.-A.} \bibnamefont{{Lin}}},
  \bibinfo{author}{\bibfnamefont{L.}~\bibnamefont{{Miller}}},
  \bibnamefont{et~al.}, \bibinfo{journal}{\aap} \textbf{\bibinfo{volume}{645}},
  \bibinfo{eid}{A105} (\bibinfo{year}{2021}), \eprint{2007.01845}.

\bibitem[{\citenamefont{{Loureiro} et~al.}(2022)\citenamefont{{Loureiro},
  {Whittaker}, {Spurio Mancini}, {Joachimi}, {Cuceu}, {Asgari}, {St{\"o}lzner},
  {Tr{\"o}ster}, {Wright}, {Bilicki} et~al.}}]{2022A&A...665A..56L}
\bibinfo{author}{\bibfnamefont{A.}~\bibnamefont{{Loureiro}}},
  \bibinfo{author}{\bibfnamefont{L.}~\bibnamefont{{Whittaker}}},
  \bibinfo{author}{\bibfnamefont{A.}~\bibnamefont{{Spurio Mancini}}},
  \bibinfo{author}{\bibfnamefont{B.}~\bibnamefont{{Joachimi}}},
  \bibinfo{author}{\bibfnamefont{A.}~\bibnamefont{{Cuceu}}},
  \bibinfo{author}{\bibfnamefont{M.}~\bibnamefont{{Asgari}}},
  \bibinfo{author}{\bibfnamefont{B.}~\bibnamefont{{St{\"o}lzner}}},
  \bibinfo{author}{\bibfnamefont{T.}~\bibnamefont{{Tr{\"o}ster}}},
  \bibinfo{author}{\bibfnamefont{A.~H.} \bibnamefont{{Wright}}},
  \bibinfo{author}{\bibfnamefont{M.}~\bibnamefont{{Bilicki}}},
  \bibnamefont{et~al.}, \bibinfo{journal}{\aap} \textbf{\bibinfo{volume}{665}},
  \bibinfo{eid}{A56} (\bibinfo{year}{2022}), \eprint{2110.06947}.

\bibitem[{\citenamefont{Amon et~al.}(2022)\citenamefont{Amon, Gruen, Troxel,
  MacCrann, Dodelson, Choi, Doux, Secco, Samuroff, Krause
  et~al.}}]{PhysRevD.105.023514}
\bibinfo{author}{\bibfnamefont{A.}~\bibnamefont{Amon}},
  \bibinfo{author}{\bibfnamefont{D.}~\bibnamefont{Gruen}},
  \bibinfo{author}{\bibfnamefont{M.~A.} \bibnamefont{Troxel}},
  \bibinfo{author}{\bibfnamefont{N.}~\bibnamefont{MacCrann}},
  \bibinfo{author}{\bibfnamefont{S.}~\bibnamefont{Dodelson}},
  \bibinfo{author}{\bibfnamefont{A.}~\bibnamefont{Choi}},
  \bibinfo{author}{\bibfnamefont{C.}~\bibnamefont{Doux}},
  \bibinfo{author}{\bibfnamefont{L.~F.} \bibnamefont{Secco}},
  \bibinfo{author}{\bibfnamefont{S.}~\bibnamefont{Samuroff}},
  \bibinfo{author}{\bibfnamefont{E.}~\bibnamefont{Krause}},
  \bibnamefont{et~al.} (\bibinfo{collaboration}{DES Collaboration}),
  \bibinfo{journal}{Phys. Rev. D} \textbf{\bibinfo{volume}{105}},
  \bibinfo{pages}{023514} (\bibinfo{year}{2022}),
  \urlprefix\url{https://link.aps.org/doi/10.1103/PhysRevD.105.023514}.

\bibitem[{\citenamefont{Secco et~al.}(2022)\citenamefont{Secco, Samuroff,
  Krause, Jain, Blazek, Raveri, Campos, Amon, Chen, Doux
  et~al.}}]{PhysRevD.105.023515}
\bibinfo{author}{\bibfnamefont{L.~F.} \bibnamefont{Secco}},
  \bibinfo{author}{\bibfnamefont{S.}~\bibnamefont{Samuroff}},
  \bibinfo{author}{\bibfnamefont{E.}~\bibnamefont{Krause}},
  \bibinfo{author}{\bibfnamefont{B.}~\bibnamefont{Jain}},
  \bibinfo{author}{\bibfnamefont{J.}~\bibnamefont{Blazek}},
  \bibinfo{author}{\bibfnamefont{M.}~\bibnamefont{Raveri}},
  \bibinfo{author}{\bibfnamefont{A.}~\bibnamefont{Campos}},
  \bibinfo{author}{\bibfnamefont{A.}~\bibnamefont{Amon}},
  \bibinfo{author}{\bibfnamefont{A.}~\bibnamefont{Chen}},
  \bibinfo{author}{\bibfnamefont{C.}~\bibnamefont{Doux}}, \bibnamefont{et~al.}
  (\bibinfo{collaboration}{DES Collaboration}), \bibinfo{journal}{Phys. Rev. D}
  \textbf{\bibinfo{volume}{105}}, \bibinfo{pages}{023515}
  (\bibinfo{year}{2022}),
  \urlprefix\url{https://link.aps.org/doi/10.1103/PhysRevD.105.023515}.

\bibitem[{\citenamefont{{Li} et~al.}(2023)\citenamefont{{Li}, {Zhang},
  {Sugiyama}, {Dalal}, {Rau}, {Mandelbaum}, {Takada}, {More}, {Strauss},
  {Miyatake} et~al.}}]{2023arXiv230400702L}
\bibinfo{author}{\bibfnamefont{X.}~\bibnamefont{{Li}}},
  \bibinfo{author}{\bibfnamefont{T.}~\bibnamefont{{Zhang}}},
  \bibinfo{author}{\bibfnamefont{S.}~\bibnamefont{{Sugiyama}}},
  \bibinfo{author}{\bibfnamefont{R.}~\bibnamefont{{Dalal}}},
  \bibinfo{author}{\bibfnamefont{M.~M.} \bibnamefont{{Rau}}},
  \bibinfo{author}{\bibfnamefont{R.}~\bibnamefont{{Mandelbaum}}},
  \bibinfo{author}{\bibfnamefont{M.}~\bibnamefont{{Takada}}},
  \bibinfo{author}{\bibfnamefont{S.}~\bibnamefont{{More}}},
  \bibinfo{author}{\bibfnamefont{M.~A.} \bibnamefont{{Strauss}}},
  \bibinfo{author}{\bibfnamefont{H.}~\bibnamefont{{Miyatake}}},
  \bibnamefont{et~al.}, \bibinfo{journal}{arXiv e-prints}
  \bibinfo{eid}{arXiv:2304.00702} (\bibinfo{year}{2023}), \eprint{2304.00702}.

\bibitem[{\citenamefont{{Blandford} and {Narayan}}(1992)}]{BN1992}
\bibinfo{author}{\bibfnamefont{R.~D.} \bibnamefont{{Blandford}}}
  \bibnamefont{and}
  \bibinfo{author}{\bibfnamefont{R.}~\bibnamefont{{Narayan}}},
  \bibinfo{journal}{\araa} \textbf{\bibinfo{volume}{30}}, \bibinfo{pages}{311}
  (\bibinfo{year}{1992}).

\bibitem[{\citenamefont{{Bartelmann} and {Narayan}}(1995)}]{BN1995}
\bibinfo{author}{\bibfnamefont{M.}~\bibnamefont{{Bartelmann}}}
  \bibnamefont{and}
  \bibinfo{author}{\bibfnamefont{R.}~\bibnamefont{{Narayan}}}, in
  \emph{\bibinfo{booktitle}{Dark Matter}}, edited by
  \bibinfo{editor}{\bibfnamefont{S.~S.} \bibnamefont{{Holt}}} \bibnamefont{and}
  \bibinfo{editor}{\bibfnamefont{C.~L.} \bibnamefont{{Bennett}}}
  (\bibinfo{year}{1995}), vol. \bibinfo{volume}{336} of
  \emph{\bibinfo{series}{American Institute of Physics Conference Series}}, pp.
  \bibinfo{pages}{307--319}, \eprint{astro-ph/9411033}.

\bibitem[{\citenamefont{{M{\'e}nard} et~al.}(2010)\citenamefont{{M{\'e}nard},
  {Scranton}, {Fukugita}, and {Richards}}}]{2010MNRAS.405.1025M}
\bibinfo{author}{\bibfnamefont{B.}~\bibnamefont{{M{\'e}nard}}},
  \bibinfo{author}{\bibfnamefont{R.}~\bibnamefont{{Scranton}}},
  \bibinfo{author}{\bibfnamefont{M.}~\bibnamefont{{Fukugita}}},
  \bibnamefont{and}
  \bibinfo{author}{\bibfnamefont{G.}~\bibnamefont{{Richards}}},
  \bibinfo{journal}{\mnras} \textbf{\bibinfo{volume}{405}},
  \bibinfo{pages}{1025} (\bibinfo{year}{2010}), \eprint{0902.4240}.

\bibitem[{\citenamefont{{Jain} and {Lima}}(2011)}]{2011MNRAS.411.2113J}
\bibinfo{author}{\bibfnamefont{B.}~\bibnamefont{{Jain}}} \bibnamefont{and}
  \bibinfo{author}{\bibfnamefont{M.}~\bibnamefont{{Lima}}},
  \bibinfo{journal}{\mnras} \textbf{\bibinfo{volume}{411}},
  \bibinfo{pages}{2113} (\bibinfo{year}{2011}), \eprint{1003.6127}.

\bibitem[{\citenamefont{Schmidt et~al.}(2012)\citenamefont{Schmidt, Leauthaud,
  Massey, Rhodes, George, Koekemoer, Finoguenov, and
  Tanaka}}]{Schmidt_Leauthaud_Massey_Rhodes_George_Koekemoer_Finoguenov_Tanaka_2012}
\bibinfo{author}{\bibfnamefont{F.}~\bibnamefont{Schmidt}},
  \bibinfo{author}{\bibfnamefont{A.}~\bibnamefont{Leauthaud}},
  \bibinfo{author}{\bibfnamefont{R.}~\bibnamefont{Massey}},
  \bibinfo{author}{\bibfnamefont{J.}~\bibnamefont{Rhodes}},
  \bibinfo{author}{\bibfnamefont{M.~R.} \bibnamefont{George}},
  \bibinfo{author}{\bibfnamefont{A.~M.} \bibnamefont{Koekemoer}},
  \bibinfo{author}{\bibfnamefont{A.}~\bibnamefont{Finoguenov}},
  \bibnamefont{and} \bibinfo{author}{\bibfnamefont{M.}~\bibnamefont{Tanaka}},
  \bibinfo{journal}{The Astrophysical Journal} \textbf{\bibinfo{volume}{744}},
  \bibinfo{pages}{L22} (\bibinfo{year}{2012}),
  \urlprefix\url{http://dx.doi.org/10.1088/2041-8205/744/2/l22}.

\bibitem[{\citenamefont{{Huff} and {Graves}}(2014)}]{2014ApJ...780L..16H}
\bibinfo{author}{\bibfnamefont{E.~M.} \bibnamefont{{Huff}}} \bibnamefont{and}
  \bibinfo{author}{\bibfnamefont{G.~J.} \bibnamefont{{Graves}}},
  \bibinfo{journal}{\apjl} \textbf{\bibinfo{volume}{780}}, \bibinfo{eid}{L16}
  (\bibinfo{year}{2014}).

\bibitem[{\citenamefont{Duncan et~al.}(2016)\citenamefont{Duncan, Heymans,
  Heavens, and Joachimi}}]{Duncan_Heymans_Heavens_Joachimi_2016}
\bibinfo{author}{\bibfnamefont{C.~A.~J.} \bibnamefont{Duncan}},
  \bibinfo{author}{\bibfnamefont{C.}~\bibnamefont{Heymans}},
  \bibinfo{author}{\bibfnamefont{A.~F.} \bibnamefont{Heavens}},
  \bibnamefont{and} \bibinfo{author}{\bibfnamefont{B.}~\bibnamefont{Joachimi}},
  \bibinfo{journal}{Monthly Notices of the Royal Astronomical Society}
  \textbf{\bibinfo{volume}{457}}, \bibinfo{pages}{764–785}
  (\bibinfo{year}{2016}),
  \urlprefix\url{http://dx.doi.org/10.1093/mnras/stw027}.

\bibitem[{\citenamefont{{Garcia-Fernandez}
  et~al.}(2018)\citenamefont{{Garcia-Fernandez}, {Sanchez}, {Sevilla-Noarbe},
  {Suchyta}, {Huff}, {Gaztanaga}, {Aleksi{\'c}}, {}, {Ponce}, {Castander}
  et~al.}}]{2018MNRAS.476.1071G}
\bibinfo{author}{\bibfnamefont{M.}~\bibnamefont{{Garcia-Fernandez}}},
  \bibinfo{author}{\bibfnamefont{E.}~\bibnamefont{{Sanchez}}},
  \bibinfo{author}{\bibfnamefont{I.}~\bibnamefont{{Sevilla-Noarbe}}},
  \bibinfo{author}{\bibfnamefont{E.}~\bibnamefont{{Suchyta}}},
  \bibinfo{author}{\bibfnamefont{E.~M.} \bibnamefont{{Huff}}},
  \bibinfo{author}{\bibfnamefont{E.}~\bibnamefont{{Gaztanaga}}},
  \bibinfo{author}{\bibnamefont{{Aleksi{\'c}}}},
  \bibinfo{author}{\bibfnamefont{J.}~\bibnamefont{{}}},
  \bibinfo{author}{\bibfnamefont{R.}~\bibnamefont{{Ponce}}},
  \bibinfo{author}{\bibfnamefont{F.~J.} \bibnamefont{{Castander}}},
  \bibnamefont{et~al.}, \bibinfo{journal}{\mnras}
  \textbf{\bibinfo{volume}{476}}, \bibinfo{pages}{1071} (\bibinfo{year}{2018}).

\bibitem[{\citenamefont{Bauer et~al.}(2014)\citenamefont{Bauer, Gaztañaga,
  Martí, and Miquel}}]{pub.1059915677}
\bibinfo{author}{\bibfnamefont{A.~H.} \bibnamefont{Bauer}},
  \bibinfo{author}{\bibfnamefont{E.}~\bibnamefont{Gaztañaga}},
  \bibinfo{author}{\bibfnamefont{P.}~\bibnamefont{Martí}}, \bibnamefont{and}
  \bibinfo{author}{\bibfnamefont{R.}~\bibnamefont{Miquel}},
  \bibinfo{journal}{Monthly Notices of the Royal Astronomical Society}
  \textbf{\bibinfo{volume}{440}}, \bibinfo{pages}{3701} (\bibinfo{year}{2014}),
  \bibinfo{note}{https://academic.oup.com/mnras/article-pdf/440/4/3701/3913172/stu530.pdf},
  \urlprefix\url{https://app.dimensions.ai/details/publication/pub.1059915677}.

\bibitem[{\citenamefont{{Bellagamba} et~al.}(2019)\citenamefont{{Bellagamba},
  {Sereno}, {Roncarelli}, {Maturi}, {Radovich}, {Bardelli}, {Puddu},
  {Moscardini}, {Getman}, {Hildebrandt} et~al.}}]{2019MNRAS.484.1598B}
\bibinfo{author}{\bibfnamefont{F.}~\bibnamefont{{Bellagamba}}},
  \bibinfo{author}{\bibfnamefont{M.}~\bibnamefont{{Sereno}}},
  \bibinfo{author}{\bibfnamefont{M.}~\bibnamefont{{Roncarelli}}},
  \bibinfo{author}{\bibfnamefont{M.}~\bibnamefont{{Maturi}}},
  \bibinfo{author}{\bibfnamefont{M.}~\bibnamefont{{Radovich}}},
  \bibinfo{author}{\bibfnamefont{S.}~\bibnamefont{{Bardelli}}},
  \bibinfo{author}{\bibfnamefont{E.}~\bibnamefont{{Puddu}}},
  \bibinfo{author}{\bibfnamefont{L.}~\bibnamefont{{Moscardini}}},
  \bibinfo{author}{\bibfnamefont{F.}~\bibnamefont{{Getman}}},
  \bibinfo{author}{\bibfnamefont{H.}~\bibnamefont{{Hildebrandt}}},
  \bibnamefont{et~al.}, \bibinfo{journal}{\mnras}
  \textbf{\bibinfo{volume}{484}}, \bibinfo{pages}{1598} (\bibinfo{year}{2019}),
  \eprint{1810.02827}.

\bibitem[{\citenamefont{{Chiu} et~al.}(2016)\citenamefont{{Chiu}, {Dietrich},
  {Mohr}, {Applegate}, {Benson}, {Bleem}, {Bayliss}, {Bocquet}, {Carlstrom},
  {Capasso} et~al.}}]{2016MNRAS.457.3050C}
\bibinfo{author}{\bibfnamefont{I.}~\bibnamefont{{Chiu}}},
  \bibinfo{author}{\bibfnamefont{J.~P.} \bibnamefont{{Dietrich}}},
  \bibinfo{author}{\bibfnamefont{J.}~\bibnamefont{{Mohr}}},
  \bibinfo{author}{\bibfnamefont{D.~E.} \bibnamefont{{Applegate}}},
  \bibinfo{author}{\bibfnamefont{B.~A.} \bibnamefont{{Benson}}},
  \bibinfo{author}{\bibfnamefont{L.~E.} \bibnamefont{{Bleem}}},
  \bibinfo{author}{\bibfnamefont{M.~B.} \bibnamefont{{Bayliss}}},
  \bibinfo{author}{\bibfnamefont{S.}~\bibnamefont{{Bocquet}}},
  \bibinfo{author}{\bibfnamefont{J.~E.} \bibnamefont{{Carlstrom}}},
  \bibinfo{author}{\bibfnamefont{R.}~\bibnamefont{{Capasso}}},
  \bibnamefont{et~al.}, \bibinfo{journal}{\mnras}
  \textbf{\bibinfo{volume}{457}}, \bibinfo{pages}{3050} (\bibinfo{year}{2016}),
  \eprint{1510.01745}.

\bibitem[{\citenamefont{{Chiu} et~al.}(2020)\citenamefont{{Chiu}, {Umetsu},
  {Murata}, {Medezinski}, and {Oguri}}}]{2020MNRAS.495..428C}
\bibinfo{author}{\bibfnamefont{I.~N.} \bibnamefont{{Chiu}}},
  \bibinfo{author}{\bibfnamefont{K.}~\bibnamefont{{Umetsu}}},
  \bibinfo{author}{\bibfnamefont{R.}~\bibnamefont{{Murata}}},
  \bibinfo{author}{\bibfnamefont{E.}~\bibnamefont{{Medezinski}}},
  \bibnamefont{and} \bibinfo{author}{\bibfnamefont{M.}~\bibnamefont{{Oguri}}},
  \bibinfo{journal}{\mnras} \textbf{\bibinfo{volume}{495}},
  \bibinfo{pages}{428} (\bibinfo{year}{2020}), \eprint{1909.02042}.

\bibitem[{\citenamefont{{Scranton} et~al.}(2005)\citenamefont{{Scranton},
  {M{\'e}nard}, {Richards}, {Nichol}, {Myers}, {Jain}, {Gray}, {Bartelmann},
  {Brunner}, {Connolly} et~al.}}]{2005ApJ...633..589S}
\bibinfo{author}{\bibfnamefont{R.}~\bibnamefont{{Scranton}}},
  \bibinfo{author}{\bibfnamefont{B.}~\bibnamefont{{M{\'e}nard}}},
  \bibinfo{author}{\bibfnamefont{G.~T.} \bibnamefont{{Richards}}},
  \bibinfo{author}{\bibfnamefont{R.~C.} \bibnamefont{{Nichol}}},
  \bibinfo{author}{\bibfnamefont{A.~D.} \bibnamefont{{Myers}}},
  \bibinfo{author}{\bibfnamefont{B.}~\bibnamefont{{Jain}}},
  \bibinfo{author}{\bibfnamefont{A.}~\bibnamefont{{Gray}}},
  \bibinfo{author}{\bibfnamefont{M.}~\bibnamefont{{Bartelmann}}},
  \bibinfo{author}{\bibfnamefont{R.~J.} \bibnamefont{{Brunner}}},
  \bibinfo{author}{\bibfnamefont{A.~J.} \bibnamefont{{Connolly}}},
  \bibnamefont{et~al.}, \bibinfo{journal}{\apj} \textbf{\bibinfo{volume}{633}},
  \bibinfo{pages}{589} (\bibinfo{year}{2005}), \eprint{astro-ph/0504510}.

\bibitem[{\citenamefont{{Bauer} et~al.}(2012)\citenamefont{{Bauer}, {Baltay},
  {Ellman}, {Jerke}, {Rabinowitz}, and {Scalzo}}}]{2012ApJ...749...56B}
\bibinfo{author}{\bibfnamefont{A.~H.} \bibnamefont{{Bauer}}},
  \bibinfo{author}{\bibfnamefont{C.}~\bibnamefont{{Baltay}}},
  \bibinfo{author}{\bibfnamefont{N.}~\bibnamefont{{Ellman}}},
  \bibinfo{author}{\bibfnamefont{J.}~\bibnamefont{{Jerke}}},
  \bibinfo{author}{\bibfnamefont{D.}~\bibnamefont{{Rabinowitz}}},
  \bibnamefont{and} \bibinfo{author}{\bibfnamefont{R.}~\bibnamefont{{Scalzo}}},
  \bibinfo{journal}{\apj} \textbf{\bibinfo{volume}{749}}, \bibinfo{eid}{56}
  (\bibinfo{year}{2012}), \eprint{1202.1371}.

\bibitem[{\citenamefont{{Morrison} et~al.}(2012)\citenamefont{{Morrison},
  {Scranton}, {M{\'e}nard}, {Schmidt}, {Tyson}, {Ryan}, {Choi}, and
  {Wittman}}}]{2012MNRAS.426.2489M}
\bibinfo{author}{\bibfnamefont{C.~B.} \bibnamefont{{Morrison}}},
  \bibinfo{author}{\bibfnamefont{R.}~\bibnamefont{{Scranton}}},
  \bibinfo{author}{\bibfnamefont{B.}~\bibnamefont{{M{\'e}nard}}},
  \bibinfo{author}{\bibfnamefont{S.~J.} \bibnamefont{{Schmidt}}},
  \bibinfo{author}{\bibfnamefont{J.~A.} \bibnamefont{{Tyson}}},
  \bibinfo{author}{\bibfnamefont{R.}~\bibnamefont{{Ryan}}},
  \bibinfo{author}{\bibfnamefont{A.}~\bibnamefont{{Choi}}}, \bibnamefont{and}
  \bibinfo{author}{\bibfnamefont{D.~M.} \bibnamefont{{Wittman}}},
  \bibinfo{journal}{\mnras} \textbf{\bibinfo{volume}{426}},
  \bibinfo{pages}{2489} (\bibinfo{year}{2012}), \eprint{1204.2830}.

\bibitem[{\citenamefont{{Tudorica} et~al.}(2017)\citenamefont{{Tudorica},
  {Hildebrandt}, {Tewes}, {Hoekstra}, {Morrison}, {Muzzin}, {Wilson}, {Yee},
  {Lidman}, {Hicks} et~al.}}]{2017A&A...608A.141T}
\bibinfo{author}{\bibfnamefont{A.}~\bibnamefont{{Tudorica}}},
  \bibinfo{author}{\bibfnamefont{H.}~\bibnamefont{{Hildebrandt}}},
  \bibinfo{author}{\bibfnamefont{M.}~\bibnamefont{{Tewes}}},
  \bibinfo{author}{\bibfnamefont{H.}~\bibnamefont{{Hoekstra}}},
  \bibinfo{author}{\bibfnamefont{C.~B.} \bibnamefont{{Morrison}}},
  \bibinfo{author}{\bibfnamefont{A.}~\bibnamefont{{Muzzin}}},
  \bibinfo{author}{\bibfnamefont{G.}~\bibnamefont{{Wilson}}},
  \bibinfo{author}{\bibfnamefont{H.~K.~C.} \bibnamefont{{Yee}}},
  \bibinfo{author}{\bibfnamefont{C.}~\bibnamefont{{Lidman}}},
  \bibinfo{author}{\bibfnamefont{A.}~\bibnamefont{{Hicks}}},
  \bibnamefont{et~al.}, \bibinfo{journal}{\aap} \textbf{\bibinfo{volume}{608}},
  \bibinfo{eid}{A141} (\bibinfo{year}{2017}), \eprint{1710.06431}.

\bibitem[{\citenamefont{{Bonavera} et~al.}(2021)\citenamefont{{Bonavera},
  {Cueli}, {Gonz{\'a}lez-Nuevo}, {Ronconi}, {Migliaccio}, {Lapi}, {Casas}, and
  {Crespo}}}]{2021A&A...656A..99B}
\bibinfo{author}{\bibfnamefont{L.}~\bibnamefont{{Bonavera}}},
  \bibinfo{author}{\bibfnamefont{M.~M.} \bibnamefont{{Cueli}}},
  \bibinfo{author}{\bibfnamefont{J.}~\bibnamefont{{Gonz{\'a}lez-Nuevo}}},
  \bibinfo{author}{\bibfnamefont{T.}~\bibnamefont{{Ronconi}}},
  \bibinfo{author}{\bibfnamefont{M.}~\bibnamefont{{Migliaccio}}},
  \bibinfo{author}{\bibfnamefont{A.}~\bibnamefont{{Lapi}}},
  \bibinfo{author}{\bibfnamefont{J.~M.} \bibnamefont{{Casas}}},
  \bibnamefont{and} \bibinfo{author}{\bibfnamefont{D.}~\bibnamefont{{Crespo}}},
  \bibinfo{journal}{\aap} \textbf{\bibinfo{volume}{656}}, \bibinfo{eid}{A99}
  (\bibinfo{year}{2021}), \eprint{2109.12413}.

\bibitem[{\citenamefont{{Crespo} et~al.}(2022)\citenamefont{{Crespo},
  {Gonz{\'a}lez-Nuevo}, {Bonavera}, {Cueli}, {Casas}, and
  {Goitia}}}]{Crespo_GonzalezNuevo_Bonavera_Cueli_Casas_Goitia_2022}
\bibinfo{author}{\bibfnamefont{D.}~\bibnamefont{{Crespo}}},
  \bibinfo{author}{\bibfnamefont{J.}~\bibnamefont{{Gonz{\'a}lez-Nuevo}}},
  \bibinfo{author}{\bibfnamefont{L.}~\bibnamefont{{Bonavera}}},
  \bibinfo{author}{\bibfnamefont{M.~M.} \bibnamefont{{Cueli}}},
  \bibinfo{author}{\bibfnamefont{J.~M.} \bibnamefont{{Casas}}},
  \bibnamefont{and} \bibinfo{author}{\bibfnamefont{E.}~\bibnamefont{{Goitia}}},
  \bibinfo{journal}{\aap} \textbf{\bibinfo{volume}{667}}, \bibinfo{eid}{A146}
  (\bibinfo{year}{2022}), \eprint{2210.17318}.

\bibitem[{\citenamefont{{Liu} et~al.}(2021)\citenamefont{{Liu}, {Liu}, {Gao},
  {Wei}, {Li}, {Fu}, {Futamase}, and
  {Fan}}}]{Liu_Liu_Gao_Wei_Li_Fu_Futamase_Fan_2021}
\bibinfo{author}{\bibfnamefont{X.}~\bibnamefont{{Liu}}},
  \bibinfo{author}{\bibfnamefont{D.}~\bibnamefont{{Liu}}},
  \bibinfo{author}{\bibfnamefont{Z.}~\bibnamefont{{Gao}}},
  \bibinfo{author}{\bibfnamefont{C.}~\bibnamefont{{Wei}}},
  \bibinfo{author}{\bibfnamefont{G.}~\bibnamefont{{Li}}},
  \bibinfo{author}{\bibfnamefont{L.}~\bibnamefont{{Fu}}},
  \bibinfo{author}{\bibfnamefont{T.}~\bibnamefont{{Futamase}}},
  \bibnamefont{and} \bibinfo{author}{\bibfnamefont{Z.}~\bibnamefont{{Fan}}},
  \bibinfo{journal}{\prd} \textbf{\bibinfo{volume}{103}}, \bibinfo{eid}{123504}
  (\bibinfo{year}{2021}), \eprint{2104.13595}.

\bibitem[{\citenamefont{{Yao} et~al.}(2023)\citenamefont{{Yao}, {Shan},
  {Zhang}, {Jullo}, {Kneib}, {Yu}, {Zu}, {Brooks}, {de la Macorra}, {Doel}
  et~al.}}]{2023MNRAS.524.6071Y}
\bibinfo{author}{\bibfnamefont{J.}~\bibnamefont{{Yao}}},
  \bibinfo{author}{\bibfnamefont{H.}~\bibnamefont{{Shan}}},
  \bibinfo{author}{\bibfnamefont{P.}~\bibnamefont{{Zhang}}},
  \bibinfo{author}{\bibfnamefont{E.}~\bibnamefont{{Jullo}}},
  \bibinfo{author}{\bibfnamefont{J.-P.} \bibnamefont{{Kneib}}},
  \bibinfo{author}{\bibfnamefont{Y.}~\bibnamefont{{Yu}}},
  \bibinfo{author}{\bibfnamefont{Y.}~\bibnamefont{{Zu}}},
  \bibinfo{author}{\bibfnamefont{D.}~\bibnamefont{{Brooks}}},
  \bibinfo{author}{\bibfnamefont{A.}~\bibnamefont{{de la Macorra}}},
  \bibinfo{author}{\bibfnamefont{P.}~\bibnamefont{{Doel}}},
  \bibnamefont{et~al.}, \bibinfo{journal}{\mnras}
  \textbf{\bibinfo{volume}{524}}, \bibinfo{pages}{6071} (\bibinfo{year}{2023}),
  \eprint{2301.13434}.

\bibitem[{\citenamefont{{Zhang} and {Pen}}(2005)}]{2005PhRvL..95x1302Z}
\bibinfo{author}{\bibfnamefont{P.}~\bibnamefont{{Zhang}}} \bibnamefont{and}
  \bibinfo{author}{\bibfnamefont{U.-L.} \bibnamefont{{Pen}}},
  \bibinfo{journal}{\prl} \textbf{\bibinfo{volume}{95}}, \bibinfo{eid}{241302}
  (\bibinfo{year}{2005}), \eprint{astro-ph/0506740}.

\bibitem[{\citenamefont{{Yang} and {Zhang}}(2011)}]{2011MNRAS.415.3485Y}
\bibinfo{author}{\bibfnamefont{X.}~\bibnamefont{{Yang}}} \bibnamefont{and}
  \bibinfo{author}{\bibfnamefont{P.}~\bibnamefont{{Zhang}}},
  \bibinfo{journal}{\mnras} \textbf{\bibinfo{volume}{415}},
  \bibinfo{pages}{3485} (\bibinfo{year}{2011}), \eprint{1105.2385}.

\bibitem[{\citenamefont{{Zhang} et~al.}(2019)\citenamefont{{Zhang}, {Zhang},
  and {Zhang}}}]{ABS}
\bibinfo{author}{\bibfnamefont{P.}~\bibnamefont{{Zhang}}},
  \bibinfo{author}{\bibfnamefont{J.}~\bibnamefont{{Zhang}}}, \bibnamefont{and}
  \bibinfo{author}{\bibfnamefont{L.}~\bibnamefont{{Zhang}}},
  \bibinfo{journal}{\mnras} \textbf{\bibinfo{volume}{484}},
  \bibinfo{pages}{1616} (\bibinfo{year}{2019}).

\bibitem[{\citenamefont{{Yang} et~al.}(2015)\citenamefont{{Yang}, {Zhang},
  {Zhang}, and {Yu}}}]{YangXJ15}
\bibinfo{author}{\bibfnamefont{X.}~\bibnamefont{{Yang}}},
  \bibinfo{author}{\bibfnamefont{P.}~\bibnamefont{{Zhang}}},
  \bibinfo{author}{\bibfnamefont{J.}~\bibnamefont{{Zhang}}}, \bibnamefont{and}
  \bibinfo{author}{\bibfnamefont{Y.}~\bibnamefont{{Yu}}},
  \bibinfo{journal}{\mnras} \textbf{\bibinfo{volume}{447}},
  \bibinfo{pages}{345} (\bibinfo{year}{2015}).

\bibitem[{\citenamefont{{Yang} et~al.}(2017)\citenamefont{{Yang}, {Zhang},
  {Yu}, and {Zhang}}}]{YangXJ17}
\bibinfo{author}{\bibfnamefont{X.}~\bibnamefont{{Yang}}},
  \bibinfo{author}{\bibfnamefont{J.}~\bibnamefont{{Zhang}}},
  \bibinfo{author}{\bibfnamefont{Y.}~\bibnamefont{{Yu}}}, \bibnamefont{and}
  \bibinfo{author}{\bibfnamefont{P.}~\bibnamefont{{Zhang}}},
  \bibinfo{journal}{\apj} \textbf{\bibinfo{volume}{845}}, \bibinfo{eid}{174}
  (\bibinfo{year}{2017}), \eprint{1703.01575}.

\bibitem[{\citenamefont{{Zhang} et~al.}(2018)\citenamefont{{Zhang}, {Yang},
  {Zhang}, and {Yu}}}]{Zhang18}
\bibinfo{author}{\bibfnamefont{P.}~\bibnamefont{{Zhang}}},
  \bibinfo{author}{\bibfnamefont{X.}~\bibnamefont{{Yang}}},
  \bibinfo{author}{\bibfnamefont{J.}~\bibnamefont{{Zhang}}}, \bibnamefont{and}
  \bibinfo{author}{\bibfnamefont{Y.}~\bibnamefont{{Yu}}},
  \bibinfo{journal}{\apj} \textbf{\bibinfo{volume}{864}}, \bibinfo{eid}{10}
  (\bibinfo{year}{2018}), \eprint{1807.00443}.

\bibitem[{\citenamefont{{Hou} et~al.}(2021)\citenamefont{{Hou}, {Yu}, and
  {Zhang}}}]{2021RAA....21..247H}
\bibinfo{author}{\bibfnamefont{S.-T.} \bibnamefont{{Hou}}},
  \bibinfo{author}{\bibfnamefont{Y.}~\bibnamefont{{Yu}}}, \bibnamefont{and}
  \bibinfo{author}{\bibfnamefont{P.-J.} \bibnamefont{{Zhang}}},
  \bibinfo{journal}{Research in Astronomy and Astrophysics}
  \textbf{\bibinfo{volume}{21}}, \bibinfo{eid}{247} (\bibinfo{year}{2021}),
  \eprint{2106.09970}.

\bibitem[{\citenamefont{{Ma} et~al.}(2024)\citenamefont{{Ma}, {Zhang}, {Yu},
  and {Qin}}}]{2024MNRAS.527.7547M}
\bibinfo{author}{\bibfnamefont{R.}~\bibnamefont{{Ma}}},
  \bibinfo{author}{\bibfnamefont{P.}~\bibnamefont{{Zhang}}},
  \bibinfo{author}{\bibfnamefont{Y.}~\bibnamefont{{Yu}}}, \bibnamefont{and}
  \bibinfo{author}{\bibfnamefont{J.}~\bibnamefont{{Qin}}},
  \bibinfo{journal}{\mnras} \textbf{\bibinfo{volume}{527}},
  \bibinfo{pages}{7547} (\bibinfo{year}{2024}), \eprint{2306.15177}.

\bibitem[{\citenamefont{{Bonoli} and {Pen}}(2009{\natexlab{a}})}]{Bonoli09}
\bibinfo{author}{\bibfnamefont{S.}~\bibnamefont{{Bonoli}}} \bibnamefont{and}
  \bibinfo{author}{\bibfnamefont{U.~L.} \bibnamefont{{Pen}}},
  \bibinfo{journal}{\mnras} \textbf{\bibinfo{volume}{396}},
  \bibinfo{pages}{1610} (\bibinfo{year}{2009}{\natexlab{a}}),
  \eprint{0810.0273}.

\bibitem[{\citenamefont{{Hamaus} et~al.}(2010)\citenamefont{{Hamaus}, {Seljak},
  {Desjacques}, {Smith}, and {Baldauf}}}]{Hamaus10}
\bibinfo{author}{\bibfnamefont{N.}~\bibnamefont{{Hamaus}}},
  \bibinfo{author}{\bibfnamefont{U.}~\bibnamefont{{Seljak}}},
  \bibinfo{author}{\bibfnamefont{V.}~\bibnamefont{{Desjacques}}},
  \bibinfo{author}{\bibfnamefont{R.~E.} \bibnamefont{{Smith}}},
  \bibnamefont{and}
  \bibinfo{author}{\bibfnamefont{T.}~\bibnamefont{{Baldauf}}},
  \bibinfo{journal}{\prd} \textbf{\bibinfo{volume}{82}}, \bibinfo{eid}{043515}
  (\bibinfo{year}{2010}), \eprint{1004.5377}.

\bibitem[{\citenamefont{{Baldauf} et~al.}(2010)\citenamefont{{Baldauf},
  {Smith}, {Seljak}, and {Mandelbaum}}}]{Baldauf10}
\bibinfo{author}{\bibfnamefont{T.}~\bibnamefont{{Baldauf}}},
  \bibinfo{author}{\bibfnamefont{R.~E.} \bibnamefont{{Smith}}},
  \bibinfo{author}{\bibfnamefont{U.}~\bibnamefont{{Seljak}}}, \bibnamefont{and}
  \bibinfo{author}{\bibfnamefont{R.}~\bibnamefont{{Mandelbaum}}},
  \bibinfo{journal}{\prd} \textbf{\bibinfo{volume}{81}}, \bibinfo{eid}{063531}
  (\bibinfo{year}{2010}), \eprint{0911.4973}.

\bibitem[{\citenamefont{{Abbott} et~al.}(2018)\citenamefont{{Abbott},
  {Abdalla}, {Allam}, {Amara}, {Annis}, {Asorey}, {Avila}, {Ballester},
  {Banerji}, {Barkhouse} et~al.}}]{2018ApJS..239...18A}
\bibinfo{author}{\bibfnamefont{T.~M.~C.} \bibnamefont{{Abbott}}},
  \bibinfo{author}{\bibfnamefont{F.~B.} \bibnamefont{{Abdalla}}},
  \bibinfo{author}{\bibfnamefont{S.}~\bibnamefont{{Allam}}},
  \bibinfo{author}{\bibfnamefont{A.}~\bibnamefont{{Amara}}},
  \bibinfo{author}{\bibfnamefont{J.}~\bibnamefont{{Annis}}},
  \bibinfo{author}{\bibfnamefont{J.}~\bibnamefont{{Asorey}}},
  \bibinfo{author}{\bibfnamefont{S.}~\bibnamefont{{Avila}}},
  \bibinfo{author}{\bibfnamefont{O.}~\bibnamefont{{Ballester}}},
  \bibinfo{author}{\bibfnamefont{M.}~\bibnamefont{{Banerji}}},
  \bibinfo{author}{\bibfnamefont{W.}~\bibnamefont{{Barkhouse}}},
  \bibnamefont{et~al.}, \bibinfo{journal}{\apjs}
  \textbf{\bibinfo{volume}{239}}, \bibinfo{eid}{18} (\bibinfo{year}{2018}),
  \eprint{1801.03181}.

\bibitem[{\citenamefont{Scranton et~al.}(2005)\citenamefont{Scranton,
  M{\'e}nard, Richards, Nichol, Myers, Jain, Gray, Bartelmann, Brunner,
  Connolly et~al.}}]{scranton2005detection}
\bibinfo{author}{\bibfnamefont{R.}~\bibnamefont{Scranton}},
  \bibinfo{author}{\bibfnamefont{B.}~\bibnamefont{M{\'e}nard}},
  \bibinfo{author}{\bibfnamefont{G.~T.} \bibnamefont{Richards}},
  \bibinfo{author}{\bibfnamefont{R.~C.} \bibnamefont{Nichol}},
  \bibinfo{author}{\bibfnamefont{A.~D.} \bibnamefont{Myers}},
  \bibinfo{author}{\bibfnamefont{B.}~\bibnamefont{Jain}},
  \bibinfo{author}{\bibfnamefont{A.}~\bibnamefont{Gray}},
  \bibinfo{author}{\bibfnamefont{M.}~\bibnamefont{Bartelmann}},
  \bibinfo{author}{\bibfnamefont{R.~J.} \bibnamefont{Brunner}},
  \bibinfo{author}{\bibfnamefont{A.~J.} \bibnamefont{Connolly}},
  \bibnamefont{et~al.}, \bibinfo{journal}{The Astrophysical Journal}
  \textbf{\bibinfo{volume}{633}}, \bibinfo{pages}{589} (\bibinfo{year}{2005}).

\bibitem[{\citenamefont{von Wietersheim Kramsta
  et~al.}(2021)\citenamefont{von Wietersheim Kramsta, Joachimi,
  van den Busch, Heymans, Hildebrandt, Asgari, Tr’oster, Unruh, and
  Wright}}]{Wietersheim-Kramsta_Joachimi_van}
\bibinfo{author}{\bibfnamefont{M.}~\bibnamefont{von Wietersheim Kramsta}},
  \bibinfo{author}{\bibfnamefont{B.}~\bibnamefont{Joachimi}},
  \bibinfo{author}{\bibfnamefont{J.~L.} \bibnamefont{van den Busch}},
  \bibinfo{author}{\bibfnamefont{C.}~\bibnamefont{Heymans}},
  \bibinfo{author}{\bibfnamefont{H.}~\bibnamefont{Hildebrandt}},
  \bibinfo{author}{\bibfnamefont{M.}~\bibnamefont{Asgari}},
  \bibinfo{author}{\bibfnamefont{T.}~\bibnamefont{Tr’oster}},
  \bibinfo{author}{\bibfnamefont{S.}~\bibnamefont{Unruh}}, \bibnamefont{and}
  \bibinfo{author}{\bibfnamefont{A.~H.} \bibnamefont{Wright}},
  \bibinfo{journal}{Monthly Notices of the Royal Astronomical Society}
  \textbf{\bibinfo{volume}{504}}, \bibinfo{pages}{1452–1465}
  (\bibinfo{year}{2021}),
  \urlprefix\url{http://dx.doi.org/10.1093/mnras/stab1000}.

\bibitem[{\citenamefont{Elvin-Poole
  et~al.}(2023{\natexlab{a}})\citenamefont{Elvin-Poole, MacCrann, Everett,
  Prat, Rykoff, De Vicente, Yanny, Herner, Ferté, Valentino
  et~al.}}]{JElvinPoole2022DarkES}
\bibinfo{author}{\bibfnamefont{J.}~\bibnamefont{Elvin-Poole}},
  \bibinfo{author}{\bibfnamefont{N.}~\bibnamefont{MacCrann}},
  \bibinfo{author}{\bibfnamefont{S.}~\bibnamefont{Everett}},
  \bibinfo{author}{\bibfnamefont{J.}~\bibnamefont{Prat}},
  \bibinfo{author}{\bibfnamefont{E.~S.} \bibnamefont{Rykoff}},
  \bibinfo{author}{\bibfnamefont{J.}~\bibnamefont{De Vicente}},
  \bibinfo{author}{\bibfnamefont{B.}~\bibnamefont{Yanny}},
  \bibinfo{author}{\bibfnamefont{K.}~\bibnamefont{Herner}},
  \bibinfo{author}{\bibfnamefont{A.}~\bibnamefont{Ferté}},
  \bibinfo{author}{\bibfnamefont{E.~D.} \bibnamefont{Valentino}},
  \bibnamefont{et~al.}, \bibinfo{journal}{Monthly Notices of the Royal
  Astronomical Society} \textbf{\bibinfo{volume}{523}}, \bibinfo{pages}{3649}
  (\bibinfo{year}{2023}{\natexlab{a}}), ISSN \bibinfo{issn}{0035-8711},
  \eprint{https://academic.oup.com/mnras/article-pdf/523/3/3649/50596748/stad1594.pdf},
  \urlprefix\url{https://doi.org/10.1093/mnras/stad1594}.

\bibitem[{\citenamefont{Elvin-Poole
  et~al.}(2023{\natexlab{b}})\citenamefont{Elvin-Poole, MacCrann, Everett,
  Prat, Rykoff, De Vicente, Yanny, Herner, Ferté, Valentino
  et~al.}}]{10.1093/mnras/stad1594}
\bibinfo{author}{\bibfnamefont{J.}~\bibnamefont{Elvin-Poole}},
  \bibinfo{author}{\bibfnamefont{N.}~\bibnamefont{MacCrann}},
  \bibinfo{author}{\bibfnamefont{S.}~\bibnamefont{Everett}},
  \bibinfo{author}{\bibfnamefont{J.}~\bibnamefont{Prat}},
  \bibinfo{author}{\bibfnamefont{E.~S.} \bibnamefont{Rykoff}},
  \bibinfo{author}{\bibfnamefont{J.}~\bibnamefont{De Vicente}},
  \bibinfo{author}{\bibfnamefont{B.}~\bibnamefont{Yanny}},
  \bibinfo{author}{\bibfnamefont{K.}~\bibnamefont{Herner}},
  \bibinfo{author}{\bibfnamefont{A.}~\bibnamefont{Ferté}},
  \bibinfo{author}{\bibfnamefont{E.~D.} \bibnamefont{Valentino}},
  \bibnamefont{et~al.}, \bibinfo{journal}{Monthly Notices of the Royal
  Astronomical Society} \textbf{\bibinfo{volume}{523}}, \bibinfo{pages}{3649}
  (\bibinfo{year}{2023}{\natexlab{b}}), ISSN \bibinfo{issn}{0035-8711},
  \eprint{https://academic.oup.com/mnras/article-pdf/523/3/3649/50596748/stad1594.pdf},
  \urlprefix\url{https://doi.org/10.1093/mnras/stad1594}.

\bibitem[{\citenamefont{{Wenzl} et~al.}(2024)\citenamefont{{Wenzl}, {Chen}, and
  {Bean}}}]{2024MNRAS.527.1760W}
\bibinfo{author}{\bibfnamefont{L.}~\bibnamefont{{Wenzl}}},
  \bibinfo{author}{\bibfnamefont{S.-F.} \bibnamefont{{Chen}}},
  \bibnamefont{and} \bibinfo{author}{\bibfnamefont{R.}~\bibnamefont{{Bean}}},
  \bibinfo{journal}{\mnras} \textbf{\bibinfo{volume}{527}},
  \bibinfo{pages}{1760} (\bibinfo{year}{2024}), \eprint{2308.05892}.

\bibitem[{\citenamefont{{Zhou} et~al.}(2023)\citenamefont{{Zhou}, {Zhang}, and
  {Chen}}}]{Zhou_Zhang_Chen_2023}
\bibinfo{author}{\bibfnamefont{S.}~\bibnamefont{{Zhou}}},
  \bibinfo{author}{\bibfnamefont{P.}~\bibnamefont{{Zhang}}}, \bibnamefont{and}
  \bibinfo{author}{\bibfnamefont{Z.}~\bibnamefont{{Chen}}},
  \bibinfo{journal}{\mnras} \textbf{\bibinfo{volume}{523}},
  \bibinfo{pages}{5789} (\bibinfo{year}{2023}), \eprint{2304.11540}.

\bibitem[{\citenamefont{{Ma} et~al.}(2023)\citenamefont{{Ma}, {Zhang}, {Yu},
  and {Qin}}}]{2023arXiv230615177M}
\bibinfo{author}{\bibfnamefont{R.}~\bibnamefont{{Ma}}},
  \bibinfo{author}{\bibfnamefont{P.}~\bibnamefont{{Zhang}}},
  \bibinfo{author}{\bibfnamefont{Y.}~\bibnamefont{{Yu}}}, \bibnamefont{and}
  \bibinfo{author}{\bibfnamefont{J.}~\bibnamefont{{Qin}}},
  \bibinfo{journal}{arXiv e-prints} \bibinfo{eid}{arXiv:2306.15177}
  (\bibinfo{year}{2023}), \eprint{2306.15177}.

\bibitem[{\citenamefont{{Planck Collaboration}
  et~al.}(2020)\citenamefont{{Planck Collaboration}, {Aghanim}, {Akrami},
  {Ashdown}, {Aumont}, {Baccigalupi}, {Ballardini}, {Banday}, {Barreiro},
  {Bartolo} et~al.}}]{Planck2018parameters}
\bibinfo{author}{\bibnamefont{{Planck Collaboration}}},
  \bibinfo{author}{\bibfnamefont{N.}~\bibnamefont{{Aghanim}}},
  \bibinfo{author}{\bibfnamefont{Y.}~\bibnamefont{{Akrami}}},
  \bibinfo{author}{\bibfnamefont{M.}~\bibnamefont{{Ashdown}}},
  \bibinfo{author}{\bibfnamefont{J.}~\bibnamefont{{Aumont}}},
  \bibinfo{author}{\bibfnamefont{C.}~\bibnamefont{{Baccigalupi}}},
  \bibinfo{author}{\bibfnamefont{M.}~\bibnamefont{{Ballardini}}},
  \bibinfo{author}{\bibfnamefont{A.~J.} \bibnamefont{{Banday}}},
  \bibinfo{author}{\bibfnamefont{R.~B.} \bibnamefont{{Barreiro}}},
  \bibinfo{author}{\bibfnamefont{N.}~\bibnamefont{{Bartolo}}},
  \bibnamefont{et~al.}, \bibinfo{journal}{\aap} \textbf{\bibinfo{volume}{641}},
  \bibinfo{eid}{A6} (\bibinfo{year}{2020}), \eprint{1807.06209}.

\bibitem[{\citenamefont{{Limber}}(1953)}]{Limber1953}
\bibinfo{author}{\bibfnamefont{D.~N.} \bibnamefont{{Limber}}},
  \bibinfo{journal}{\apj} \textbf{\bibinfo{volume}{117}}, \bibinfo{pages}{134}
  (\bibinfo{year}{1953}).

\bibitem[{\citenamefont{Yang et~al.}(2021)\citenamefont{Yang, Xu, He, Gu,
  Katsianis, Meng, Shi, Zou, Zhang, Liu et~al.}}]{yang2021extended}
\bibinfo{author}{\bibfnamefont{X.}~\bibnamefont{Yang}},
  \bibinfo{author}{\bibfnamefont{H.}~\bibnamefont{Xu}},
  \bibinfo{author}{\bibfnamefont{M.}~\bibnamefont{He}},
  \bibinfo{author}{\bibfnamefont{Y.}~\bibnamefont{Gu}},
  \bibinfo{author}{\bibfnamefont{A.}~\bibnamefont{Katsianis}},
  \bibinfo{author}{\bibfnamefont{J.}~\bibnamefont{Meng}},
  \bibinfo{author}{\bibfnamefont{F.}~\bibnamefont{Shi}},
  \bibinfo{author}{\bibfnamefont{H.}~\bibnamefont{Zou}},
  \bibinfo{author}{\bibfnamefont{Y.}~\bibnamefont{Zhang}},
  \bibinfo{author}{\bibfnamefont{C.}~\bibnamefont{Liu}}, \bibnamefont{et~al.},
  \bibinfo{journal}{The Astrophysical Journal} \textbf{\bibinfo{volume}{909}},
  \bibinfo{pages}{143} (\bibinfo{year}{2021}).

\bibitem[{\citenamefont{Zhou et~al.}(2021)\citenamefont{Zhou, Newman, Mao,
  Meisner, Moustakas, Myers, Prakash, Zentner, Brooks, Duan
  et~al.}}]{zhou2021clustering}
\bibinfo{author}{\bibfnamefont{R.}~\bibnamefont{Zhou}},
  \bibinfo{author}{\bibfnamefont{J.~A.} \bibnamefont{Newman}},
  \bibinfo{author}{\bibfnamefont{Y.-Y.} \bibnamefont{Mao}},
  \bibinfo{author}{\bibfnamefont{A.}~\bibnamefont{Meisner}},
  \bibinfo{author}{\bibfnamefont{J.}~\bibnamefont{Moustakas}},
  \bibinfo{author}{\bibfnamefont{A.~D.} \bibnamefont{Myers}},
  \bibinfo{author}{\bibfnamefont{A.}~\bibnamefont{Prakash}},
  \bibinfo{author}{\bibfnamefont{A.~R.} \bibnamefont{Zentner}},
  \bibinfo{author}{\bibfnamefont{D.}~\bibnamefont{Brooks}},
  \bibinfo{author}{\bibfnamefont{Y.}~\bibnamefont{Duan}}, \bibnamefont{et~al.},
  \bibinfo{journal}{Monthly Notices of the Royal Astronomical Society}
  \textbf{\bibinfo{volume}{501}}, \bibinfo{pages}{3309} (\bibinfo{year}{2021}).

\bibitem[{\citenamefont{{Xu} et~al.}(2023)\citenamefont{{Xu}, {Zhang}, {Peng},
  {Yu}, {Zhang}, {Yao}, {Qin}, {Sun}, {He}, and {Yang}}}]{HaojieXu2022UsingAT}
\bibinfo{author}{\bibfnamefont{H.}~\bibnamefont{{Xu}}},
  \bibinfo{author}{\bibfnamefont{P.}~\bibnamefont{{Zhang}}},
  \bibinfo{author}{\bibfnamefont{H.}~\bibnamefont{{Peng}}},
  \bibinfo{author}{\bibfnamefont{Y.}~\bibnamefont{{Yu}}},
  \bibinfo{author}{\bibfnamefont{L.}~\bibnamefont{{Zhang}}},
  \bibinfo{author}{\bibfnamefont{J.}~\bibnamefont{{Yao}}},
  \bibinfo{author}{\bibfnamefont{J.}~\bibnamefont{{Qin}}},
  \bibinfo{author}{\bibfnamefont{Z.}~\bibnamefont{{Sun}}},
  \bibinfo{author}{\bibfnamefont{M.}~\bibnamefont{{He}}}, \bibnamefont{and}
  \bibinfo{author}{\bibfnamefont{X.}~\bibnamefont{{Yang}}},
  \bibinfo{journal}{\mnras} \textbf{\bibinfo{volume}{520}},
  \bibinfo{pages}{161} (\bibinfo{year}{2023}), \eprint{2209.03967}.

\bibitem[{\citenamefont{Chaussidon et~al.}(2021)\citenamefont{Chaussidon,
  Yèche, Palanque-Delabrouille, de Mattia, Myers, Rezaie, Ross, Seo, Brooks,
  Gaztañaga et~al.}}]{10.1093/mnras/stab3252}
\bibinfo{author}{\bibfnamefont{E.}~\bibnamefont{Chaussidon}},
  \bibinfo{author}{\bibfnamefont{C.}~\bibnamefont{Yèche}},
  \bibinfo{author}{\bibfnamefont{N.}~\bibnamefont{Palanque-Delabrouille}},
  \bibinfo{author}{\bibfnamefont{A.}~\bibnamefont{de Mattia}},
  \bibinfo{author}{\bibfnamefont{A.~D.} \bibnamefont{Myers}},
  \bibinfo{author}{\bibfnamefont{M.}~\bibnamefont{Rezaie}},
  \bibinfo{author}{\bibfnamefont{A.~J.} \bibnamefont{Ross}},
  \bibinfo{author}{\bibfnamefont{H.-J.} \bibnamefont{Seo}},
  \bibinfo{author}{\bibfnamefont{D.}~\bibnamefont{Brooks}},
  \bibinfo{author}{\bibfnamefont{E.}~\bibnamefont{Gaztañaga}},
  \bibnamefont{et~al.}, \bibinfo{journal}{Monthly Notices of the Royal
  Astronomical Society} \textbf{\bibinfo{volume}{509}}, \bibinfo{pages}{3904}
  (\bibinfo{year}{2021}), ISSN \bibinfo{issn}{0035-8711},
  \eprint{https://academic.oup.com/mnras/article-pdf/509/3/3904/41446828/stab3252.pdf},
  \urlprefix\url{https://doi.org/10.1093/mnras/stab3252}.

\bibitem[{\citenamefont{Percival et~al.}(2014)\citenamefont{Percival, Ross,
  S{\'a}nchez, Samushia, Burden, Crittenden, Cuesta, Magana, Manera, Beutler
  et~al.}}]{percival2014clustering}
\bibinfo{author}{\bibfnamefont{W.~J.} \bibnamefont{Percival}},
  \bibinfo{author}{\bibfnamefont{A.~J.} \bibnamefont{Ross}},
  \bibinfo{author}{\bibfnamefont{A.~G.} \bibnamefont{S{\'a}nchez}},
  \bibinfo{author}{\bibfnamefont{L.}~\bibnamefont{Samushia}},
  \bibinfo{author}{\bibfnamefont{A.}~\bibnamefont{Burden}},
  \bibinfo{author}{\bibfnamefont{R.}~\bibnamefont{Crittenden}},
  \bibinfo{author}{\bibfnamefont{A.~J.} \bibnamefont{Cuesta}},
  \bibinfo{author}{\bibfnamefont{M.~V.} \bibnamefont{Magana}},
  \bibinfo{author}{\bibfnamefont{M.}~\bibnamefont{Manera}},
  \bibinfo{author}{\bibfnamefont{F.}~\bibnamefont{Beutler}},
  \bibnamefont{et~al.}, \bibinfo{journal}{Monthly Notices of the Royal
  Astronomical Society} \textbf{\bibinfo{volume}{439}}, \bibinfo{pages}{2531}
  (\bibinfo{year}{2014}).

\bibitem[{\citenamefont{Wang et~al.}(2020)\citenamefont{Wang, Zhao, Zhao,
  Philcox, Alam, Tamone, de Mattia, Ross, Raichoor, Burtin
  et~al.}}]{Wang_Zhao_Zhao_Philcox_Alam_Tamone_deMattia_Ross_Raichoor_Burtin2020}
\bibinfo{author}{\bibfnamefont{Y.}~\bibnamefont{Wang}},
  \bibinfo{author}{\bibfnamefont{G.-B.} \bibnamefont{Zhao}},
  \bibinfo{author}{\bibfnamefont{C.}~\bibnamefont{Zhao}},
  \bibinfo{author}{\bibfnamefont{O.~H.~E.} \bibnamefont{Philcox}},
  \bibinfo{author}{\bibfnamefont{S.}~\bibnamefont{Alam}},
  \bibinfo{author}{\bibfnamefont{A.}~\bibnamefont{Tamone}},
  \bibinfo{author}{\bibfnamefont{A.}~\bibnamefont{de Mattia}},
  \bibinfo{author}{\bibfnamefont{A.~J.} \bibnamefont{Ross}},
  \bibinfo{author}{\bibfnamefont{A.}~\bibnamefont{Raichoor}},
  \bibinfo{author}{\bibfnamefont{E.}~\bibnamefont{Burtin}},
  \bibnamefont{et~al.}, \bibinfo{journal}{Monthly Notices of the Royal
  Astronomical Society} \textbf{\bibinfo{volume}{498}},
  \bibinfo{pages}{3470–3483} (\bibinfo{year}{2020}),
  \urlprefix\url{http://dx.doi.org/10.1093/mnras/staa2593}.

\bibitem[{\citenamefont{{Chisari} et~al.}(2019)\citenamefont{{Chisari},
  {Alonso}, {Krause}, {Leonard}, {Bull}, {Neveu}, {Villarreal}, {Singh},
  {McClintock}, {Ellison} et~al.}}]{pyccl}
\bibinfo{author}{\bibfnamefont{N.~E.} \bibnamefont{{Chisari}}},
  \bibinfo{author}{\bibfnamefont{D.}~\bibnamefont{{Alonso}}},
  \bibinfo{author}{\bibfnamefont{E.}~\bibnamefont{{Krause}}},
  \bibinfo{author}{\bibfnamefont{C.~D.} \bibnamefont{{Leonard}}},
  \bibinfo{author}{\bibfnamefont{P.}~\bibnamefont{{Bull}}},
  \bibinfo{author}{\bibfnamefont{J.}~\bibnamefont{{Neveu}}},
  \bibinfo{author}{\bibfnamefont{A.}~\bibnamefont{{Villarreal}}},
  \bibinfo{author}{\bibfnamefont{S.}~\bibnamefont{{Singh}}},
  \bibinfo{author}{\bibfnamefont{T.}~\bibnamefont{{McClintock}}},
  \bibinfo{author}{\bibfnamefont{J.}~\bibnamefont{{Ellison}}},
  \bibnamefont{et~al.}, \bibinfo{journal}{\apjs}
  \textbf{\bibinfo{volume}{242}}, \bibinfo{eid}{2} (\bibinfo{year}{2019}),
  \eprint{1812.05995}.

\bibitem[{\citenamefont{{Smith} et~al.}(2003)\citenamefont{{Smith}, {Peacock},
  {Jenkins}, {White}, {Frenk}, {Pearce}, {Thomas}, {Efstathiou}, and
  {Couchman}}}]{Smith2003}
\bibinfo{author}{\bibfnamefont{R.~E.} \bibnamefont{{Smith}}},
  \bibinfo{author}{\bibfnamefont{J.~A.} \bibnamefont{{Peacock}}},
  \bibinfo{author}{\bibfnamefont{A.}~\bibnamefont{{Jenkins}}},
  \bibinfo{author}{\bibfnamefont{S.~D.~M.} \bibnamefont{{White}}},
  \bibinfo{author}{\bibfnamefont{C.~S.} \bibnamefont{{Frenk}}},
  \bibinfo{author}{\bibfnamefont{F.~R.} \bibnamefont{{Pearce}}},
  \bibinfo{author}{\bibfnamefont{P.~A.} \bibnamefont{{Thomas}}},
  \bibinfo{author}{\bibfnamefont{G.}~\bibnamefont{{Efstathiou}}},
  \bibnamefont{and} \bibinfo{author}{\bibfnamefont{H.~M.~P.}
  \bibnamefont{{Couchman}}}, \bibinfo{journal}{\mnras}
  \textbf{\bibinfo{volume}{341}}, \bibinfo{pages}{1311} (\bibinfo{year}{2003}),
  \eprint{astro-ph/0207664}.

\bibitem[{\citenamefont{{Takahashi} et~al.}(2012)\citenamefont{{Takahashi},
  {Sato}, {Nishimichi}, {Taruya}, and {Oguri}}}]{Takahashi2012}
\bibinfo{author}{\bibfnamefont{R.}~\bibnamefont{{Takahashi}}},
  \bibinfo{author}{\bibfnamefont{M.}~\bibnamefont{{Sato}}},
  \bibinfo{author}{\bibfnamefont{T.}~\bibnamefont{{Nishimichi}}},
  \bibinfo{author}{\bibfnamefont{A.}~\bibnamefont{{Taruya}}}, \bibnamefont{and}
  \bibinfo{author}{\bibfnamefont{M.}~\bibnamefont{{Oguri}}},
  \bibinfo{journal}{\apj} \textbf{\bibinfo{volume}{761}}, \bibinfo{eid}{152}
  (\bibinfo{year}{2012}), \eprint{1208.2701}.

\bibitem[{\citenamefont{{Mead} et~al.}(2015)\citenamefont{{Mead}, {Peacock},
  {Heymans}, {Joudaki}, and {Heavens}}}]{Mead2015}
\bibinfo{author}{\bibfnamefont{A.~J.} \bibnamefont{{Mead}}},
  \bibinfo{author}{\bibfnamefont{J.~A.} \bibnamefont{{Peacock}}},
  \bibinfo{author}{\bibfnamefont{C.}~\bibnamefont{{Heymans}}},
  \bibinfo{author}{\bibfnamefont{S.}~\bibnamefont{{Joudaki}}},
  \bibnamefont{and} \bibinfo{author}{\bibfnamefont{A.~F.}
  \bibnamefont{{Heavens}}}, \bibinfo{journal}{\mnras}
  \textbf{\bibinfo{volume}{454}}, \bibinfo{pages}{1958} (\bibinfo{year}{2015}),
  \eprint{1505.07833}.

\bibitem[{\citenamefont{{Zhou} and
  {Zhang}}(2024{\natexlab{a}})}]{2024arXiv240603018Z}
\bibinfo{author}{\bibfnamefont{S.}~\bibnamefont{{Zhou}}} \bibnamefont{and}
  \bibinfo{author}{\bibfnamefont{P.}~\bibnamefont{{Zhang}}},
  \bibinfo{journal}{arXiv e-prints} \bibinfo{eid}{arXiv:2406.03018}
  (\bibinfo{year}{2024}{\natexlab{a}}), \eprint{2406.03018}.

\bibitem[{\citenamefont{{Tegmark} and {Bromley}}(1999)}]{1999ApJ...518L..69T}
\bibinfo{author}{\bibfnamefont{M.}~\bibnamefont{{Tegmark}}} \bibnamefont{and}
  \bibinfo{author}{\bibfnamefont{B.~C.} \bibnamefont{{Bromley}}},
  \bibinfo{journal}{\apjl} \textbf{\bibinfo{volume}{518}}, \bibinfo{pages}{L69}
  (\bibinfo{year}{1999}), \eprint{astro-ph/9809324}.

\bibitem[{\citenamefont{{Bonoli} and
  {Pen}}(2009{\natexlab{b}})}]{2009MNRAS.396.1610B}
\bibinfo{author}{\bibfnamefont{S.}~\bibnamefont{{Bonoli}}} \bibnamefont{and}
  \bibinfo{author}{\bibfnamefont{U.~L.} \bibnamefont{{Pen}}},
  \bibinfo{journal}{\mnras} \textbf{\bibinfo{volume}{396}},
  \bibinfo{pages}{1610} (\bibinfo{year}{2009}{\natexlab{b}}),
  \eprint{0810.0273}.

\bibitem[{\citenamefont{{Zhou} and
  {Zhang}}(2024{\natexlab{b}})}]{2024PhRvD.110f3551Z}
\bibinfo{author}{\bibfnamefont{S.}~\bibnamefont{{Zhou}}} \bibnamefont{and}
  \bibinfo{author}{\bibfnamefont{P.}~\bibnamefont{{Zhang}}},
  \bibinfo{journal}{\prd} \textbf{\bibinfo{volume}{110}}, \bibinfo{eid}{063551}
  (\bibinfo{year}{2024}{\natexlab{b}}), \eprint{2409.01954}.

\bibitem[{\citenamefont{{Foreman-Mackey}
  et~al.}(2013)\citenamefont{{Foreman-Mackey}, {Hogg}, {Lang}, and
  {Goodman}}}]{emcee}
\bibinfo{author}{\bibfnamefont{D.}~\bibnamefont{{Foreman-Mackey}}},
  \bibinfo{author}{\bibfnamefont{D.~W.} \bibnamefont{{Hogg}}},
  \bibinfo{author}{\bibfnamefont{D.}~\bibnamefont{{Lang}}}, \bibnamefont{and}
  \bibinfo{author}{\bibfnamefont{J.}~\bibnamefont{{Goodman}}},
  \bibinfo{journal}{PASP} \textbf{\bibinfo{volume}{125}}, \bibinfo{pages}{306}
  (\bibinfo{year}{2013}), \eprint{1202.3665}.

\end{thebibliography}

\appendix

\section{The imaging systematics mitigation on the measured correction functions}\label{sec:imgweight}

\begin{figure} 
    \centering
    \includegraphics[width=\columnwidth]{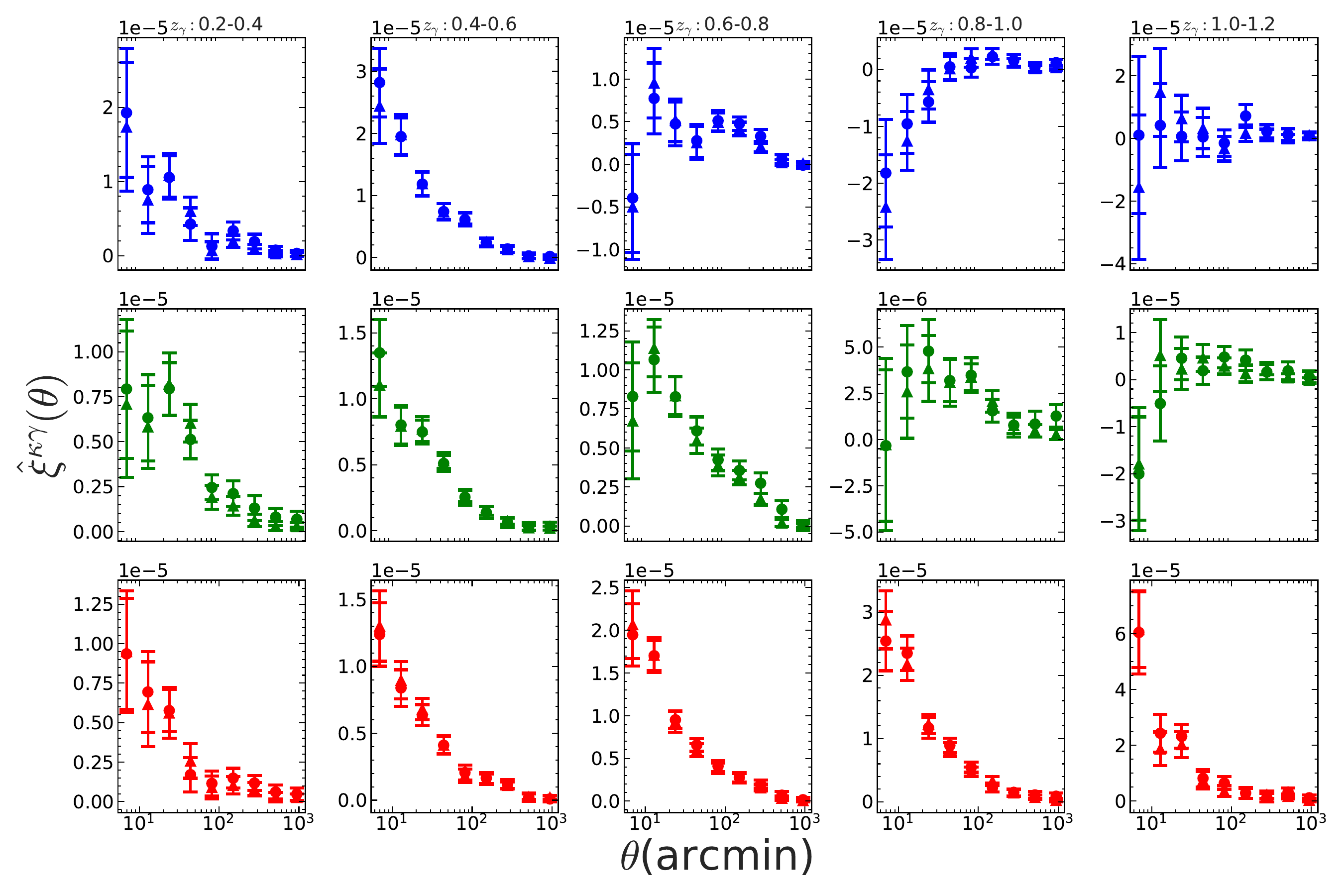}
    \caption{Comparison of the measured convergence-shear cross-correlation before and after the imaging systematics mitigation.
    The panels display the results for the three photo-$z$ bins of the convergence from top to bottom, 
    and for the five photo-$z$ bins of the shear from left to right.
    Triangular and circular points indicate measurements before and after addressing imaging systematics, respectively.
    The differences are too small to be visible in most cases.
    The convergence-shear cross-correlation is less affected by the imaging systematics than the convergence-convergence correlation.
    \label{fig:xi_kg_imgW}}
    \end{figure}

\reffig{fig:xi_kg_imgW} shows the measured convergence-shear cross-correlation function before and after mitigating imaging systematics.
The impact of this mitigation on the convergence-shear cross-correlation is minor.
The reconstruction of the convergence relies on galaxy luminosity, while the shear is measured from galaxy shapes.
As a result, they are sensitive to different imaging systematics, which explains the minor impact observed.

\begin{figure}
\centering
\includegraphics[width=0.51\columnwidth]{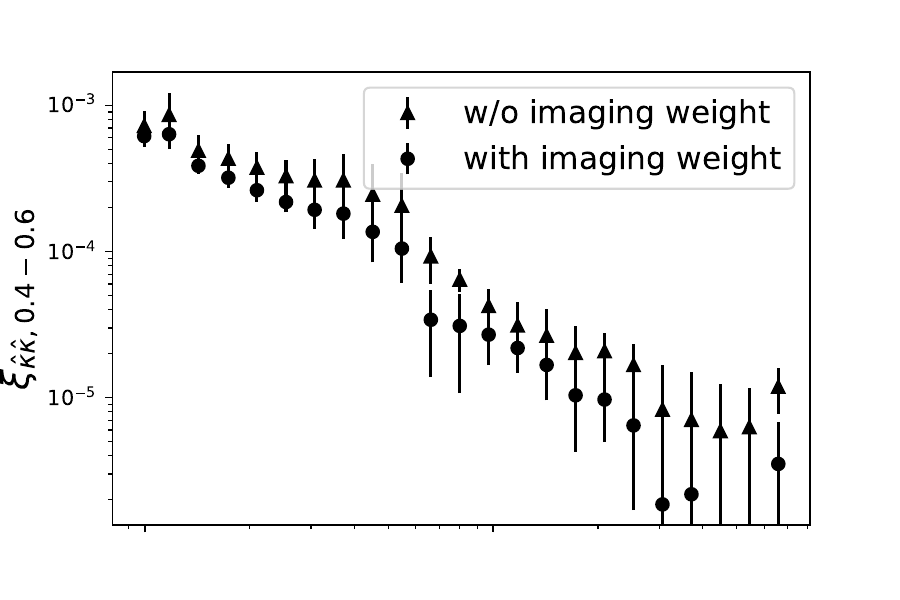}
\hspace{-0.45cm}
\includegraphics[width=0.51\columnwidth]
{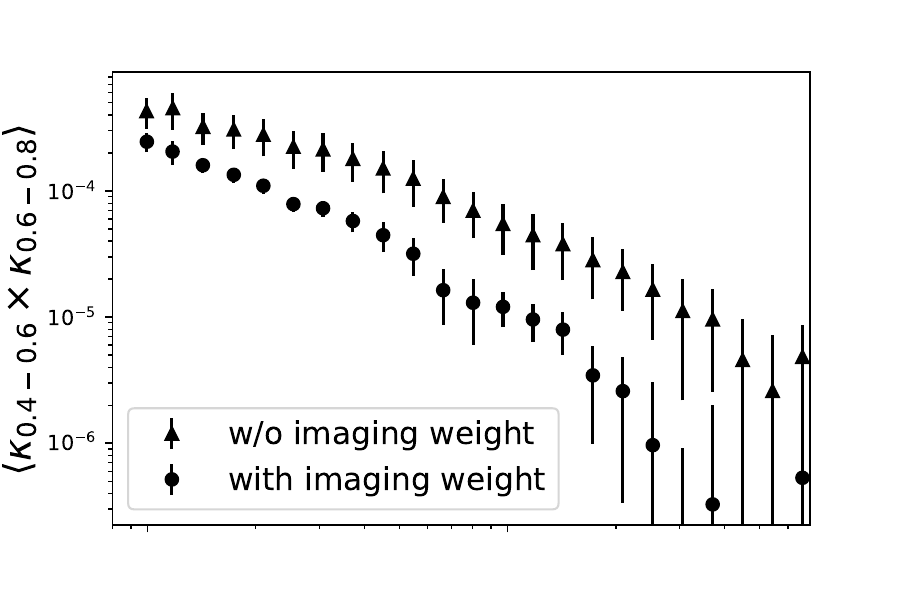}

\vspace{-0.35cm}
\centering
\vspace{-0.35cm}
\includegraphics[width=0.51\columnwidth]{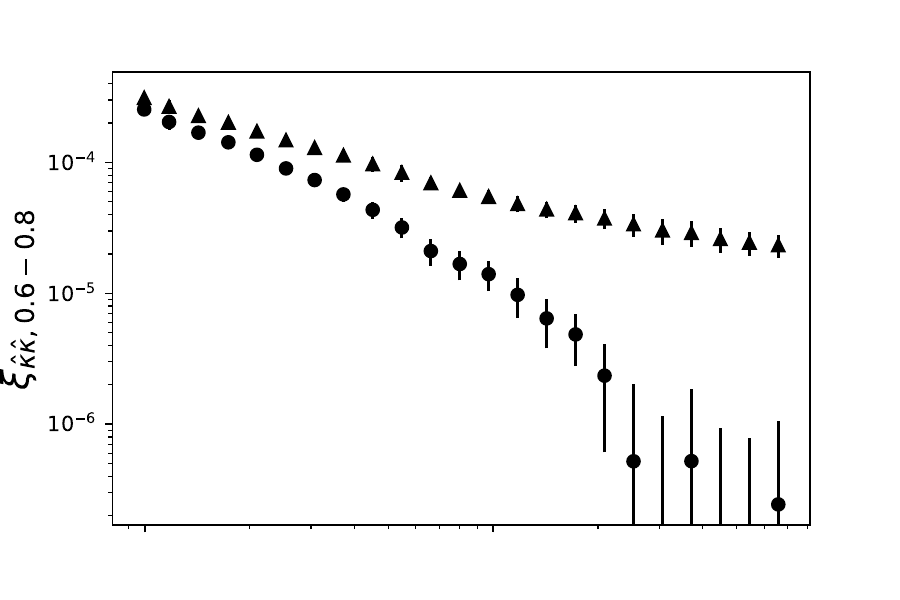}
\hspace{-0.45cm}
\includegraphics[width=0.51\columnwidth]{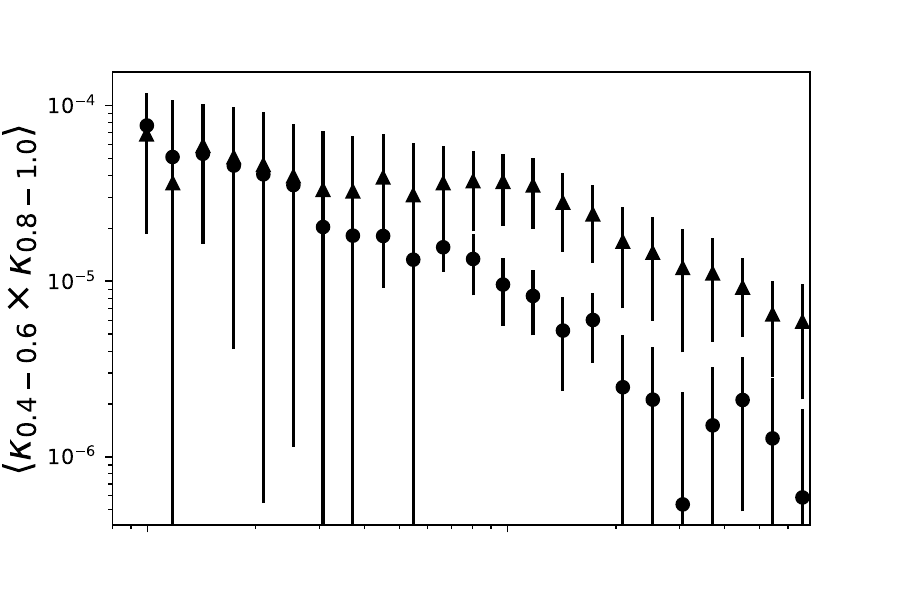}

\vspace{-0.35cm}
\centering
\vspace{-0.35cm}
\includegraphics[width=0.51\columnwidth]{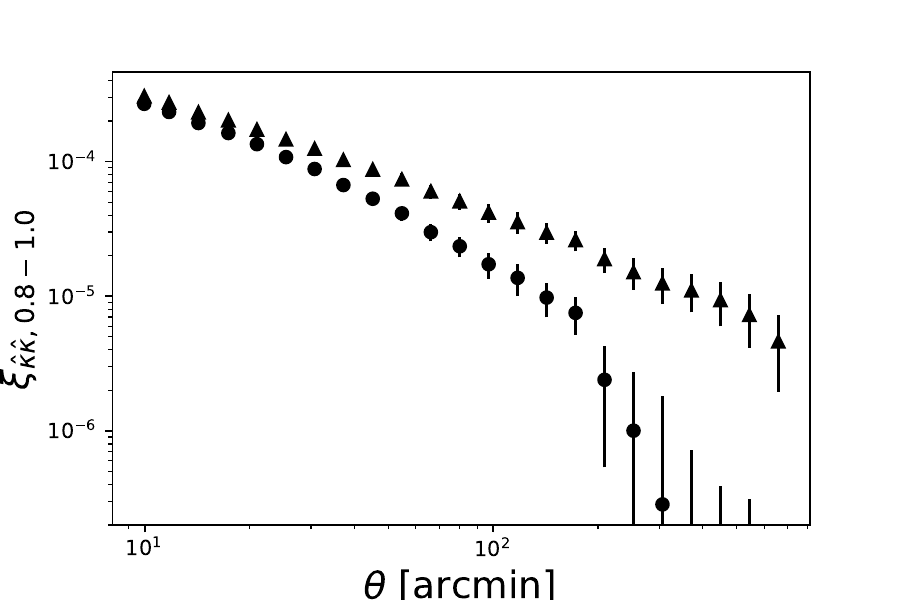}
\hspace{-0.45cm}
\includegraphics[width=0.51\columnwidth]{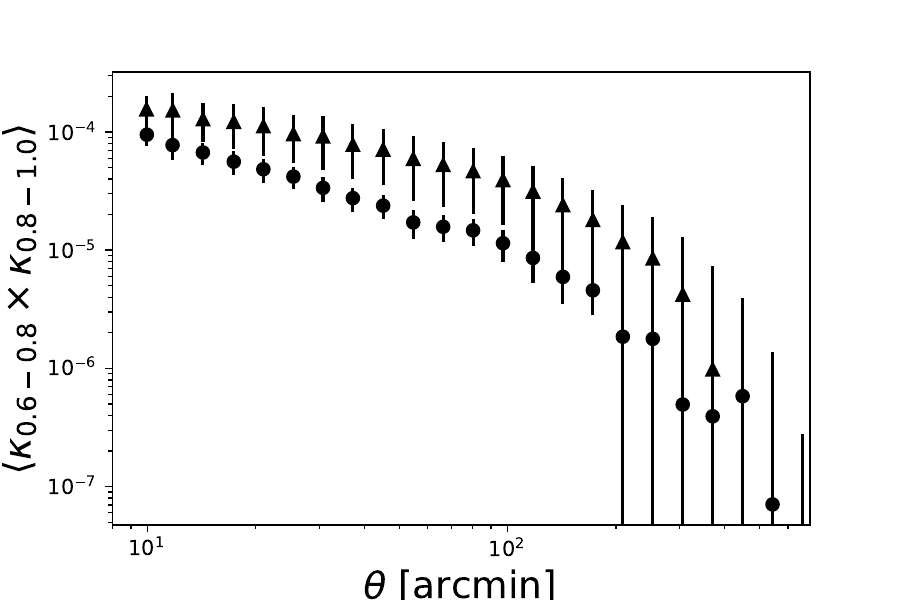}
\caption{The convergence-convergence correlation measurements of the reconstructed lensing convergence, before and after the imaging systematics mitigation.
The left/right panels show the auto-/cross-correlation, while  panels from top to bottom display the results for different photo-$z$ bins. 
The errors of the measurements are estimated by the jackknife method.
The triangular/circular data points are the measurements before/after the imaging systematics mitigation.
The imaging systematics mitigation significantly reduces the amplitude of the measurements.
}
\label{fig:clkk_imgW}
\end{figure}

The convergence-convergence corrections are shown in \reffig{fig:clkk_imgW}.
In all cases, mitigating imaging systematics significantly reduces the amplitude of the measurements.
The galaxy luminosity at two photo-$z$ bins is sensitive to the same imaging systematics.
Therefore, unlike the convergence-shear correlation, the cross-correlation between these bins cannot efficiently eliminate the systematics.
The discrepancy between the measurements and the predictions is significantly reduced after mitigation (\reffig{fig:clkk} and \reffig{fig:clk1k2}).

\section{Constraints from ignoring stochastic terms}\label{sec:approx_model}
\begin{figure} 
\includegraphics[width=\columnwidth]{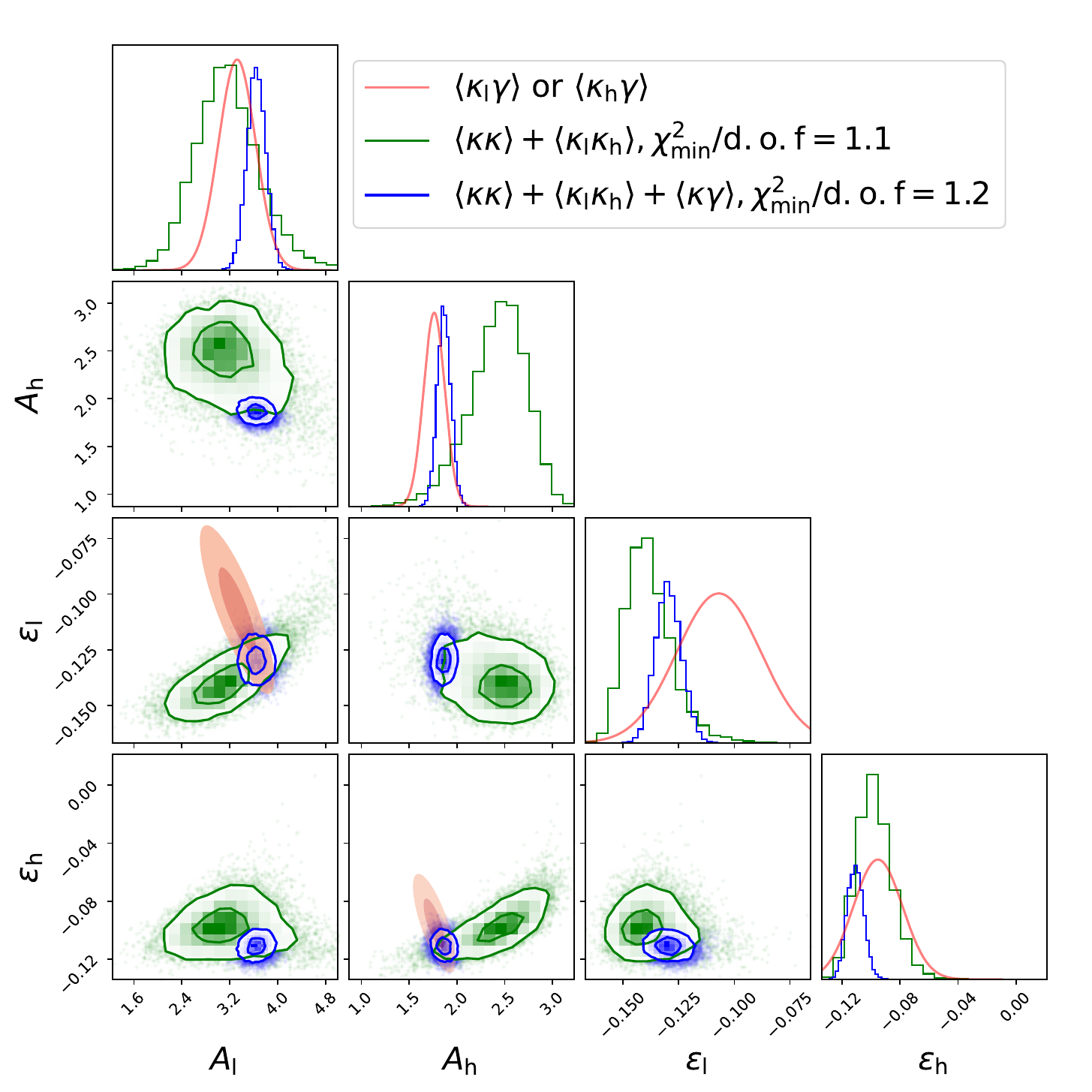}
\caption{Results of the convergence-convergence correlation analysis for $0.4 < z_\kappa < 0.6$ and $0.6 < z_\kappa < 0.8$. 
The model for the convergence-convergence correlation is approximated by the deterministic terms.
Parameters $A$ and $\epsilon$ are constrained by jointly fitting the auto- and cross-correlations of the two convergence maps, 
shown by green contours.
The indices $l$ and $h$ denote the photo-$z$ bins $0.4 < z_\kappa < 0.6$ and $0.6 < z_\kappa < 0.8$, respectively.
As a comparison, the red contours show the fitting results from the convergence-shear cross-correlation analysis (see Sec.\ref{sec:result-kap-shear}, where only $A_l-\epsilon_l$ and $A_h-\epsilon_h$ planes are obtained).
The $\lesssim 2\sigma$ discrepancy between the two analyses can be attributed to the approximations in modeling the convergence-convergence correlations.
The blue contours shows the constraints obtained by combining the convergence-convergence and convergence-shear cross-correlations.
The figure of merit improves compared to analyzing convergence-shear or convergence-convergence alone.
The two convergence maps are selected for this analysis, as their convergence-convergence correlations are in good agreement with the predictions from the deterministic terms (see \reffig{fig:clkk} and \reffig{fig:clk1k2}). 
}
\label{fig:mcmc}
\end{figure}

For the lensing convergence reconstructed at $0.4 < z_{\kappa} < 0.6$ and $0.6 < z_{\kappa} < 0.8$, the measured convergence-convergence correlations align well with the predictions. This suggests that, for these redshift bins, the modeling of the convergence-convergence correlations can be approximated using deterministic terms (see Eq.\eqref{eq:xik1k2_decom})
\be\label{eq:xik1k2_decom_appro}
{\xi}^{\kappa\kappa}_{ij,\rm th}\approx
\xi_{ij}^{\rm D}\ .
\ee
Using this approximation, we can establish constraints on the parameters $A$ and $\epsilon$ based solely on convergence-convergence correlations.
These constraints can then be compared with results from the convergence-shear cross-correlation analysis.
The $\chi^2$ for jointly fitting the convergence-convergence (auto- and cross-) correlations of the two convergence maps is approximated by:
\begin{equation}
\label{eq:chi_1}
\begin{aligned}
    \chi^2 &\approx \sum_{i\geq j}\sum\limits_{j\alpha\beta}
    \big[\hat{\xi}^{\kappa\kappa}_{ij}(\theta_\alpha)-{\xi}_{ij,\rm th}^{\kappa\kappa}(\theta_\alpha)\big] \times \textbf{Cov}_{ij}^{-1} \\ 
    &\times \big[\hat{\xi}^{\kappa\kappa}_{ij}(\theta_\beta)-{\xi}_{ij,\rm th}^{\kappa\kappa}(\theta_\beta)\big] \ .
\end{aligned}
\end{equation}
Here the summation over indices $i,j$ contains three terms: $i=j=l$, $i=j=h$, and $i=l,j=h$,
corresponding to the auto-correlation of the two convergence maps at $0.4 < z_{\kappa} < 0.6\ (l)$ and $0.6 < z_{\kappa} < 0.8\ (h)$, and their cross-correlation. 
The likelihood for this case is non-Gaussian. 
To obtain constraints on $A_l,A_h,\epsilon_l,\epsilon_h$, we use the publicly available code \texttt{emcee} \citep{emcee} to sample the posterior distribution of the parameters $A$ and $\epsilon$.
The results of the posterior and the comparison with constraints from the convergence-shear cross-correlation analysis are shown in \reffig{fig:mcmc}.

For $A_l$ and $\epsilon_h$, the discrepancy in constraints between the two analyses is $\lesssim 1\sigma$, whereas for $A_h$ and $\epsilon_l$, the discrepancy is $\lesssim 2\sigma$.
The parameter degeneracy in the $A_l-\epsilon_h$ and $A_h-\epsilon_l$ planes differs in direction between the two analyses.
This indicates that a joint fitting of the convergence-convergence and convergence-shear correlations can break these degeneracies.
The $\chi^2$ for the joint fitting is approximated by:
\begin{equation}
\begin{aligned}
\chi^2 &\approx \sum_{i\geq j}\sum\limits_{j\alpha\beta}
    \big[\hat{\xi}^{\kappa\kappa}_{ij}(\theta_\alpha)-{\xi}_{ij,\rm th}^{\kappa\kappa}(\theta_\alpha)\big]
    \times \textbf{Cov}_{ij}^{-1} \\
    &\times \big[\hat{\xi}^{\kappa\kappa}_{ij}(\theta_\beta)-{\xi}_{ij,\rm th}^{\kappa\kappa}(\theta_\beta)\big] \\ 
    &+ \sum\limits_{ij\alpha\beta}
    \big[\hat{\xi}^{\kappa\gamma}_{ij}(\theta_\alpha)-{\xi}_{ij,\rm th}(\theta_\alpha)\big]
    \times \textbf{Cov}_{ij}^{-1} \\
    &\times \big[\hat{\xi}^{\kappa\gamma}_{ij}(\theta_\beta)-{\xi}_{ij,\rm th}(\theta_\beta)\big] \ .
\end{aligned}
\end{equation}
In the second term of the $\chi^2$, the summation over index $i$ encompasses the two convergence maps, while the sum over $j$ covers the shear photo-$z$ bins.
As presented in \reffig{fig:mcmc}, the degeneracy is reduced, and the figure of merit shows improvement compared to analyzing convergence-shear or convergence-convergence alone.
This highlights the potential of combining magnification and shear measurements to enhance cosmological constraints.
Note that throughout this work, we have neglected correlations between $\xi^{\kappa\kappa}_{ij}$ or $\xi^{\kappa\gamma}_{ij}$ of different convergence or shear redshifts ($i_1\neq i_2, j_1\neq j_2$) arising from four-point correlations.
For the current data, these correlations are negligible compared to the shape measurement error in $\gamma$ and the shot noise in $\kappa$.

\label{lastpage}
\end{document}